\documentclass[%
 reprint, 
 superscriptaddress,
 amsmath,amssymb,
 aps, 
 prapplied,
 final,
]{revtex4-2}

\usepackage{graphicx}
\usepackage{dcolumn}
\usepackage{bm}
\usepackage{hyperref}

\usepackage{siunitx}
\DeclareSIUnit\corehour{\text{core-hours}}
\DeclareSIUnit[quantity-product = ]\percent{\char`\%}

\usepackage[nameinlink,capitalize]{cleveref}

\usepackage{upgreek}
\usepackage{nicematrix}
\usepackage{derivative} 
\usepackage{mathtools} 
\usepackage[utopia]{mathdesign} 
\usepackage{xspace}

\DeclareMathAlphabet{\mathoms}{OMS}{cmsy}{m}{n}

\newcommand{\mymatht}[1]{\mbox{\ensuremath{#1}}}
\newcommand{\mymathtv}[1]{\ensuremath{#1}}

\newcommand{\myOrdernum}[1]{\ensuremath{\mathoms{O}(\num{#1})}}
\newcommand{\myOrderqty}[2]{\ensuremath{\mathoms{O}(\num{#1})\,\unit{#2}}}

\newcommand{\myincrement}{\ensuremath{\Updelta}}
\newcommand{\myvect}[1]{\ensuremath{\boldsymbol{\mathrm{#1}}}}

\newcommand{\mymaxp}[1]{\ensuremath{\operatorname*{max}{#1}}}

\newcommand{\myintcramp}[2]{\smashoperator{\int\limits_{#1}^{#2}}}

\newcommand{\myiintcramp}[4]{\smashoperator{\int\limits_{#1}^{#2}}\smashoperator{\int\limits_{#3}^{#4}}}

\newcommand{\mysum}[2]{\sum\limits_{#1}^{#2}}
\newcommand{\mysuum}[4]{\sum\limits_{#1}^{#2}\sum\limits_{#3}^{#4}}

\newcommand{\myodd}[2][]{\ensuremath{\odif[#1]{#2}}}
\newcommand{\myodv}[3][]{\odv[#1]{#2}{#3}}
\newcommand{\myodvt}[2]{\mymatht{\mathrm{d}_{#2}{#1}}}
\newcommand{\mypdv}[3][]{\pdv[#1]{#2}{#3}}

\newcommand{\mypdvvt}[3]{\mymatht{\partial_{#2}\partial_{#3}{#1}}}

\newcommand{\myhadpower}[1]{\ensuremath{\circ{#1}}}

\newcommand{\myNat}{\mathbb{N}}
\newcommand{\myReal}{\mathbb{R}}
\newcommand{\myNatl}[1]{\mathbb{N}_{#1}}
\newcommand{\myReall}[1]{\mathbb{R}_{#1}}
\newcommand{\myNatu}[1]{\mathbb{N}^{#1}}
\newcommand{\myRealu}[1]{\mathbb{R}^{#1}}

\newcommand{\myRealul}[2]{\mathbb{R}^{#1}_{#2}}

\newcommand{\mycode}[1]{\texttt{#1}}
\newcommand{\myversion}[1]{{#1}}

\newcommand{\myonlinecite}[1]{Ref.~\onlinecite{#1}}
\newcommand{\myonlinecitepl}[1]{Refs.~\onlinecite{#1}}

\begin{document}

\title{Full-spectrum modeling of mobile gamma-ray spectrometry systems in scattering media
}%

\author{David Breitenmoser}
\email{Lead and contact author: david.breitenmoser@psi.ch}
\affiliation{%
Department of Radiation Safety and Security, Paul~Scherrer~Institute (PSI), Forschungsstrasse~111, 5232 Villigen~PSI, Switzerland
}%
\affiliation{%
Department of Nuclear Engineering \& Radiological Sciences, University of Michigan, 2355 Bonisteel Blvd., Ann Arbor, MI 48109-2104, United States of America
}%
\author{Alberto Stabilini}%
\author{Malgorzata Magdalena Kasprzak}%
\author{Sabine Mayer}%
\affiliation{%
Department of Radiation Safety and Security, Paul~Scherrer~Institute (PSI), Forschungsstrasse~111, 5232 Villigen~PSI, Switzerland
}%

\date{\today}

\begin{abstract}

Mobile gamma-ray spectrometry (MGRS) systems are essential for localizing, identifying, and quantifying gamma-ray sources in complex environments. Full-spectrum template matching offers the highest accuracy and sensitivity for these tasks but is limited by the computational cost of generating the required spectral templates. Here, we present a generalized full-spectrum modeling framework for MGRS systems in scattering media, enabling near real-time template generation through dynamic, anisotropic instrument response functions. Benchmarked against high-fidelity brute-force Monte Carlo simulations, our method yields a computational speedup by a factor of \myOrdernum{1d7}, while achieving comparable accuracy with median spectral deviations below \qty{6}{\percent}. The methodology presented is platform-agnostic and applicable across marine, terrestrial, and airborne domains, unlocking new capabilities for MGRS in a variety of applications, such as environmental monitoring, geophysical exploration, nuclear safeguards, and radiological emergency response.

\end{abstract}

\maketitle
\newpage

\section{\label{sec:Introduction}Introduction}

By deploying gamma-ray spectrometers on ground-based, airborne, marine, and spaceborne platforms, mobile gamma-ray spectrometry (MGRS) enables the localization, identification, and quantification of gamma-ray sources across a diverse range of environments, from Earth’s oceans to outer space. These systems support a broad spectrum of applications, from environmental monitoring in marine ecosystems \cite{Jones2001a,Lee2019a,Lee2023c} and mapping radioactive contamination after nuclear accidents \cite{Rosenthal1991a,Drovnikov1997a,Lyons2012,Torii2013a,Sanada2014a,Sanada2015a,Nishizawa2016a,PradeepKumar2020b} to source localization in nuclear security and nonproliferation contexts \cite{Deal1972a,Prieto2020a,Hellfeld2021a,Bandstra2021a,Curtis2020a}. They have also been used to investigate terrestrial gamma-ray flashes \cite{Fishman1994,Briggs2010,Tavani2011,Smith2011a,Gjesteland2015,Neubert2020a}, track atmospheric radionuclide transport \cite{Appleton2008a,Sinclair2011a,Baldoncini2017a}, and remotely analyze the elemental composition of planetary-mass objects and small solar system bodies \cite{Prettyman2006a,Hahn2007a,Kobayashi2010a,Prettyman2011a,Peplowski2011b,Peplowski2016c,Prettyman2017,Prettyman2019a}. Recent advancements in unmanned mobile platforms, exemplified by the successful flight of the autonomous helicopter \textit{Ingenuity} on Mars in 2021 \cite{Balaram2021,Tzanetos2022,Grip2022}, have prompted renewed efforts to advance MGRS technologies, opening new possibilities to deploy these systems in previously inaccessible or difficult-to-reach environments.

In contrast to laboratory and \textit{in-situ} gamma-ray spectroscopy settings, most MGRS systems operate in a near real-time acquisition bin mode, enabling the spatial localization of gamma-ray sources with typical sampling frequencies of the order of \myOrderqty{1}{\Hz}. Furthermore, characteristic source-detector distances in MGRS applications are significantly larger than those encountered in stationary laboratory or \textit{in-situ} measurements, typically ranging from \myOrderqty{1d2}{\m} in terrestrial surveys to at least \myOrderqty{1d5}{\m} for spaceborne platforms. As a result, gamma-ray spectra acquired during single-pass MGRS surveys are inherently sparse, with the number of counts per pulse-height channel typically of the order of \myOrdernum{1d1}, leading to increased statistical uncertainties and requiring dedicated data reduction techniques to extract the spectral information encoded in the recorded pulse-height spectra.

For that purpose, template-matching techniques, commonly referred to as full-spectrum analysis (FSA) algorithms, have proven very effective for both terrestrial \cite{Grasty1985a,Minty1998d,Hendriks2001a,Bare2011a,Caciolli2012a} and space-based MGRS systems \cite{Prettyman2006a,Prettyman2011a,Peplowski2016c}, offering significantly improved accuracy and sensitivity compared to traditional peak-fitting methods typically employed in stationary gamma-ray spectroscopy. In FSA, the expected gamma-ray event rates \mymatht{\dot{\myvect{N}}\in\myRealul{N_{E'}}{\geq0}} for a given number of spectral channels \mymatht{N_{E'}\in\myNatl{\geq1}} are modeled as a superposition of spectral templates \mymatht{\myvect{\uppsi}\in\myRealul{N_{E'}}{\geq0}}, each representing the characteristic full-spectrum response of the MGRS system to a specific gamma-ray source---normalized by the source strength---at a given time \mymatht{t\in\myReal} during the survey \cite{Paradis2020a,Andre2021a}:

\begin{equation}
    \dot{\myvect{N}}\left(t\right) = %
    \mysum{s \in \mathfrak{I}\left(t\right)}{}a_{s}\left(t\right)\myvect{\uppsi}_{s}\left(t\right)\label{eq:FSA}%
\end{equation}

\noindent with \mymatht{\mathfrak{I} = \{ s \in \myNat \mid 1 \leq s \leq N_{\text{src}}, N_{\text{src}} \in \myNat \}} denoting the set of gamma-ray sources with associated source strengths \mymatht{a_s\in\myReall{\geq0}} giving rise to the expected gamma-ray event rates at time \mymatht{t}. 

Consequently, to reliably infer the source properties from MGRS measurements, specifically the source sets \mymatht{\mathfrak{I}} and associated source strengths \mymatht{\myvect{a}\in\myNatu{N_{\text{src}}}}, accurate predictions of the spectral templates \mymatht{\myvect{\uppsi}} for a given MGRS survey are needed. Traditionally, empirical radiation measurements using dedicated calibration sources have been employed for this purpose \cite{Dickson1981a,Minty1990a,Grasty1991a,Hendriks2001a}. While straightforward to conduct, these methods are limited to a small set of gamma-ray emitting radionuclides, often require days of measurement time to account for all relevant source-detector geometries, and can involve significant costs for source procurement and decommissioning. Over the past two decades, with the increasing computational capabilities and advancements in high-fidelity Monte Carlo radiation transport codes \cite{Allison2016,Goorley2016,Ahdida2022a,Sato2024}, simulation-based calibration has emerged as a more effective alternative. These codes enable accurate, high-fidelity estimates of spectral templates \mymatht{\myvect{\uppsi}} for arbitrary gamma-ray source sets and source-detector geometries, thereby overcoming the key limitations of empirical methods \cite{Prettyman2006a,Prettyman2011a,VanderGraaf2011a,Kulisek2018a,Androulakaki2016b,Androulakaki2016b,Peplowski2016c,Breitenmoser2025a}.

While numerical derivation of spectral templates has proven superior to traditional empirical methods, performing Monte Carlo simulations for MGRS systems remains computationally challenging due to the complexity of the model and high computational cost. These difficulties arise mainly from the large simulation domain discussed above and the need for an accurate mass model of the MGRS platform. As a result, achieving the precision needed for MGRS applications typically requires \ensuremath{\Updelta{t}_{\mathrm{MC}}=\myOrderqty{1d4}{\corehour}} for a single spectral template \mymatht{\myvect{\uppsi}} on a modern computer cluster \cite{Breitenmoser2025a}.

Depending on the environment and the spatial discretization of extended, inhomogeneous gamma-ray sources, the number of superposed source signatures \mymatht{N_{\text{src}}} ranges from \mymatht{\myOrdernum{1d1}} for background surveys of uniform gamma-ray sources to \mymatht{\myOrdernum{1d5}} for high-resolution contamination mapping in radiological emergency response scenarios. In addition, the dynamic source-detector geometry, combined with variations in both the MGRS platform state and the surrounding material composition, requires computing spectral templates for each recorded pulse-height spectrum. Given that typical MGRS surveys record at least \mymatht{N_{\mathrm{spec}}\geq\myOrdernum{1d4}} spectra, the total computational cost of evaluating such a survey via brute-force Monte Carlo simulations scales as \mymatht{N_{\mathrm{src}}N_{\mathrm{spec}}\Updelta{t}_{\mathrm{MC}}}, yielding characteristic runtimes between \ensuremath{\myOrderqty{1d9}{\corehour}} and \ensuremath{\myOrderqty{1d13}{\corehour}}, depending on the application.

Given these numbers, it is evident that deriving spectral templates for MGRS surveys using brute-force Monte Carlo simulations is computationally prohibitive with the current infrastructure. To mitigate this, previous investigators proposed a two-stage decomposition of the computation \cite{Reedy1973a,Knoll2010a,Kluson2010c,Kaastra2016a}: first, the gamma-ray transport through the environment, and second, the interaction of the propagated gamma rays within the MGRS platform, followed by their energy deposition and subsequent signal processing in the gamma-ray spectrometer. For MGRS, the latter stage constitutes the most computationally intensive part of the simulation. By precomputing this second stage, the two-stage approach avoids redundant high-cost computations, leading to substantial reductions in computational expenses. In this framework, the differential spectral template \myodvt{\psi}{E'} is formally expressed as the convolution of the source-strength-normalized differential gamma-ray flux \myodvt{\phi_{\gamma}}{E_{\gamma}} at the MGRS system's location at time \mymatht{t} (obtained from the first-stage computations) with the differential instrument response function (IRF) \myodvt{R}{E'}, which results from the second-stage computations and characterizes the system's intrinsic response to incident gamma rays:

\begin{equation}
    \myodv{\psi}{E'}\left(E',t\right) = %
    \myintcramp{0}{\infty}\myodv{R}{E'}\left(E',E_{\gamma}\right)%
    \myodv{\phi_{\gamma} }{E_{\gamma}}\left(E_{\gamma},t\right)%
    \myodd{E_{\gamma}}\label{eq:IRF0}%
\end{equation}

\noindent with \mymatht{E_{\gamma}} and \mymatht{E'} denoting the gamma-ray energy and spectral energy related to the pulse-height in the gamma-ray spectrometer, respectively. 

In \cref{eq:IRF0}, the differential IRF is assumed to be static, which is a reasonable approximation for laboratory-based spectroscopy with fixed source-detector geometries. However, this assumption generally does not hold for MGRS platforms. Since the IRF in MGRS is typically highly anisotropic due to the platform's complex geometry and inhomogeneous composition, the continuous variation in the source-detector geometry during a survey results in a dynamic instrument response. Additionally, the IRF itself may evolve over time due to changes in platform state variables, such as fuel depletion, relocation of crew members, and the extension of gears or solar sails, among other factors. If unaccounted for, these dynamic effects can introduce substantial biases in spectral template predictions.

To mitigate these biases, prior studies have extended the detector response model in \cref{eq:IRF0} for spaceborne applications by leveraging the analytically tractable gamma-ray transport in vacuum in conjunction with anisotropic IRFs \cite{Prettyman2006a,Kobayashi2010a,Peplowski2011b,Prettyman2011a,Prettyman2017,Peplowski2016c,Prettyman2019a}. While these protocols have proven effective for spaceborne MGRS, they are not directly applicable to terrestrial MGRS systems, where the presence of scattering media---such as water in marine applications or air in ground-based and airborne surveys---modulate the gamma-ray field. Additionally, since terrestrial MGRS platforms typically have much shorter endurance than robotic space probes, dynamic evolution of the IRF itself during a single-run survey can be much more substantial, making dynamic modeling of the IRF essential. Because of these additional model complexities, the accurate and numerically efficient generation of spectral templates \mymatht{\myvect{\uppsi}} for terrestrial MGRS remains an open problem. 

In this work, we introduce a generalized full-spectrum modeling framework for MGRS systems in scattering media. The capabilities of this framework are demonstrated through its implementation for a real-world airborne MGRS system. By leveraging multithreaded evaluation via vectorized convolution operations, combined with precomputed IRFs and double-differential gamma-ray flux banks, we achieve near real-time spectral template generation with computation times of \myOrderqty{1}{\s} on a local workstation. We benchmark the methodology against high-fidelity brute-force Monte Carlo simulations, revealing excellent accuracy with median spectral deviations of less than \qty{6}{\percent}, translating to at least a twofold improvement over conventional isotropic models. Finally, we leverage the dynamic modeling approach to analyze temporal dynamics in the IRF, providing conservative constraints on expected IRF variations during MGRS surveys with crewed airborne platforms.

The remainder of this paper is organized as follows. \cref{sec:IRFtheory} introduces the methodology for generating spectral templates using dynamic anisotropic IRFs. In \cref{sec:Implementation}, we discuss its implementation for a real-world MGRS system. \cref{sec:Results} examines the spectral and angular dispersion of the modeled MGRS system, as uncovered by the computed IRFs, and investigates the expected dynamic IRF variations during an MGRS survey. In this section, we also present the results of a benchmark study comparing the methodology against high-fidelity brute-force Monte Carlo simulations. Finally, \cref{sec:Discussion} summarizes the key findings and explores their implications for MGRS, along with future research directions. Additional details on specific aspects of our work are provided in \cref{app:SwissAGRS,app:UQ,app:Coordinates}.

\section{\label{sec:IRFtheory}Spectral template generation using dynamic anisotropic instrument response functions}

To account for the anisotropy and the dynamics in both the IRF and the gamma-ray field, we generalize the differential spectral template computation in \cref{eq:IRF0} as the convolution of the differential dynamic anisotropic IRF \mymatht{\myodvt{R}{E'}(E',E_{\gamma},\myvect{\Upomega}',t)} and the source-strength-normalized double-differential gamma-ray flux \mymatht{\mypdvvt{\phi_{\gamma}}{E_{\gamma}}{\varOmega'}(E_{\gamma},\myvect{\Upomega}',t)}:

\begin{align}
     \myodv{\psi}{E'}\left(E',t\right) =& %
     \myiintcramp{0}{\infty}{0}{4\pi}\myodv{R}{E'}\left(E',E_{\gamma},\myvect{\Upomega}',t\right)\nonumber\\%
     &\times\mypdv{\phi_{\gamma} }{E_{\gamma},\varOmega'}\left(E_{\gamma},\myvect{\Upomega}',t\right)%
      \myodd{\varOmega',E_{\gamma}}\label{eq:IRFc}%
\end{align}

\noindent with \mymatht{\varOmega'} and \mymatht{\myvect{\Upomega}'} denoting the solid angle and the direction unit vector in the local noninertial platform-fixed coordinate system, respectively. In \cref{eq:IRFc}, both the differential IRF and the double-differential gamma-ray flux are continuous functions. However, due to the complex geometry and heterogeneous composition of MGRS platforms, the surrounding media, as well as the gamma-ray sources themselves, no closed-form solution generally exists for either function. Consequently, numerical approximations using dedicated radiation transport codes are required.

Building upon the methodology outlined in previous studies \cite{Prettyman2006a,Kobayashi2010a,Peplowski2011b,Peplowski2016c,Prettyman2017,Prettyman2019a}, we use dedicated Monte Carlo simulations to evaluate the dynamic anisotropic IRF on a discrete grid of spectral energies \mymatht{\{ E'_{i} \mid i \in \myNatl{\geq1}, i \leq N_{E'} \}}, corresponding to the center pulse heights of the spectrometers' spectral channels, gamma-ray energies \mymatht{\{E_{\gamma,j} \mid j\in\myNatl{\geq1},j\leq N_{E_{\gamma}}\}}, directions \mymatht{\{\myvect{\Upomega}'_{k} \mid k\in\myNatl{\geq1},k\leq N_{\myvect{\Upomega}'}\}}, and time instances \mymatht{\{t_l \mid l \in \myNatl{\geq1},l \leq N_{t}\}}, where \mymatht{N_{E'}}, \mymatht{N_{E_{\gamma}}}, \mymatht{N_{\myvect{\Upomega}'}}, and \mymatht{N_{t}} represent the number of pulse-height channels, evaluated gamma-ray energies, directions, and time instances, respectively. Simultaneously, the double-differential gamma-ray flux \mypdvvt{\phi_{\gamma}}{E_{\gamma}}{\varOmega'} is evaluated on the same discrete set of gamma-ray energies, directions, and time instances as the dynamic anisotropic IRF. This can be accomplished using standard built-in functionalities available in both deterministic \cite{Sjoden1996,Wareing1996,Alcouffe2001} and Monte Carlo based \cite{Allison2016,Goorley2016,Ahdida2022a,Sato2024} numerical codes. Following a quadrature approach, we rewrite \cref{eq:IRFc} in its discretized form as

\begin{align}
     \psi\left(E'_i,t_l\right) \approx&%
     \mysuum{j=1\vphantom{k=1}}{N_{E_{\gamma}}\vphantom{N_{\myvect{\Upomega}'}}}{k=1\vphantom{j=1}}{N_{\myvect{\Upomega}'}\vphantom{N_{E_{\gamma}}}}%
     \myodv{R}{E'}\left(E'_i,E_{\gamma,j},\myvect{\Upomega}'_k,t_l\right)\myincrement{E'_{i}}\nonumber\\%
     &\times\mypdv{\phi_{\gamma}}{E_{\gamma},\varOmega'}\left(E_{\gamma,j},\myvect{\Upomega}'_k,t_l\right)%
     \myincrement{E_{\gamma,j}}\myincrement{\varOmega'_k}%
     \label{eq:IRFn}%
\end{align}

\noindent with \mymatht{\myincrement{E'_{i}}}, \mymatht{\myincrement{E_{\gamma,j}}}, and \mymatht{\myincrement{\varOmega'_k}} denoting the variable bin widths in spectral energy \mymatht{E'}, gamma-ray energy \mymatht{E_{\gamma}}, and solid angle \mymatht{\varOmega'}, respectively. 

To enable efficient spectral template computations using multithreaded evaluation, \cref{eq:IRFn} may be recast in matrix notation as

\begin{equation}
    \myvect{\uppsi}\left(t_l\right) \approx  
    \sum_{k=1}^{N_{\myvect{\Upomega}'}} \myvect{R}\left(\myvect{\Upomega}'_k, t_l\right) \myvect{\upphi}_{\gamma}\left(\myvect{\Upomega}'_k, t_l\right) \label{eq:IRFm}
\end{equation}

\noindent where \mymatht{\myvect{\uppsi} \in \myRealul{N_{E'}}{\geq 0}} denotes the spectral template vector at time \mymatht{t_l}. Each term in \cref{eq:IRFm} involves a direction- and time-specific instrument response matrix \mymathtv{\myvect{R}(\myvect{\Upomega}'_k, t_l) \in \myRealul{N_{E'} \times N_{E_{\gamma}}}{\geq 0}} and the corresponding gamma-ray flux vector \mymathtv{\myvect{\upphi}_{\gamma}(\myvect{\Upomega}'_k, t_l) \in \myRealul{N_{E_{\gamma}}}{\geq 0}}. The summation thus accumulates directional matrix-vector products at fixed time \mymatht{t_l}. The individual elements of the instrument response matrix are estimated by scaling the differential instrument response components with the corresponding spectral energy bin width, i.e., \mymathtv{R(E'_i,E_{\gamma,j},\myvect{\Upomega}'_k,t_l) \approx \myodvt{R}{E'}(E'_i,E_{\gamma,j}, \myvect{\Upomega}'_k, t_l)\myincrement{E'_i}}. Similarly, we compute the individual elements of the gamma-ray flux vector by scaling the double-differential gamma-ray flux components with the corresponding energy and solid angle bin widths: \mymathtv{\phi_{\gamma}(E_{\gamma,j}, \myvect{\Upomega}'_k, t_l) \approx \mypdvvt{\phi_{\gamma}}{E_{\gamma}}{\varOmega'}(E_{\gamma,j}, \myvect{\Upomega}'_k, t_l) \myincrement{E_{\gamma,j}}\myincrement{\varOmega'_k}}.

\section{\label{sec:Implementation}Implementation for a real-world MGRS system}

To demonstrate the capabilities of the proposed methodology, we describe here its implementation for a real-world MGRS system, detailing the IRF and gamma-ray flux computations as well as the numerical convolution operations. Given the diversity of MGRS platforms, we focus on a class exhibiting both strong anisotropy and dynamic instrument response, specifically crewed airborne MGRS systems \cite{Breitenmoser2024}. As a representative case, we selected the Swiss Airborne Gamma-Ray Spectrometry (SAGRS) system, for which a well-characterized and validated Monte Carlo mass model is available \cite{Breitenmoser2022e,Breitenmoser2025a}. For completeness, a brief overview of the SAGRS system is given in \cref{app:SwissAGRS}, with additional information available in \myonlinecite{Breitenmoser2025a}.

\subsection{\label{subsec:IRFmethod}Instrument response function}

\begin{figure}[b]
\includegraphics[width=8.6cm]{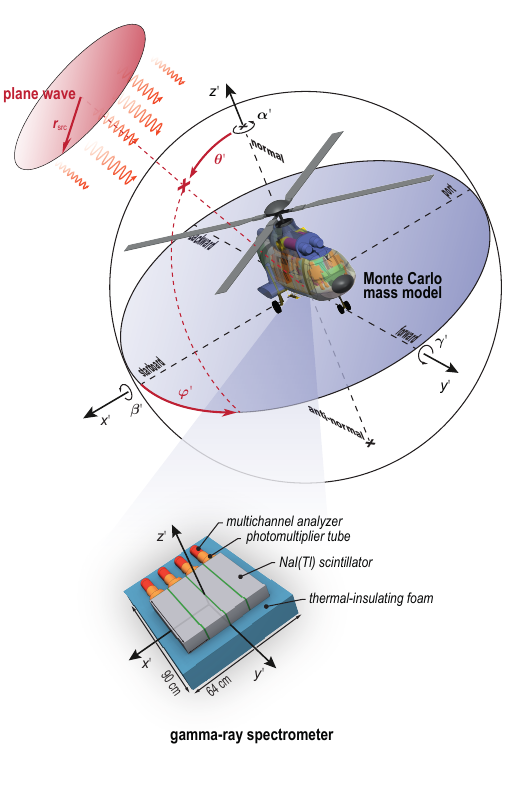}
\caption{\label{fig:IRFscheme}Graphical illustration of the configuration used to derive instrument response functions (IRFs) for the SAGRS system via Monte Carlo simulations. The main elements in the setup include the Monte Carlo mass model of the MGRS platform and the circular gamma-ray plane wave characterized by the source radius \mymatht{r_{\text{src}}} as well as the angular variables (\mymatht{\phi'}, \mymatht{\theta'}) in the local platform-fixed coordinate system. The local coordinate system is defined by the platform's principal axes \mymatht{x'}, \mymatht{y'}, \mymatht{z'} and parameterized by the Tait-Bryan angles \mymatht{\alpha'} (yaw), \mymatht{\beta'} (pitch), and \mymatht{\gamma'} (roll) relative to the global inertial reference frame. The inset highlights the mass model of the gamma-ray spectrometer integrated into the SAGRS system, with the local frame's origin positioned at the scintillator array's center of mass. All mass model visualizations were generated using the graphical interface \mycode{FLAIR} \citep{Vlachoudis2009a} for the SAGRS Monte Carlo model derived in our previous work \citep{Breitenmoser2022e,Breitenmoser2025a}. For improved visibility and interpretability, semitransparent false colors were applied.}
\end{figure}

Similar to previous work on spaceborne MGRS systems \citep{Prettyman2006a,Kobayashi2010a,Peplowski2011b,Prettyman2011a} as well as gamma-ray space telescopes \citep{Calura2000,Chen2013a,Ackermann2012a}, we estimate the dynamic anisotropic IRF using a series of dedicated Monte Carlo simulations. These simulations follow the standard far-field approximation, assuming a homogeneous double-differential photon flux across the characteristic volume of the mobile platform. Under this assumption, the IRF can be estimated by simulating the platform’s response to a circular monoenergetic gamma-ray plane wave for a predefined set of gamma-ray energies \mymatht{E_{\gamma}} and incident wave directions \mymatht{\myvect{\Upomega}'}, parameterized by the azimuth \mymatht{\phi'} and polar angle \mymatht{\theta'} in the local noninertial platform-fixed coordinate system, as schematically depicted in \cref{fig:IRFscheme}. For each configuration, the number of events \mymatht{N_{\text{event}}} registered in the spectral channel \mymatht{\myincrement{E'_i}} for a given number of simulated source primaries \mymatht{N_{\text{src}}} is used to estimate the expected instrument response as

\begin{equation}
R\left(E'_i,E_{\gamma,j},\myvect{\Upomega}'_k,\myvect{\xi}_l\right) \approx%
\pi r_{\text{src}}^2\frac{N_{\text{event}}\left(E'_i,E_{\gamma,j},\myvect{\Upomega}'_k,\myvect{\xi}_l\right)}{ N_{\text{prim}}\left(E_{\gamma,j},\myvect{\Upomega}'_k,\myvect{\xi}_l\right)}\label{eq:IRFmc}\end{equation}

\noindent where \mymatht{\myvect{\xi}\in\myRealu{N_{\xi}}} with \mymatht{N_{\xi}\in\myNat} represents the set of state variables required to describe the temporal evolution of the IRF as \mymathtv{t \mapsto \myvect{\xi}(t)}, while \mymatht{r_{\text{src}}\in\myReall{>0}} denotes the radius characterizing the extension of the circular plane wave. The source radius needs to be sufficiently large so that the IRF remains invariant to the source extension. Based on a sensitivity analysis, we set the source radius to \qty{2}{\m} (cf. Fig.~S1 in the Supplemental Material \cite{zotero-item-4600}). We note that, due to the far-field approximation, this approach is most accurate when the gamma-ray source is at distances significantly larger than the platform’s characteristic dimensions. For near-field sources, such as traversed plumes, computationally efficient one-stage Monte Carlo simulations may be adopted to fully capture spatial inhomogeneities in the gamma-ray flux \cite{Sinclair2011a,Breitenmoser2025a}.

With terrestrial and airborne radionuclide gamma-ray emitters as primary sources of interest for the SAGRS system, we adopted a gamma-ray energy grid with variable spacing ranging from \qty{50}{\keV} to \qty{3}{\MeV} and an angular grid covering the entire \mymatht{4\pi} solid angle with a regular \qty{30}{\degree} spacing in \mymatht{\phi'} and \qty{15}{\degree} spacing in \mymatht{\theta'} to account for the expected pronounced anisotropy in the polar direction. In total, 30~gamma-ray energies were combined with 134~unique directions, resulting in 4020~individual Monte Carlo simulations.

As indicated in \cref{eq:IRFmc}, the temporal evolution of the instrument response is parameterized by a state vector \mymatht{\myvect{\xi}\in\myRealu{N_{\xi}}}, encoding all state variables that may modify the SAGRS mass model and thus alter the instrument response during or between MGRS surveys. Given the numerous dynamic systems of the SAGRS platform, we focus in this work on key state variables that are expected to dominate the temporal evolution of the IRF, specifically fuel depletion (\mymatht{\varrho_{\text{fuel}}=\qtyrange{0}{100}{\percent}} fuel volume fraction in \qty{25}{\percent} increments), crew configuration (no crew, minimum, and maximum capacity), as well as landing gear position (extended \& retracted). Starting from a fiducial state \mymatht{\myvect{\xi}_{\text{fid}}} (\mymatht{\varrho_{\text{fuel}}=\qty{50}{\percent}}, minimum crew, landing gear extended), a series of IRFs was computed by changing one state variable at the time, resulting in eight distinct mass model states. 

To perform the required high-fidelity Monte Carlo simulations, we adopted the \mycode{FLUKA} code, Version \myversion{4-2.2}, maintained by the FLUKA.CERN Collaboration \citep{Bohlen2014a,Battistoni2015a,Ahdida2022a}, together with the graphical interface \mycode{FLAIR} \citep{Vlachoudis2009a}, Version \myversion{3.2-4.5}. All simulations were conducted on a local computer cluster at the Paul Scherrer Institute, featuring \num{520} cores at a nominal clock speed of \qty{2.6}{\giga\Hz}, with each simulation running a fixed number of \mymatht{N_{\text{prim}}=\num{1d7}} primaries distributed across 20 parallel runs. The total computation time for the eight mass model states was approximately \qty{5.5d4}{\corehour}. 

To ensure high-fidelity radiation transport, we employed the \mycode{precisio} physics model set in \mycode{FLUKA}, enabling high-fidelity fully coupled photon, electron, and positron transport with secondary electron production, Landau energy fluctuations, and X-ray fluorescence \citep{Battistoni2015a}. To account for the gamma-ray spectrometer's non-proportional scintillation response, we implemented a non-proportional scintillation model recently published by \citet{Payne2009a,Payne2011a} using the user routines \mycode{comscw} and \mycode{usrglo} \citep{Breitenmoser2023i,Breitenmoser2023f}. Transport thresholds were set to \qty{1}{\keV} for the scintillation crystals and surrounding components, including the reflector, optical window, and aluminum casing. Motivated by the range of the transported particles, a higher threshold of \qty{10}{\keV} was selected for the remaining parts of the simulation domain to optimize the balance between model fidelity and computational efficiency. Benchmark tests indicate that increasing the threshold from \qty{1}{\keV} to \qty{10}{\keV} for the respective mass elements resulted in median relative response deviations below \qty{1}{\percent} (cf. Fig.~S2 in the Supplemental Material \cite{zotero-item-4600}), confirming that the reduction in model fidelity remains marginal. Additional information on the physics models, in particular the custom calibrated non-proportional scintillation model, is available in \myonlinecitepl{Breitenmoser2023f,Breitenmoser2025a}. 

As in previous studies \citep{Breitenmoser2023f,Breitenmoser2022e,Breitenmoser2025a}, to simulate the spectral response of the gamma-ray spectrometer, we scored the energy deposition events in each scintillation crystal individually on an event-by-event basis using the custom user routine \mycode{usreou} in conjunction with the \mycode{detect} card. Data reduction of the scored energy deposition events to the expected pulse-height spectral counts \mymatht{N_{\text{event}}} in \cref{eq:IRFmc}, along with the associated statistical and systematic uncertainties, was performed using the \mycode{NPScinMC} pipeline described in \myonlinecite{Breitenmoser2025a}.

\subsection{\label{subsec:FluxMethod}Gamma-ray flux}

As discussed in \cref{sec:IRFtheory}, double-differential gamma-ray flux banks can be generated numerically using standard built-in functionalities in both deterministic \cite{Sjoden1996,Wareing1996,Alcouffe2001} and Monte Carlo based \cite{Allison2016,Goorley2016,Ahdida2022a,Sato2024} nuclear codes. Since this study focuses on demonstrating full-spectrum modeling using dynamic anisotropic IRFs, rather than generating extensive gamma-ray flux banks, we restricted our gamma-ray flux computations to simplified flux templates, whose instrument response is also tractable by brute-force Monte Carlo simulation for direct benchmark verification. Specifically, we consecutively evaluated the double-differential gamma-ray flux at a fixed distance of \qty{30}{\m} from two monoenergetic isotropic gamma-ray point sources with \mymatht{E_{\gamma}=\{\qty{120}{\keV},\qty{2700}{\keV}}\}, covering the relevant spectral range for airborne MGRS systems. Both sources were embedded in homogeneous humid air (temperature, \qty{20}{\degreeCelsius}; pressure, \qty{1013.25}{\hecto\Pa}; relative humidity, \qty{50}{\percent}), with elemental composition data sourced from \myonlinecite{Mcconn2011a}. 

For consistency, we again employed the \mycode{FLUKA} code \citep{Bohlen2014a,Battistoni2015a,Ahdida2022a}, together with the \mycode{FLAIR} graphical interface \citep{Vlachoudis2009a}, to compute the expected double-differential gamma-ray flux and associated uncertainties for the specified gamma-ray sources, using the built-in \mycode{USRYIELD} card. Exploiting the azimuthal symmetry, we adopted a custom binning scheme with a bin width of \qty{3}{\keV} for the gamma-ray energy \mymatht{E_{\gamma}} and \qty{5}{\degree} for the polar angle \mymatht{\theta} in a global inertial coordinate system. The accumulated computation time for the flux computations was about \qty{30}{\corehour}, with \num{5d10} primaries distributed across \num{1d3} independent runs. The gamma-ray flux signatures obtained from this analysis are presented in Fig.~S3 within the Supplemental Material \cite{zotero-item-4600}.

\subsection{\label{subsec:ConvolutionMethod}Spectral template computation}

To compute the spectral templates \mymatht{\myvect{\uppsi}} as defined in \cref{eq:IRFm}, we adopted the \mycode{pagemtimes} function implemented in \mycode{MATLAB}, Version \myversion{R2024a}, allowing efficient multithreaded convolution of the IRFs derived in \cref{subsec:IRFmethod} with the gamma-ray flux banks obtained in \cref{subsec:FluxMethod}. The uncertainty in the spectral templates is quantified by propagating statistical and systematic uncertainty contributions from both the gamma-ray flux and the IRF through the convolution pipeline (see \cref{app:UQ}).

Since the IRF and gamma-ray flux bank use distinct discretization schemes optimized for their respective dispersion characteristics, their spectral and angular grids must be aligned prior to the convolution in \cref{eq:IRFm}. Spectral alignment is achieved by interpolating the IRF onto the regular energy grid used in the double-differential gamma-ray flux simulation \citep{Jandel2004a}. For angular alignment, we first expand the gamma-ray flux and associated angular grid in the azimuth by \qty{30}{\degree} increments to account for the IRF's azimuthal anisotropy. We then map the expanded angular grid to the local coordinate system using a sequence of intrinsic rotations parameterized by the Tait-Bryan angles, which define the platform's instantaneous attitude relative to the global frame (see \cref{app:Coordinates}). As a last step, we perform angular interpolation of the IRF to match the transformed angular grid of the double-differential gamma-ray flux in the local frame. The total spectral template generation time, including uncertainty estimation, was benchmarked at \myOrderqty{1}{\s} on a local workstation with 8 cores running at \qty{2.1}{\GHz}.

\section{\label{sec:Results}Results}

In this section, we present the characteristics and performance of the full-spectrum model for the SAGRS system described in \cref{sec:Implementation}. Specifically, we analyze the system's intrinsic energy and angular dispersion responses, revealed by the numerically estimated IRF, explore the IRF's dynamics in relation to various state variables \mymatht{\myvect{\xi}} discussed in \cref{subsec:IRFmethod}, and compare the methodology against high-fidelity brute-force Monte Carlo simulations in a dedicated benchmark study.

\subsection{\label{subsec:EnergyDispersion}Instrument Response Anisotropy}

As discussed in \cref{sec:Introduction}, the IRFs of MGRS systems are typically highly anisotropic, which motivates the extended response modeling detailed in \cref{sec:IRFtheory}. To quantify this effect, we analyze the intrinsic spectral-angular dispersion of the SAGRS system using the numerically estimated IRF from \cref{sec:Implementation}. This analysis not only illustrates the significance of directional effects in response modeling but also provides key insights for the benchmark simulations in \cref{subsec:Benchmark}.

\begin{figure}
\includegraphics[width=7.5cm]{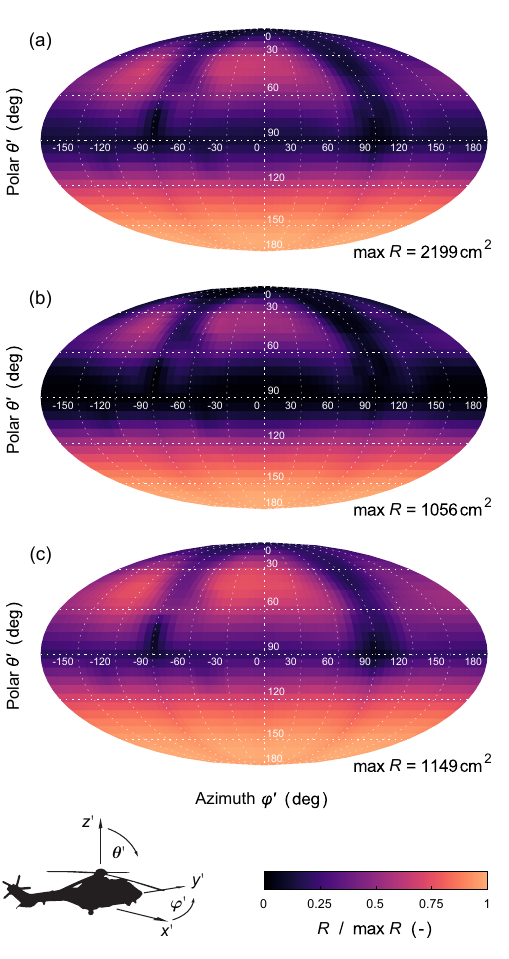}
\caption{Angular dispersion of the SAGRS system's instrument response \mymatht{R} as a function of the polar angle \mymatht{\theta'} and the azimuthal angle \mymatht{\varphi'} with respect to the local noninertial platform-fixed coordinate system $x'$, $y'$, $z'$. The angular dispersion was computed for a gamma-ray energy \mymatht{E_{\gamma}=\qty{662}{\keV}} and normalized by its maximum, \mymatht{R/\mymaxp{R}}, with \mymatht{\mymaxp{R}} indicated in each graph. Three different spectral bands were evaluated: (a)~full-spectrum band (\mymatht{\mathoms{B}_{\mathrm{tot}}}); (b)~full-energy band (\mymatht{\mathoms{B}_{\gamma}}); (c)~Compton band (\mymatht{\mathoms{B}_{\mathrm{C}}}). The angular dispersion graphs are interpolated on a regular $\qty{6}{\degree}\times\qty{6}{\degree}$ angular grid and displayed using the Mollweide projection.}
\label{fig:ADispersion}
\end{figure}

\begin{figure*}
\includegraphics[width=1\textwidth]{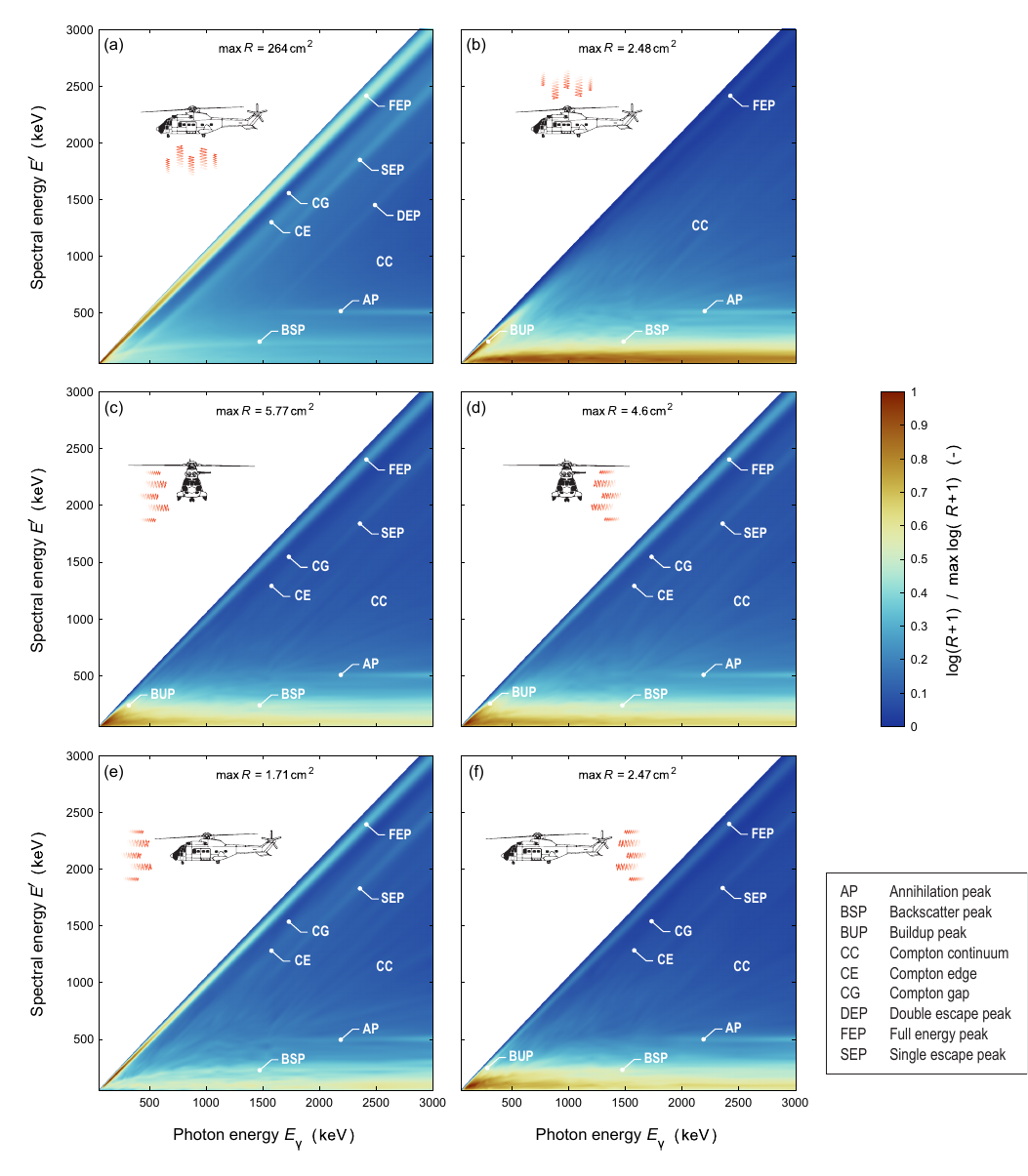}
\caption{Spectral dispersion of the SAGRS system's instrument response \mymatht{R} as a function of the spectral energy \mymatht{E'} and the gamma-ray energy \mymatht{E_{\gamma}} with a constant spectral energy bin width \mymathtv{\myincrement{E'}\sim\qty{3}{\keV}}. The IRF was computed for the six cardinal incident directions (see also \cref{fig:IRFscheme}): (a)~antinormal (\mymathtv{\theta'=\qty{180}{\degree}}); (b)~normal (\mymathtv{\theta'=\qty{0}{\degree}}); (c)~starboard (\mymathtv{\theta'=\qty{90}{\degree}}, \mymathtv{\phi'=\qty{0}{\degree}}); (d)~port (\mymathtv{\theta'=\qty{90}{\degree}}, \mymathtv{\phi'=\qty{180}{\degree}}); (e)~forward (\mymathtv{\theta'=\qty{90}{\degree}}, \mymathtv{\phi'=\qty{90}{\degree}}); (f)~backward (\mymathtv{\theta'=\qty{90}{\degree}}, \mymathtv{\phi'=\qty{-90}{\degree}}). Characteristic spectral features typical of inorganic scintillator responses are indicated \citep{Knoll2010a,Breitenmoser2024}. For visualization purposes, the IRF is constrained to the spectral energy range \mymathtv{E'<E_{\gamma}+3\sigma_{E}} with \mymatht{\sigma_{E}} being the spectral resolution standard deviation at \mymatht{E_{\gamma}}. The line drawing of the AS332 helicopter displayed in the individual subfigures was adapted from Jetijones, \href{https://creativecommons.org/licenses/by/3.0}{\texttt{CC~BY~3.0}}, via Wikimedia Commons.}
\label{fig:EDispersion}
\end{figure*}

We start our analysis by projecting the numerically estimated IRF in its fiducial state \mymatht{R(E',E_{\gamma},\myvect{\Upomega}',\myvect{\xi}_{\text{fid}})} to the angular space \mymatht{(\mymatht{\theta'},\mymatht{\phi'})} in \cref{fig:ADispersion} as well as to the spectral space \mymatht{(\mymatht{E'},\mymatht{E_{\gamma}})} in \cref{fig:EDispersion}. For the angular projection, we iterated the projection for three spectral bands at a given characteristic gamma-ray energy \mymathtv{E_{\gamma}=\qty{662}{\keV}}: full-spectrum band \mymathtv{\mathoms{B}_{\mathrm{tot}}=\{E' \in \myReall{+} \mid E'\leq N_{E'}\myincrement{E'}\}}, full-energy band \mymathtv{\mathoms{B}_{\gamma}=\{E' \in \myReall{+} \mid -3\sigma_{E}\leq E'-E_{\gamma}\leq 3\sigma_{E}\}}, and Compton band \mymathtv{\mathoms{B}_{\mathrm{C}}=\{E' \in \myReall{+} \mid E'-E_{\gamma}< 3\sigma_{E}\}}, with \mymatht{\sigma_{E}} denoting the spectral resolution standard deviation at \mymatht{E_{\gamma}}. For the spectral projection, we repeat the projection for the six cardinal incident directions: normal (\mymathtv{\theta'=\qty{0}{\degree}}), antinormal (\mymathtv{\theta'=\qty{180}{\degree}}), starboard (\mymathtv{\theta'=\qty{90}{\degree}}, \mymathtv{\phi'=\qty{0}{\degree}}), port (\mymathtv{\theta'=\qty{90}{\degree}}, \mymathtv{\phi'=\qty{180}{\degree}}), forward (\mymathtv{\theta'=\qty{90}{\degree}}, \mymathtv{\phi'=\qty{90}{\degree}}), and backward (\mymathtv{\theta'=\qty{90}{\degree}}, \mymathtv{\phi'=\qty{-90}{\degree}}).

From the individual projections shown in \cref{fig:EDispersion,fig:ADispersion}, we find that the instrument response shows a complex anisotropic profile across the entire spectral range with \myOrdernum{1d2} relative variations between the highest and lowest responses. This strong anisotropy can be mainly attributed to the increased attenuation by the aircraft structure obstructing the incoming gamma rays for certain directions, as well as the positioning of the scintillation crystals within the aircraft, which are placed to maximize sensitivity to terrestrial gamma-ray sources (cf. \cref{app:SwissAGRS} and Figs.~S12--S14 within the Supplemental Material \cite{zotero-item-4600}). From these findings, it is evident that ignoring directional dependencies in the response function would introduce substantial biases in event rate predictions, as formulated in \cref{eq:FSA}. This strongly motivates the anisotropic generalization of the full-spectrum modeling framework developed in \cref{sec:IRFtheory} for MGRS systems. To support this conclusion, we provide additional projections in Figs.~S4 and S5 within the Supplemental Material \cite{zotero-item-4600}, along with a comprehensive uncertainty analysis in Figs.~S6–S9 (see also \cref{app:UQ}).

\subsection{\label{subsec:IRFDynamics}Instrument Response Dynamics}

\begin{figure*}[bth!]
\includegraphics[width=1\textwidth]{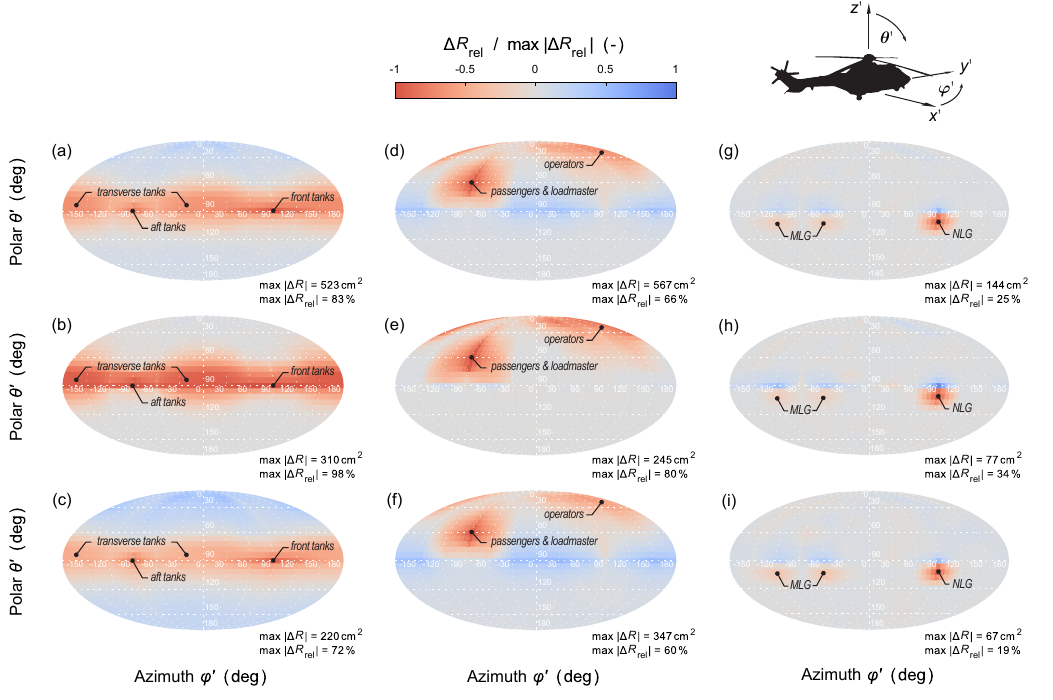}
\caption{Relative deviation in the angular dispersion of the SAGRS system’s instrument response, defined as \mymathtv{\myincrement{R}_{\mathrm{rel}}=\myincrement{R}/R(\myvect{\xi}_0)} with \mymathtv{\myincrement{R}\coloneqq R(\myvect{\xi}_1)-R(\myvect{\xi}_0)}, shown as a function of selected state variable changes \mymathtv{\myvect{\xi}_0 \rightarrow \myvect{\xi}_1}. Three perturbation scenarios are considered: (a--c) fuel state from empty (\mymatht{\myvect{\xi}_0}) to full (\mymatht{\myvect{\xi}_1}); (d--f) crew loading from unoccupied (\mymatht{\myvect{\xi}_0}) to maximum capacity (\mymatht{\myvect{\xi}_1}); (g--i) landing gear deployment from retracted (\mymatht{\myvect{\xi}_0}) to extended (\mymatht{\myvect{\xi}_1}). All remaining parameters are kept at their fiducial value \mymatht{\myvect{\xi}_{\text{fid}}} (see \cref{subsec:IRFmethod}). For each state variable change, the instrument response is evaluated in three distinct spectral bands at a gamma-ray energy of \mymatht{E_{\gamma}=\qty{662}{\keV}}: (a, d, g)~full-spectrum band (\mymatht{\mathoms{B}_{\mathrm{tot}}}); (b, e, h)~full-energy band (\mymatht{\mathoms{B}_{\gamma}}); (c, f, i)~Compton band (\mymatht{\mathoms{B}_{\mathrm{C}}}). All graphs are interpolated on a regular $\qty{6}{\degree}\times\qty{6}{\degree}$ angular grid and displayed using the Mollweide projection. Additionally, characteristic dynamic subsystems are annotated for reference, for example the main landing gear (MLG) and the nose landing gear (NLG) in (g--i) (see also \cref{app:SwissAGRS}).}
\label{fig:RDynamics}
\end{figure*}

Next, we investigate how the SAGRS system’s IRF dynamically evolves due to changes in its internal state variables \mymatht{\myvect{\xi}} during MGRS surveys, as discussed in \cref{sec:Implementation}. We focus this analysis on three key state variables expected to dominate the temporal evolution of the IRF: fuel depletion, crew configuration, and landing gear position.

In \cref{fig:RDynamics}, we present upper bounds on the expected relative deviation of the IRF at a gamma-ray energy of \mymatht{E_{\gamma}=\qty{662}{\keV}} for the aforementioned state variables, evaluated in the full-spectrum (\mymatht{\mathoms{B}_{\mathrm{tot}}}), full-energy (\mymatht{\mathoms{B}_{\gamma}}), and Compton (\mymatht{\mathoms{B}_{\mathrm{C}}}) bands. We find statistically significant results, with peak relative deviations ranging from \qty{\sim25}{\percent} for the landing gear position up to \qty{\sim83}{\percent} for the fuel depletion dynamics. Additionally, while response changes associated with landing gear motion are confined to localized angular sectors, changes in the crew configuration and fuel level affect broader solid angles \mymatht{\geq2\pi\,\unit{\steradian}}. Similar trends are observed at other gamma-ray energies, where relative deviations in the full-energy band \mymatht{\mathoms{B}_{\gamma}} systematically decrease with increasing gamma-ray energy, while those in the Compton band \mymatht{\mathoms{B}_{\mathrm{C}}} show an increase, consistent with the associated differential cross-sections (see Figs.~S10 and S11 in the Supplemental Material \cite{zotero-item-4600}). 

From these results, we conclude that time-dependent changes in MGRS systems can induce substantial changes in the IRF. Neglecting these dynamics would introduce systematic biases in gamma-ray event predictions, particularly when integrated over extended survey durations.

\subsection{\label{subsec:Benchmark}Monte Carlo Benchmark}

\begin{figure*}[bth!]
\includegraphics[width=1\textwidth]{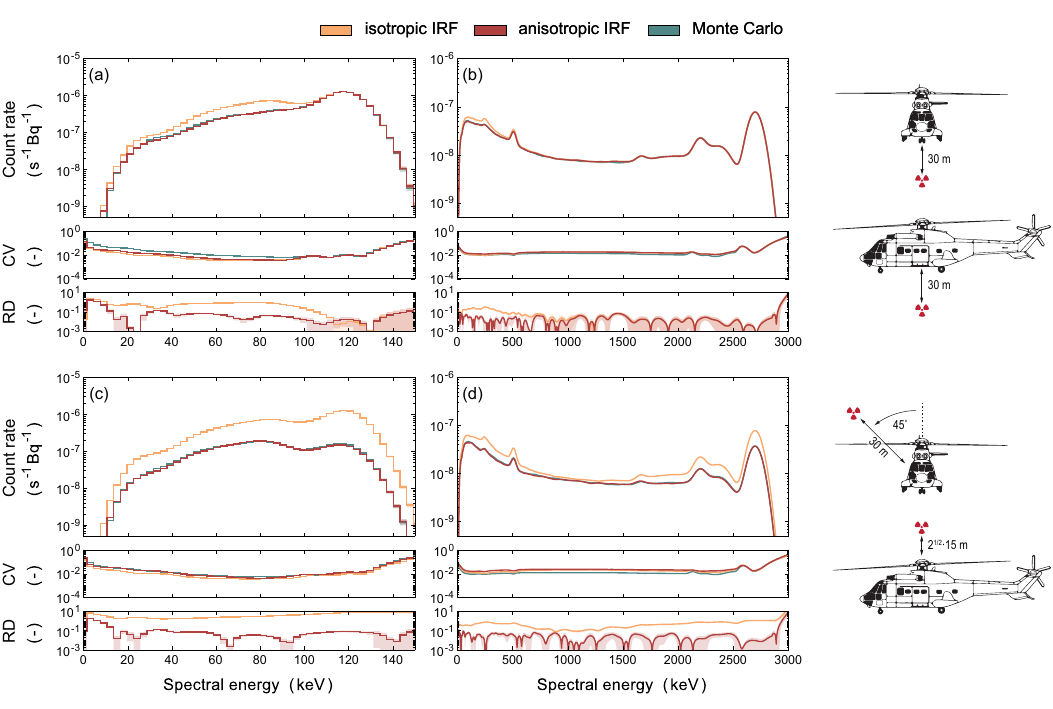}
\caption{Spectral signature computation results obtained by isotropic IRFs, anisotropic IRFs, as well as brute-force Monte Carlo simulations for four selected source-detector configurations, adopting a monoenergetic, isotropic gamma-ray source with energy \mymatht{E_\gamma} and located at (\mymatht{r=\qty{30}{\m}, \phi_\gamma=\qty{0}{\degree}, \theta_\gamma}): (a)~\mymatht{E_\gamma=\qty{120}{\keV}}, \mymatht{\theta_\gamma=\qty{180}{\degree}}; (b)~\mymatht{E_\gamma=\qty{2.7}{\MeV}}, \mymatht{\theta_\gamma=\qty{180}{\degree}}; (c)~\mymatht{E_\gamma=\qty{120}{\keV}}, \mymatht{\theta_\gamma=\qty{45}{\degree}}; (d)~\mymatht{E_\gamma=\qty{2.7}{\MeV}}, \mymatht{\theta_\gamma=\qty{45}{\degree}}. For each configuration, the mean spectral template \mymatht{\myvect{\uppsi}} is displayed as a function of the spectral energy \mymatht{E'} with a spectral bin width \mymatht{\myincrement E'\sim\qty{3}{\keV}}. Uncertainties are provided as 1~standard deviation shaded areas. In addition, the coefficient of variation (CV) as well as the relative deviation (RD) between the isotropic/anisotropic IRF and the brute-force Monte Carlo simulation results are provided (see also \cref{app:UQ}). The displayed line drawings of the AS332 helicopter were adapted from Jetijones, \href{https://creativecommons.org/licenses/by/3.0}{\texttt{CC~BY~3.0}}, via Wikimedia Commons.}\label{fig:benchmark}
\end{figure*}

To verify the generalized full-spectrum methodology outlined in \cref{sec:IRFtheory}, we benchmark spectral templates generated by the SAGRS system's dynamic anisotropic IRF against high-fidelity brute-force Monte Carlo simulations. Due to the high computational cost of brute-force Monte Carlo simulations for MGRS systems, we limit this verification study to four simplified benchmark cases that remain computationally tractable on a computer cluster. Specifically, we generated spectral templates for two monoenergetic isotropic gamma-ray sources with energies \mymathtv{E_{\gamma}=\{\qty{120}{\keV},\qty{2.7}{\MeV}\}}, each placed at a distance of \qty{30}{m} from the SAGRS spectrometer, with azimuthal angle \mymatht{\phi_\gamma=\qty{0}{\degree}} and polar angle \mymathtv{\theta_\gamma=\{\qty{45}{\degree},\qty{180}{\degree}\}}, respectively. For all evaluations, we adopted the Monte Carlo mass model in its fiducial state embedded in homogeneous humid air (see \cref{subsec:IRFmethod,subsec:FluxMethod}). Together, the selected benchmark cases probe the SAGRS system’s response to both low- and high-energy gamma rays in two representative airborne MGRS scenarios: a typical flatland survey and a more complex geometry relevant in alpine or urban environments.

\cref{fig:benchmark} shows the spectral templates for the four benchmark configurations, computed using brute-force Monte Carlo simulations, the dynamic anisotropic IRF method introduced in \cref{sec:IRFtheory}, and the conventional isotropic IRF model based on \cref{eq:IRF0}. Following standard practice \cite{Kluson2010c}, the isotropic IRF was referenced to the anti-normal direction. The anisotropic IRF method shows excellent agreement with the Monte Carlo benchmark, yielding median relative deviations of less than \qty{6}{\percent} across the spectral domain of interest (SDOI), as defined in \myonlinecite{Breitenmoser2025a}

In contrast, the conventional isotropic IRF approach shows significant systematic deviations, even in the best-case configurations shown in \cref{fig:benchmark}(a,b), with median relative deviations \qty{>50}{\percent} and \qty{>16}{\percent} below \mymathtv{E'=\qty{120}{\keV}} for low- and high-energy gamma-ray sources, respectively. These discrepancies primarily reflect the anisotropic nature of the incident gamma-ray flux, wherein a significant fraction of low-energy photons impinge off-axis due to Compton scattering in the air (see Fig.~S3 in the Supplemental Material \cite{zotero-item-4600}). As expected, systematic errors increase substantially for the configurations in \cref{fig:benchmark}(c,d), where the source location no longer aligns with the isotropic IRF's reference direction, resulting in relative deviations \qty{>250}{\percent} and \qty{>40}{\percent} for the respective low- and high-energy sources across the SDOI. 

These benchmark results demonstrate that the generalized anisotropic IRF methodology accurately reproduces high-fidelity Monte Carlo simulations, while the conventional isotropic IRF approach introduces substantial systematic biases, even under idealized conditions. This underscores the importance of accounting for the anisotropy in both the double-differential gamma-ray flux and the IRF to generate accurate spectral templates for airborne MGRS systems.

\section{\label{sec:Discussion}Conclusion}

Here, we have shown that the generation of spectral templates for FSA by MGRS systems in scattering media can be accurately achieved using a generalized response methodology based on dynamic anisotropic IRFs and double-differential gamma-ray flux banks. The resulting spectral templates reproduce the spectral response with accuracy comparable to high-fidelity brute-force Monte Carlo simulations across the full spectral range, while reducing the computational cost by a factor \myOrdernum{1d7}, with typical generation times of \myOrderqty{1}{\s} per template on a local workstation. In contrast, conventional methods relying on isotropic IRFs and single-differential flux banks exhibit at least twofold larger systematic spectral errors, even under idealized conditions. 

The framework further enables systematic exploration of IRF sensitivity to platform-specific state variables, revealing that perturbations in these variables can induce significant variations in the IRF over a large fraction of the full \mymatht{4\pi} solid angle and across the full spectral range of the MGRS system. Combined with the verification results, these findings underscore the critical importance of accounting for both the dynamics in the state variables and the anisotropy in the IRF as well as the double-differential gamma-ray flux to generate accurate spectral templates for MGRS systems operating in scattering media.

While the implementation presented in this work focused on airborne systems, the formalism is system-agnostic and extends naturally to ground-based and marine MGRS platforms for arbitrary far-field gamma-ray sources. Future work will leverage this methodology to enhance near real-time FSA capabilities for MGRS systems across a broad range of applied domains, including geophysical studies, mineral exploration, nuclear safeguards, and radiological emergency response.

\appendix

\section{\label{app:SwissAGRS}Swiss Airborne Gamma-Ray Spectrometry System}

In this study, we demonstrated the applicability and performance of the proposed generalized full-spectrum modeling framework using a real-world MGRS system. Specifically, we applied it to the Swiss Airborne Gamma-Ray Spectrometry (SAGRS) system, for which a well-characterized and validated Monte Carlo mass model is available \cite{Breitenmoser2022e,Breitenmoser2025a}. For completeness, we provide here a brief overview of the SAGRS system's technical specifications, with further details available in \myonlinecite{Breitenmoser2025a}.

The SAGRS system consists of four \mymathtv{\qty{10.2}{\cm}\times\qty{10.2}{\cm}\times\qty{40.6}{\cm}} prismatic NaI(Tl) scintillation crystals (Saint-Gobain 4*4H16/3.5-X), enclosed in individual aluminum casings. Each crystal is coupled to a Hamamatsu R10755 photomultiplier tube with associated electronics, enabling independent readout for each scintillation crystal. The four spectrometers are embedded in thermally insulating and vibration-damping polyethylene foam, within a rugged aluminum box with outer dimensions of \mymathtv{\qty{90}{\cm}\times\qty{64}{\cm}\times\qty{35}{\cm}}. The system is mounted in the cargo bay of an A\'{e}rospatiale AS332M1 Super Puma helicopter, with its bottom aligned to the helicopter's underside to maximize sensitivity to terrestrial gamma-ray sources. It integrates aircraft sensor data, including pressure, temperature, global navigation satellite system positioning, and radar altimeter readings, via an ARINC~429 avionics data bus.

The system employs a bin-mode data acquisition scheme with a sampling time of \qty{1}{\s} per detector channel. In addition to the four individual channels (\texttt{\#DET1}--\texttt{\#DET4}), a fifth channel (\texttt{\#DET5}) records the summed pulse-height spectrum. Each channel consists of \num{1024} pulse-height channels, covering a spectral range from \qty{\sim30}{\keV} to \qty{\sim3.072}{\MeV}. The spectrometer also incorporates automatic gain stabilization, spectrum linearization with offset correction, and live-time recording, simplifying data postprocessing as discussed in \myonlinecite{Breitenmoser2022e}. Data acquisition during surveys is managed by two operators within the crew cabin, using rugged client computers and a central data server, with the total mass of all support systems being \qty{\sim290}{\kg}.

\section{\label{app:UQ}Uncertainty Quantification}

We quantify the combined uncertainty (statistical \& systematic) in the estimated spectral template vector \mymatht{\myvect{\uppsi}} by propagating the individual contributions from the IRF and gamma-ray flux through \cref{eq:IRFn} using the standard error propagation formalism for independent variables \citep{Bevington2003a,JointCommitteeForGuidesInMetrology2008}: 

\begin{align}
    \sigma_{\psi}&\left(E'_i,t_l\right) \approx%
    \left\{\mysuum{j=1\vphantom{k=1}}{N_{E_{\gamma}}\vphantom{N_{\myvect{\Upomega}'}}}{k=1\vphantom{j=1}}{N_{\myvect{\Upomega}'}\vphantom{N_{E_{\gamma}}}}\left[\left(\myodvt{R_{ijkl}}{E'}\mypdvvt{\sigma_{\phi_{\gamma},jkl}}{E_{\gamma}}{\varOmega'}\right)^2\right.\right.\nonumber\\%
    &+\left.\left.\left(\myodvt{\sigma_{R_{ijkl}}}{E'}\mypdvvt{\phi_{\gamma,jkl}}{E_{\gamma}}{\varOmega'}\right)^2\right]\myincrement{E'^2_{i}}\myincrement{E_{\gamma,j}^2}\myincrement{\varOmega'^{2}_{k}}\vphantom{\mysuum{j=1\vphantom{k=1}}{N_{E_{\gamma}}\vphantom{N_{\myvect{\Upomega}'}}}{k=1\vphantom{j=1}}{N_{\myvect{\Upomega}'}\vphantom{N_{E_{\gamma}}}}}\right\}^{1/2}\label{eq:IRFunc1}%
\end{align}

\noindent with \mymatht{R_{ijkl} \coloneqq R(E'_i,E_{\gamma,j},\myvect{\Upomega}'_k,t_l)} and \mymatht{\phi_{\gamma,jkl} \coloneqq \phi_{\gamma}(E_{\gamma,j},\myvect{\Upomega}'_k,t_l)} standing for the evaluated IRF and gamma-ray flux, while \mymatht{\sigma_{*}} denotes the combined statistical and systematic uncertainty of the corresponding quantities ($*$) characterized by the standard error of the mean. 

In full analogy to \cref{sec:IRFtheory}, we may recast \cref{eq:IRFunc1} in matrix notation to compute the standard error of the mean for the spectral template vector \mymatht{\myvect{\upsigma}_{\myvect{\uppsi}}\in\myRealul{N_{E'}}{\geq0}} as

\begin{align}
    \myvect{\upsigma}_{\myvect{\uppsi}}\left(t_l\right) \approx%
    \left\{\mysum{k=1}{N_{\myvect{\Upomega}'}}%
    \left[\left(\myvect{R}_{kl}\myvect{\upsigma}_{\myvect{\upphi}_{\gamma},kl}\right)^{\myhadpower{2}}%
    +\left(\myvect{\upsigma}_{\myvect{R},kl}\myvect{\upphi}_{\gamma,kl}\right)^{\myhadpower{2}}\right]\vphantom{\mysum{k=1}{N_{\myvect{\Upomega}'}}}\right\}^{\mathrlap{\myhadpower{1/2}}}\label{eq:IRFunc2}%
\end{align}

\noindent with \mymathtv{\myvect{R}_{kl} \coloneqq \myvect{R}(\myvect{\Upomega}'_k,t_l) \in \myRealul{ N_{E'} \times N_{E_{\gamma}}}{\geq0}} and \mymathtv{\myvect{\upphi}_{\gamma,kl} \coloneqq \myvect{\upphi}_{\gamma}(\myvect{\Upomega}'_k,t_l) \in \myRealul{N_{E'}}{\geq0}} standing for the instrument response matrix and the gamma-ray flux vector defined in \cref{sec:IRFtheory}, while \mymatht{\myhadpower{}} represents the Hadamard (elementwise) power. Additionally, \mymathtv{\myvect{\upsigma}_{\myvect{R},kl} \in \myRealul{ N_{E'} \times N_{E_{\gamma}}}{\geq0}} and \mymathtv{\myvect{\upsigma}_{\myvect{\upphi}_{\gamma},kl} \in \myRealul{N_{E'}}{\geq0}} denote the combined statistical and systematic uncertainties of the corresponding quantities, both characterized again by the standard error of the mean. 

Similarly to \cref{eq:IRFm}, the matrix notation in \cref{eq:IRFunc2} enables the efficient multithreaded evaluation of the spectral template vector's standard error of the mean. To perform the multithreaded computations, we adopted the \mycode{pagemtimes} function provided in \mycode{MATLAB}. The individual statistical and systematic contributions of the instrument response function and gamma-ray flux are quantified using the \mycode{NPScinMC} pipeline \citep{Breitenmoser2025a} and the built-in \mycode{USRYIELD} card provided by the \mycode{FLUKA} code, respectively. Given the simplified flux banks considered in this study, systematic uncertainties in the gamma-ray flux are assumed to be negligible. The combined uncertainties are estimated by adding the statistical and systematic components in quadrature, following the formalism recommended by the Particle Data Group \cite{ParticleDataGroup2022} and the Joint Committee for Guides in Metrology \cite{JointCommitteeForGuidesInMetrology2008}.

\section{\label{app:Coordinates}Global-to-instrument transformation}

As discussed in \cref{subsec:ConvolutionMethod}, to account for the orientation of the MGRS platform, it is necessary to transform the incident gamma-ray flux, initially expressed in the global inertial reference frame, into the local noninertial platform-fixed frame. For that purpose, we perform a coordinate transformation of the double differential gamma-ray flux using a sequence of intrinsic rotations parameterized by the Tait-Bryan angles, which define the platform's instantaneous attitude relative to the reference frame.

Consider a unit direction vector \mymatht{\myvect{\Upomega}} in a global Cartesian coordinate system, parameterized by the azimuthal angle \mymatht{\phi} and the polar angle \mymatht{\theta}:

\begin{equation}
    \myvect{\Upomega}\left(\theta,\varphi\right) =\begin{pmatrix}
        \sin{\theta}\cos{\varphi}\\
        \sin{\theta}\sin{\varphi}\\
        \cos{\theta}   
    \end{pmatrix}.
    \label{eq:APPdir2detDirection}
\end{equation}

\noindent Next, we specify the orientation of the MGRS platform using the Tait-Bryan formalism with the three Cardan angles \mymatht{\alpha'} (yaw), \mymatht{\beta'} (pitch), and \mymatht{\gamma'} (roll), commonly adopted in aerospace applications to describe the attitude of mobile systems (cf. \cref{fig:IRFscheme}). Using this formalism, the transformation of the unit direction vector from the global frame to the local platform-fixed frame is then expressed as a linear mapping,

\begin{equation}
    \label{eq:APPdir2detTrafo}
    \myvect{\Upomega}'\left(\theta',\varphi'\right) = \myvect{\mathfrak{R}}\left(\alpha',\beta',\gamma'\right)\myvect{\Upomega}\left(\theta,\varphi\right), 
\end{equation}

\noindent where \mymatht{\myvect{\mathfrak{R}} \in \myRealu{3\times3}} denotes the rotation matrix associated with the prescribed sequence of intrinsic rotations. Specifically, \mymatht{\myvect{\mathfrak{R}}} is given by the product of three elemental rotation matrices, representing a sequence $z'$-$x'$-$y'$ of three intrinsic rotations about the principal aircraft axes $z'$ (yaw), $x'$ (pitch), and $y'$-axis (roll) \cite{Roithmayr2016}:

\begin{subequations}
    \begin{align}
    \label{eq:APPdir2detRotMatrixA}
    \myvect{\mathfrak{R}} &= %
    \begin{pNiceMatrix}[columns-width=2mm]%
        \text{c}_{\gamma'} & 0 & -\text{s}_{\gamma'} \\
        0 & 1 & 0 \\
        \text{s}_{\gamma'} & 0 & \text{c}_{\gamma'}
    \end{pNiceMatrix}\,%
    \begin{pNiceMatrix}[columns-width=2mm]%
        1 & 0 & 0 \\
        0 & \text{c}_{\beta'} & \text{s}_{\beta'} \\
        0 & -\text{s}_{\beta'} & \text{c}_{\beta'}
    \end{pNiceMatrix}\,%
    \begin{pNiceMatrix}[columns-width=2mm]%
        \text{c}_{\alpha'} & \text{s}_{\alpha'} & 0 \\
        -\text{s}_{\alpha'} & \text{c}_{\alpha'} & 0 \\
        0 & 0 & 1
    \end{pNiceMatrix}\,\\
    &= \begin{pNiceMatrix}[columns-width=2mm]%
        \text{c}_{\alpha'}\text{c}_{\gamma'}-\text{s}_{\alpha'}\text{s}_{\beta'}\text{s}_{\gamma'} & \text{c}_{\gamma'}\text{s}_{\alpha'}+\text{c}_{\alpha'}\text{s}_{\beta'}\text{s}_{\gamma'} & -\text{c}_{\beta'}\text{s}_{\gamma'}\\
        -\text{c}_{\beta'}\text{s}_{\alpha'} & \text{c}_{\alpha'}\text{c}_{\beta'} & \text{s}_{\beta'}\\
        \text{c}_{\alpha'}\text{s}_{\gamma'}+\text{c}_{\gamma'}\text{s}_{\alpha'}\text{s}_{\beta'} & \text{s}_{\alpha'}\text{s}_{\gamma'}-\text{c}_{\alpha'}\text{c}_{\gamma'}\text{s}_{\beta'} & \text{c}_{\beta'}\text{c}_{\gamma'}\\
    \end{pNiceMatrix}\label{eq:APPdir2detRotMatrixB}
\end{align}
\end{subequations}

\noindent with \mymatht{\text{s}_{*}} and \mymatht{\text{c}_{*}} standing for the sine \mymatht{\sin(*)} and cosine \mymatht{\cos(*)} functions, respectively. Substituting \eqref{eq:APPdir2detDirection} and \eqref{eq:APPdir2detRotMatrixB} into \eqref{eq:APPdir2detTrafo}, the transformed direction unit vector \mymatht{\myvect{\Upomega}'} in the local platform-fixed coordinate system is obtained as

\begin{equation}
    \myvect{\Upomega}'%
    =\begin{pmatrix}%
    \scriptstyle[\text{c}_{\alpha'}\text{c}_{\gamma'}-\text{s}_{\alpha'}\text{s}_{\beta'}\text{s}_{\gamma'}]{c}_{\varphi}{s}_{\theta}+[\text{c}_{\gamma'}\text{s}_{\alpha'}+\text{c}_{\alpha'}\text{s}_{\beta'}\text{s}_{\gamma'}]{s}_{\varphi}{s}_{\theta}-\text{c}_{\beta'}\text{c}_{\theta}\text{s}_{\gamma'} \\
    \scriptstyle-\text{c}_{\beta'}\text{c}_{\varphi}\text{s}_{\alpha'}\text{s}_{\theta}+\text{c}_{\alpha'}\text{c}_{\beta'}{s}_{\varphi}{s}_{\theta}+\text{c}_{\theta}\text{s}_{\beta'} \\
    \scriptstyle[\text{c}_{\alpha'}\text{s}_{\gamma'}+\text{c}_{\gamma'}\text{s}_{\alpha'}\text{s}_{\beta'}]\text{c}_{\varphi}\text{s}_{\theta}+[\text{s}_{\alpha'}\text{s}_{\gamma'}-\text{c}_{\alpha'}\text{c}_{\gamma'}\text{s}_{\beta'}]\text{s}_{\varphi}\text{s}_{\theta}+\text{c}_{\beta'}\text{c}_{\gamma'}\text{c}_{\theta}
    \end{pmatrix}.\label{eq:APPdir2detDirectionPrime}
\end{equation}

\noindent As a last step, the azimuthal angle \mymatht{\varphi'} and the polar angle \mymatht{\theta'} of the direction unit vector \mymatht{\myvect{\Upomega}'} in the local frame are obtained by transforming the Cartesian coordinates derived in \cref{eq:APPdir2detDirectionPrime} into spherical coordinates as

\begin{equation}
    \begin{pmatrix}
        \varphi' \\
        \theta'
    \end{pmatrix}=
    \begin{pmatrix}
    \operatorname{arctan2}\left[ \varOmega'_{y},\varOmega'_{x}\right] \\
    \arccos{\left[\varOmega'_{z}\right]} 
    \end{pmatrix}\label{eq:APPdir2detSphPrime}
\end{equation}

\noindent where \mymatht{\varOmega'_{x}}, \mymatht{\varOmega'_{y}}, and \mymatht{\varOmega'_{z}} represent the Cartesian coordinates in \cref{eq:APPdir2detDirectionPrime}, and \mymatht{\operatorname{arctan2}\left[\cdot,\cdot\right]} denotes the two-argument arctangent function.

\begin{acknowledgments}
This work was partly supported by the Swiss Federal Nuclear Safety Inspectorate (Grant No. CTR00836 \& CTR00491). We sincerely thank Gernot Butterweck for his invaluable scientific guidance and supervision. We also gratefully acknowledge Dominik Werthmüller for his technical assistance with the Monte Carlo simulations conducted on the Paul Scherrer Institute’s computing cluster.
\end{acknowledgments}

\section*{\label{sec:CompetingInterest}Declaration of competing interest}
The authors declare that they have no known competing financial interests or personal relationships that could have appeared to influence the work reported in this paper.

\section*{\label{sec:AuthorContr}Author Contributions}

\noindent\textbf{D.~Breitenmoser:} Conceptualization, Investigation, Methodology, Formal analysis, Project administration, Software, Supervision, Validation, Visualization, Writing original draft, Review and editing, Data curation. \textbf{A.~Stabilini:} Review, Data curation. \textbf{M.~M.~Kasprzak:} Review. \textbf{S.~Mayer:} Review, Funding acquisition.


%


\begin{thebibliography}{80}%
\makeatletter
\providecommand \@ifxundefined [1]{%
 \@ifx{#1\undefined}
}%
\providecommand \@ifnum [1]{%
 \ifnum #1\expandafter \@firstoftwo
 \else \expandafter \@secondoftwo
 \fi
}%
\providecommand \@ifx [1]{%
 \ifx #1\expandafter \@firstoftwo
 \else \expandafter \@secondoftwo
 \fi
}%
\providecommand \natexlab [1]{#1}%
\providecommand \enquote  [1]{``#1''}%
\providecommand \bibnamefont  [1]{#1}%
\providecommand \bibfnamefont [1]{#1}%
\providecommand \citenamefont [1]{#1}%
\providecommand \href@noop [0]{\@secondoftwo}%
\providecommand \href [0]{\begingroup \@sanitize@url \@href}%
\providecommand \@href[1]{\@@startlink{#1}\@@href}%
\providecommand \@@href[1]{\endgroup#1\@@endlink}%
\providecommand \@sanitize@url [0]{\catcode `\\12\catcode `\$12\catcode `\&12\catcode `\#12\catcode `\^12\catcode `\_12\catcode `\%12\relax}%
\providecommand \@@startlink[1]{}%
\providecommand \@@endlink[0]{}%
\providecommand \url  [0]{\begingroup\@sanitize@url \@url }%
\providecommand \@url [1]{\endgroup\@href {#1}{\urlprefix }}%
\providecommand \urlprefix  [0]{URL }%
\providecommand \Eprint [0]{\href }%
\providecommand \doibase [0]{https://doi.org/}%
\providecommand \selectlanguage [0]{\@gobble}%
\providecommand \bibinfo  [0]{\@secondoftwo}%
\providecommand \bibfield  [0]{\@secondoftwo}%
\providecommand \translation [1]{[#1]}%
\providecommand \BibitemOpen [0]{}%
\providecommand \bibitemStop [0]{}%
\providecommand \bibitemNoStop [0]{.\EOS\space}%
\providecommand \EOS [0]{\spacefactor3000\relax}%
\providecommand \BibitemShut  [1]{\csname bibitem#1\endcsname}%
\let\auto@bib@innerbib\@empty
\bibitem [{\citenamefont {Jones}(2001)}]{Jones2001a}%
  \BibitemOpen
  \bibfield  {author} {\bibinfo {author} {\bibfnamefont {D.~G.}\ \bibnamefont {Jones}},\ }\bibfield  {title} {\bibinfo {title} {Development and application of marine gamma-ray measurements: {{A}} review},\ }\href {https://doi.org/10.1016/S0265-931X(00)00139-9} {\bibfield  {journal} {\bibinfo  {journal} {J. Environ. Radioact.}\ }\textbf {\bibinfo {volume} {53}},\ \bibinfo {pages} {313} (\bibinfo {year} {2001})}\BibitemShut {NoStop}%
\bibitem [{\citenamefont {Lee}\ and\ \citenamefont {Kim}(2019)}]{Lee2019a}%
  \BibitemOpen
  \bibfield  {author} {\bibinfo {author} {\bibfnamefont {C.}~\bibnamefont {Lee}}\ and\ \bibinfo {author} {\bibfnamefont {H.~R.}\ \bibnamefont {Kim}},\ }\bibfield  {title} {\bibinfo {title} {Conceptual {{Development}} of {{Sensing Module Applied}} to {{Autonomous Radiation Monitoring System}} for {{Marine Environment}}},\ }\href {https://doi.org/10.1109/JSEN.2019.2921550} {\bibfield  {journal} {\bibinfo  {journal} {IEEE Sens. J.}\ }\textbf {\bibinfo {volume} {19}},\ \bibinfo {pages} {8920} (\bibinfo {year} {2019})}\BibitemShut {NoStop}%
\bibitem [{\citenamefont {Lee}\ \emph {et~al.}(2023)\citenamefont {Lee}, \citenamefont {Kim}, \citenamefont {Jang}, \citenamefont {Cha}, \citenamefont {Seo}, \citenamefont {Baek},\ and\ \citenamefont {Lim}}]{Lee2023c}%
  \BibitemOpen
  \bibfield  {author} {\bibinfo {author} {\bibfnamefont {M.~S.}\ \bibnamefont {Lee}}, \bibinfo {author} {\bibfnamefont {S.~M.}\ \bibnamefont {Kim}}, \bibinfo {author} {\bibfnamefont {M.}~\bibnamefont {Jang}}, \bibinfo {author} {\bibfnamefont {H.}~\bibnamefont {Cha}}, \bibinfo {author} {\bibfnamefont {J.~M.}\ \bibnamefont {Seo}}, \bibinfo {author} {\bibfnamefont {S.}~\bibnamefont {Baek}},\ and\ \bibinfo {author} {\bibfnamefont {J.~M.}\ \bibnamefont {Lim}},\ }\bibfield  {title} {\bibinfo {title} {Real-time wireless marine radioactivity monitoring system using a {{SiPM-based}} mobile gamma spectroscopy mounted on an unmanned marine vehicle},\ }\href {https://doi.org/10.1016/J.NET.2023.03.017} {\bibfield  {journal} {\bibinfo  {journal} {Nucl. Eng. Technol.}\ }\textbf {\bibinfo {volume} {55}},\ \bibinfo {pages} {2158} (\bibinfo {year} {2023})}\BibitemShut {NoStop}%
\bibitem [{\citenamefont {Rosenthal}\ \emph {et~al.}(1991)\citenamefont {Rosenthal}, \citenamefont {{de Almeidat}},\ and\ \citenamefont {Mendonca}}]{Rosenthal1991a}%
  \BibitemOpen
  \bibfield  {author} {\bibinfo {author} {\bibfnamefont {J.~J.}\ \bibnamefont {Rosenthal}}, \bibinfo {author} {\bibfnamefont {C.~E.}\ \bibnamefont {{de Almeidat}}},\ and\ \bibinfo {author} {\bibfnamefont {A.~H.}\ \bibnamefont {Mendonca}},\ }\bibfield  {title} {\bibinfo {title} {The {{Radiological Accident}} in {{Goiania}}},\ }\href {https://doi.org/10.1097/00004032-199101000-00001} {\bibfield  {journal} {\bibinfo  {journal} {Health Phys.}\ }\textbf {\bibinfo {volume} {60}},\ \bibinfo {pages} {7} (\bibinfo {year} {1991})}\BibitemShut {NoStop}%
\bibitem [{\citenamefont {Drovnikov}\ \emph {et~al.}(1997)\citenamefont {Drovnikov}, \citenamefont {Egorov}, \citenamefont {Kovalenko}, \citenamefont {Serboulov},\ and\ \citenamefont {Zadorozhny}}]{Drovnikov1997a}%
  \BibitemOpen
  \bibfield  {author} {\bibinfo {author} {\bibfnamefont {V.~V.}\ \bibnamefont {Drovnikov}}, \bibinfo {author} {\bibfnamefont {N.~Y.}\ \bibnamefont {Egorov}}, \bibinfo {author} {\bibfnamefont {V.~V.}\ \bibnamefont {Kovalenko}}, \bibinfo {author} {\bibfnamefont {Y.~A.}\ \bibnamefont {Serboulov}},\ and\ \bibinfo {author} {\bibfnamefont {Y.~A.}\ \bibnamefont {Zadorozhny}},\ }\bibfield  {title} {\bibinfo {title} {Some results of the airborne high energy resolution gamma-spectrometry application for the research of the {{USSR European}} territory radioactive contamination in 1986 caused by the {{Chernobyl}} accident},\ }\href {https://doi.org/10.1016/S0265-931X(96)00093-8} {\bibfield  {journal} {\bibinfo  {journal} {J. Environ. Radioact.}\ }\textbf {\bibinfo {volume} {37}},\ \bibinfo {pages} {223} (\bibinfo {year} {1997})}\BibitemShut {NoStop}%
\bibitem [{\citenamefont {Lyons}\ and\ \citenamefont {Colton}(2012)}]{Lyons2012}%
  \BibitemOpen
  \bibfield  {author} {\bibinfo {author} {\bibfnamefont {C.}~\bibnamefont {Lyons}}\ and\ \bibinfo {author} {\bibfnamefont {D.}~\bibnamefont {Colton}},\ }\bibfield  {title} {\bibinfo {title} {Aerial {{Measuring System}} in {{Japan}}},\ }\href {https://doi.org/10.1097/HP.0b013e31824d0056} {\bibfield  {journal} {\bibinfo  {journal} {Health Phys.}\ }\textbf {\bibinfo {volume} {102}},\ \bibinfo {pages} {509} (\bibinfo {year} {2012})}\BibitemShut {NoStop}%
\bibitem [{\citenamefont {Torii}\ \emph {et~al.}(2013)\citenamefont {Torii}, \citenamefont {Sugita}, \citenamefont {Okada}, \citenamefont {Reed},\ and\ \citenamefont {Blumenthal}}]{Torii2013a}%
  \BibitemOpen
  \bibfield  {author} {\bibinfo {author} {\bibfnamefont {T.}~\bibnamefont {Torii}}, \bibinfo {author} {\bibfnamefont {T.}~\bibnamefont {Sugita}}, \bibinfo {author} {\bibfnamefont {C.~E.}\ \bibnamefont {Okada}}, \bibinfo {author} {\bibfnamefont {M.~S.}\ \bibnamefont {Reed}},\ and\ \bibinfo {author} {\bibfnamefont {D.~J.}\ \bibnamefont {Blumenthal}},\ }\bibfield  {title} {\bibinfo {title} {Enhanced {{Analysis Methods}} to {{Derive}} the {{Spatial Distribution}} of {{131I Deposition}} on the {{Ground}} by {{Airborne Surveys}} at an {{Early Stage}} after the {{Fukushima Daiichi Nuclear Power Plant Accident}}},\ }\href {https://doi.org/10.1097/HP.0b013e318294444e} {\bibfield  {journal} {\bibinfo  {journal} {Health Phys.}\ }\textbf {\bibinfo {volume} {105}},\ \bibinfo {pages} {192} (\bibinfo {year} {2013})}\BibitemShut {NoStop}%
\bibitem [{\citenamefont {Sanada}\ \emph {et~al.}(2014)\citenamefont {Sanada}, \citenamefont {Sugita}, \citenamefont {Nishizawa}, \citenamefont {Kondo},\ and\ \citenamefont {Torii}}]{Sanada2014a}%
  \BibitemOpen
  \bibfield  {author} {\bibinfo {author} {\bibfnamefont {Y.}~\bibnamefont {Sanada}}, \bibinfo {author} {\bibfnamefont {T.}~\bibnamefont {Sugita}}, \bibinfo {author} {\bibfnamefont {Y.}~\bibnamefont {Nishizawa}}, \bibinfo {author} {\bibfnamefont {A.}~\bibnamefont {Kondo}},\ and\ \bibinfo {author} {\bibfnamefont {T.}~\bibnamefont {Torii}},\ }\bibfield  {title} {\bibinfo {title} {The aerial radiation monitoring in {{Japan}} after the {{Fukushima Daiichi}} nuclear power plant accident},\ }\href {https://doi.org/10.15669/pnst.4.76} {\bibfield  {journal} {\bibinfo  {journal} {Nucl. Sci. Technol.}\ }\textbf {\bibinfo {volume} {4}},\ \bibinfo {pages} {76} (\bibinfo {year} {2014})}\BibitemShut {NoStop}%
\bibitem [{\citenamefont {Sanada}\ and\ \citenamefont {Torii}(2015)}]{Sanada2015a}%
  \BibitemOpen
  \bibfield  {author} {\bibinfo {author} {\bibfnamefont {Y.}~\bibnamefont {Sanada}}\ and\ \bibinfo {author} {\bibfnamefont {T.}~\bibnamefont {Torii}},\ }\bibfield  {title} {\bibinfo {title} {Aerial radiation monitoring around the {{Fukushima Dai-ichi}} nuclear power plant using an unmanned helicopter},\ }\href {https://doi.org/10.1016/j.jenvrad.2014.06.027} {\bibfield  {journal} {\bibinfo  {journal} {J. Environ. Radioact.}\ }\textbf {\bibinfo {volume} {139}},\ \bibinfo {pages} {294} (\bibinfo {year} {2015})}\BibitemShut {NoStop}%
\bibitem [{\citenamefont {Nishizawa}\ \emph {et~al.}(2016)\citenamefont {Nishizawa}, \citenamefont {Yoshida}, \citenamefont {Sanada},\ and\ \citenamefont {Torii}}]{Nishizawa2016a}%
  \BibitemOpen
  \bibfield  {author} {\bibinfo {author} {\bibfnamefont {Y.}~\bibnamefont {Nishizawa}}, \bibinfo {author} {\bibfnamefont {M.}~\bibnamefont {Yoshida}}, \bibinfo {author} {\bibfnamefont {Y.}~\bibnamefont {Sanada}},\ and\ \bibinfo {author} {\bibfnamefont {T.}~\bibnamefont {Torii}},\ }\bibfield  {title} {\bibinfo {title} {Distribution of the {\textsuperscript{134}} {{Cs}}/ {\textsuperscript{137}} {{Cs}} ratio around the {{Fukushima Daiichi}} nuclear power plant using an unmanned helicopter radiation monitoring system},\ }\href {https://doi.org/10.1080/00223131.2015.1071721} {\bibfield  {journal} {\bibinfo  {journal} {J. Nucl. Sci. Technol.}\ }\textbf {\bibinfo {volume} {53}},\ \bibinfo {pages} {468} (\bibinfo {year} {2016})}\BibitemShut {NoStop}%
\bibitem [{\citenamefont {Pradeep~Kumar}\ \emph {et~al.}(2020)\citenamefont {Pradeep~Kumar}, \citenamefont {Shanmugha~Sundaram}, \citenamefont {Sharma}, \citenamefont {Venkatesh},\ and\ \citenamefont {Thiruvengadathan}}]{PradeepKumar2020b}%
  \BibitemOpen
  \bibfield  {author} {\bibinfo {author} {\bibfnamefont {K.~A.}\ \bibnamefont {Pradeep~Kumar}}, \bibinfo {author} {\bibfnamefont {G.~A.}\ \bibnamefont {Shanmugha~Sundaram}}, \bibinfo {author} {\bibfnamefont {B.~K.}\ \bibnamefont {Sharma}}, \bibinfo {author} {\bibfnamefont {S.}~\bibnamefont {Venkatesh}},\ and\ \bibinfo {author} {\bibfnamefont {R.}~\bibnamefont {Thiruvengadathan}},\ }\bibfield  {title} {\bibinfo {title} {Advances in gamma radiation detection systems for emergency radiation monitoring},\ }\href {https://doi.org/10.1016/j.net.2020.03.014} {\bibfield  {journal} {\bibinfo  {journal} {Nucl. Eng. Technol.}\ }\textbf {\bibinfo {volume} {52}},\ \bibinfo {pages} {2151} (\bibinfo {year} {2020})}\BibitemShut {NoStop}%
\bibitem [{\citenamefont {Deal}\ \emph {et~al.}(1972)\citenamefont {Deal}, \citenamefont {Doyle}, \citenamefont {Burson},\ and\ \citenamefont {Boyns}}]{Deal1972a}%
  \BibitemOpen
  \bibfield  {author} {\bibinfo {author} {\bibfnamefont {L.~J.}\ \bibnamefont {Deal}}, \bibinfo {author} {\bibfnamefont {J.~F.}\ \bibnamefont {Doyle}}, \bibinfo {author} {\bibfnamefont {Z.~G.}\ \bibnamefont {Burson}},\ and\ \bibinfo {author} {\bibfnamefont {P.~K.}\ \bibnamefont {Boyns}},\ }\bibfield  {title} {\bibinfo {title} {Locating the lost athena missile in mexico by the aerial radiological measuring system ({{ARMS}})},\ }\href {https://doi.org/10.1097/00004032-197207000-00013} {\bibfield  {journal} {\bibinfo  {journal} {Health Phys.}\ }\textbf {\bibinfo {volume} {23}},\ \bibinfo {pages} {95} (\bibinfo {year} {1972})}\BibitemShut {NoStop}%
\bibitem [{\citenamefont {Prieto}\ \emph {et~al.}(2020)\citenamefont {Prieto}, \citenamefont {Jabaloyas}, \citenamefont {Casanovas}, \citenamefont {Rovira},\ and\ \citenamefont {Salvad{\'o}}}]{Prieto2020a}%
  \BibitemOpen
  \bibfield  {author} {\bibinfo {author} {\bibfnamefont {E.}~\bibnamefont {Prieto}}, \bibinfo {author} {\bibfnamefont {E.}~\bibnamefont {Jabaloyas}}, \bibinfo {author} {\bibfnamefont {R.}~\bibnamefont {Casanovas}}, \bibinfo {author} {\bibfnamefont {C.}~\bibnamefont {Rovira}},\ and\ \bibinfo {author} {\bibfnamefont {M.}~\bibnamefont {Salvad{\'o}}},\ }\bibfield  {title} {\bibinfo {title} {Set up of a gamma spectrometry mobile unit equipped with {{LaBr3}}({{Ce}}) detectors for radioactivity monitoring},\ }\href {https://doi.org/10.1016/J.RADPHYSCHEM.2019.108600} {\bibfield  {journal} {\bibinfo  {journal} {Radiat. Phys. Chem.}\ }\textbf {\bibinfo {volume} {168}},\ \bibinfo {pages} {108600} (\bibinfo {year} {2020})}\BibitemShut {NoStop}%
\bibitem [{\citenamefont {Hellfeld}\ \emph {et~al.}(2021)\citenamefont {Hellfeld}, \citenamefont {Bandstra}, \citenamefont {Vavrek}, \citenamefont {Gunter}, \citenamefont {Curtis}, \citenamefont {Salathe}, \citenamefont {Pavlovsky}, \citenamefont {Negut}, \citenamefont {Barton}, \citenamefont {Cates}, \citenamefont {Quiter}, \citenamefont {Cooper}, \citenamefont {Vetter},\ and\ \citenamefont {Joshi}}]{Hellfeld2021a}%
  \BibitemOpen
  \bibfield  {author} {\bibinfo {author} {\bibfnamefont {D.}~\bibnamefont {Hellfeld}}, \bibinfo {author} {\bibfnamefont {M.~S.}\ \bibnamefont {Bandstra}}, \bibinfo {author} {\bibfnamefont {J.~R.}\ \bibnamefont {Vavrek}}, \bibinfo {author} {\bibfnamefont {D.~L.}\ \bibnamefont {Gunter}}, \bibinfo {author} {\bibfnamefont {J.~C.}\ \bibnamefont {Curtis}}, \bibinfo {author} {\bibfnamefont {M.}~\bibnamefont {Salathe}}, \bibinfo {author} {\bibfnamefont {R.}~\bibnamefont {Pavlovsky}}, \bibinfo {author} {\bibfnamefont {V.}~\bibnamefont {Negut}}, \bibinfo {author} {\bibfnamefont {P.~J.}\ \bibnamefont {Barton}}, \bibinfo {author} {\bibfnamefont {J.~W.}\ \bibnamefont {Cates}}, \bibinfo {author} {\bibfnamefont {B.~J.}\ \bibnamefont {Quiter}}, \bibinfo {author} {\bibfnamefont {R.~J.}\ \bibnamefont {Cooper}}, \bibinfo {author} {\bibfnamefont {K.}~\bibnamefont {Vetter}},\ and\ \bibinfo {author} {\bibfnamefont {T.~H.}\ \bibnamefont {Joshi}},\ }\bibfield  {title} {\bibinfo {title} {Free-moving {{Quantitative Gamma-ray
  Imaging}}},\ }\href {https://doi.org/10.1038/s41598-021-99588-z} {\bibfield  {journal} {\bibinfo  {journal} {Sci. Rep.}\ }\textbf {\bibinfo {volume} {11}},\ \bibinfo {pages} {20515} (\bibinfo {year} {2021})},\ \Eprint {https://arxiv.org/abs/2107.04080} {arXiv:2107.04080} \BibitemShut {NoStop}%
\bibitem [{\citenamefont {Bandstra}\ \emph {et~al.}(2021)\citenamefont {Bandstra}, \citenamefont {Hellfeld}, \citenamefont {Vavrek}, \citenamefont {Quiter}, \citenamefont {Meehan}, \citenamefont {Barton}, \citenamefont {Cates}, \citenamefont {Moran}, \citenamefont {Negut}, \citenamefont {Pavlovsky},\ and\ \citenamefont {Joshi}}]{Bandstra2021a}%
  \BibitemOpen
  \bibfield  {author} {\bibinfo {author} {\bibfnamefont {M.~S.}\ \bibnamefont {Bandstra}}, \bibinfo {author} {\bibfnamefont {D.}~\bibnamefont {Hellfeld}}, \bibinfo {author} {\bibfnamefont {J.~R.}\ \bibnamefont {Vavrek}}, \bibinfo {author} {\bibfnamefont {B.~J.}\ \bibnamefont {Quiter}}, \bibinfo {author} {\bibfnamefont {K.}~\bibnamefont {Meehan}}, \bibinfo {author} {\bibfnamefont {P.~J.}\ \bibnamefont {Barton}}, \bibinfo {author} {\bibfnamefont {J.~W.}\ \bibnamefont {Cates}}, \bibinfo {author} {\bibfnamefont {A.}~\bibnamefont {Moran}}, \bibinfo {author} {\bibfnamefont {V.}~\bibnamefont {Negut}}, \bibinfo {author} {\bibfnamefont {R.}~\bibnamefont {Pavlovsky}},\ and\ \bibinfo {author} {\bibfnamefont {T.~H.~Y.}\ \bibnamefont {Joshi}},\ }\bibfield  {title} {\bibinfo {title} {Improved {{Gamma-Ray Point Source Quantification}} in {{Three Dimensions}} by {{Modeling Attenuation}} in the {{Scene}}},\ }\href {https://doi.org/10.1109/TNS.2021.3113588} {\bibfield  {journal} {\bibinfo  {journal} {IEEE Trans. Nucl. Sci.}\
  }\textbf {\bibinfo {volume} {68}},\ \bibinfo {pages} {2637} (\bibinfo {year} {2021})},\ \Eprint {https://arxiv.org/abs/2104.11318v2} {arXiv:2104.11318v2} \BibitemShut {NoStop}%
\bibitem [{\citenamefont {Curtis}\ \emph {et~al.}(2020)\citenamefont {Curtis}, \citenamefont {Cooper}, \citenamefont {Joshi}, \citenamefont {Cosofret}, \citenamefont {Schmit}, \citenamefont {Wright}, \citenamefont {Rameau}, \citenamefont {Konno}, \citenamefont {Brown}, \citenamefont {Otsuka}, \citenamefont {Rappeport}, \citenamefont {Marshall},\ and\ \citenamefont {Speicher}}]{Curtis2020a}%
  \BibitemOpen
  \bibfield  {author} {\bibinfo {author} {\bibfnamefont {J.~C.}\ \bibnamefont {Curtis}}, \bibinfo {author} {\bibfnamefont {R.~J.}\ \bibnamefont {Cooper}}, \bibinfo {author} {\bibfnamefont {T.~H.}\ \bibnamefont {Joshi}}, \bibinfo {author} {\bibfnamefont {B.}~\bibnamefont {Cosofret}}, \bibinfo {author} {\bibfnamefont {T.}~\bibnamefont {Schmit}}, \bibinfo {author} {\bibfnamefont {J.}~\bibnamefont {Wright}}, \bibinfo {author} {\bibfnamefont {J.}~\bibnamefont {Rameau}}, \bibinfo {author} {\bibfnamefont {D.}~\bibnamefont {Konno}}, \bibinfo {author} {\bibfnamefont {D.}~\bibnamefont {Brown}}, \bibinfo {author} {\bibfnamefont {F.}~\bibnamefont {Otsuka}}, \bibinfo {author} {\bibfnamefont {E.}~\bibnamefont {Rappeport}}, \bibinfo {author} {\bibfnamefont {M.}~\bibnamefont {Marshall}},\ and\ \bibinfo {author} {\bibfnamefont {J.}~\bibnamefont {Speicher}},\ }\bibfield  {title} {\bibinfo {title} {Simulation and validation of the {{Mobile Urban Radiation Search}} ({{MURS}}) gamma-ray detector response},\ }\href
  {https://doi.org/10.1016/j.nima.2018.08.087} {\bibfield  {journal} {\bibinfo  {journal} {Nucl. Instrum. Methods Phys. Res. Sect. Accel. Spectrometers Detect. Assoc. Equip.}\ }\textbf {\bibinfo {volume} {954}},\ \bibinfo {pages} {161128} (\bibinfo {year} {2020})}\BibitemShut {NoStop}%
\bibitem [{\citenamefont {Fishman}\ \emph {et~al.}(1994)\citenamefont {Fishman}, \citenamefont {Bhat}, \citenamefont {Mallozzi}, \citenamefont {Horack}, \citenamefont {Koshut}, \citenamefont {Kouveliotou}, \citenamefont {Pendleton}, \citenamefont {Meegan}, \citenamefont {Wilson}, \citenamefont {Paciesas}, \citenamefont {Goodman},\ and\ \citenamefont {Christian}}]{Fishman1994}%
  \BibitemOpen
  \bibfield  {author} {\bibinfo {author} {\bibfnamefont {G.~J.}\ \bibnamefont {Fishman}}, \bibinfo {author} {\bibfnamefont {P.~N.}\ \bibnamefont {Bhat}}, \bibinfo {author} {\bibfnamefont {R.}~\bibnamefont {Mallozzi}}, \bibinfo {author} {\bibfnamefont {J.~M.}\ \bibnamefont {Horack}}, \bibinfo {author} {\bibfnamefont {T.}~\bibnamefont {Koshut}}, \bibinfo {author} {\bibfnamefont {C.}~\bibnamefont {Kouveliotou}}, \bibinfo {author} {\bibfnamefont {G.~N.}\ \bibnamefont {Pendleton}}, \bibinfo {author} {\bibfnamefont {C.~A.}\ \bibnamefont {Meegan}}, \bibinfo {author} {\bibfnamefont {R.~B.}\ \bibnamefont {Wilson}}, \bibinfo {author} {\bibfnamefont {W.~S.}\ \bibnamefont {Paciesas}}, \bibinfo {author} {\bibfnamefont {S.~J.}\ \bibnamefont {Goodman}},\ and\ \bibinfo {author} {\bibfnamefont {H.~J.}\ \bibnamefont {Christian}},\ }\bibfield  {title} {\bibinfo {title} {Discovery of {{Intense Gamma-Ray Flashes}} of {{Atmospheric Origin}}},\ }\href {https://doi.org/10.1126/science.264.5163.1313} {\bibfield  {journal} {\bibinfo
  {journal} {Science}\ }\textbf {\bibinfo {volume} {264}},\ \bibinfo {pages} {1313} (\bibinfo {year} {1994})}\BibitemShut {NoStop}%
\bibitem [{\citenamefont {Briggs}\ \emph {et~al.}(2010)\citenamefont {Briggs}, \citenamefont {Fishman}, \citenamefont {Connaughton}, \citenamefont {Bhat}, \citenamefont {Paciesas}, \citenamefont {Preece}, \citenamefont {{Wilson-Hodge}}, \citenamefont {Chaplin}, \citenamefont {Kippen}, \citenamefont {{von Kienlin}}, \citenamefont {Meegan}, \citenamefont {Bissaldi}, \citenamefont {Dwyer}, \citenamefont {Smith}, \citenamefont {Holzworth}, \citenamefont {Grove},\ and\ \citenamefont {Chekhtman}}]{Briggs2010}%
  \BibitemOpen
  \bibfield  {author} {\bibinfo {author} {\bibfnamefont {M.~S.}\ \bibnamefont {Briggs}}, \bibinfo {author} {\bibfnamefont {G.~J.}\ \bibnamefont {Fishman}}, \bibinfo {author} {\bibfnamefont {V.}~\bibnamefont {Connaughton}}, \bibinfo {author} {\bibfnamefont {P.~N.}\ \bibnamefont {Bhat}}, \bibinfo {author} {\bibfnamefont {W.~S.}\ \bibnamefont {Paciesas}}, \bibinfo {author} {\bibfnamefont {R.~D.}\ \bibnamefont {Preece}}, \bibinfo {author} {\bibfnamefont {C.}~\bibnamefont {{Wilson-Hodge}}}, \bibinfo {author} {\bibfnamefont {V.~L.}\ \bibnamefont {Chaplin}}, \bibinfo {author} {\bibfnamefont {R.~M.}\ \bibnamefont {Kippen}}, \bibinfo {author} {\bibfnamefont {A.}~\bibnamefont {{von Kienlin}}}, \bibinfo {author} {\bibfnamefont {C.~A.}\ \bibnamefont {Meegan}}, \bibinfo {author} {\bibfnamefont {E.}~\bibnamefont {Bissaldi}}, \bibinfo {author} {\bibfnamefont {J.~R.}\ \bibnamefont {Dwyer}}, \bibinfo {author} {\bibfnamefont {D.~M.}\ \bibnamefont {Smith}}, \bibinfo {author} {\bibfnamefont {R.~H.}\ \bibnamefont {Holzworth}},
  \bibinfo {author} {\bibfnamefont {J.~E.}\ \bibnamefont {Grove}},\ and\ \bibinfo {author} {\bibfnamefont {A.}~\bibnamefont {Chekhtman}},\ }\bibfield  {title} {\bibinfo {title} {First results on terrestrial gamma ray flashes from the {{Fermi Gamma-ray Burst Monitor}}},\ }\bibfield  {journal} {\bibinfo  {journal} {J. Geophys. Res. Space Phys.}\ }\textbf {\bibinfo {volume} {115}},\ \href {https://doi.org/10.1029/2009JA015242} {10.1029/2009JA015242} (\bibinfo {year} {2010})\BibitemShut {NoStop}%
\bibitem [{\citenamefont {Tavani}\ \emph {et~al.}(2011)\citenamefont {Tavani}, \citenamefont {Marisaldi}, \citenamefont {Labanti}, \citenamefont {Fuschino}, \citenamefont {Argan}, \citenamefont {Trois}, \citenamefont {Giommi}, \citenamefont {Colafrancesco}, \citenamefont {Pittori}, \citenamefont {Palma}, \citenamefont {Trifoglio}, \citenamefont {Gianotti}, \citenamefont {Bulgarelli}, \citenamefont {Vittorini}, \citenamefont {Verrecchia}, \citenamefont {Salotti}, \citenamefont {Barbiellini}, \citenamefont {Caraveo}, \citenamefont {Cattaneo}, \citenamefont {Chen}, \citenamefont {Contessi}, \citenamefont {Costa}, \citenamefont {D'Ammando}, \citenamefont {Del~Monte}, \citenamefont {De~Paris}, \citenamefont {Di~Cocco}, \citenamefont {Di~Persio}, \citenamefont {Donnarumma}, \citenamefont {Evangelista}, \citenamefont {Feroci}, \citenamefont {Ferrari}, \citenamefont {Galli}, \citenamefont {Giuliani}, \citenamefont {Giusti}, \citenamefont {Lapshov}, \citenamefont {Lazzarotto}, \citenamefont {Lipari}, \citenamefont
  {Longo}, \citenamefont {Mereghetti}, \citenamefont {Morelli}, \citenamefont {Moretti}, \citenamefont {Morselli}, \citenamefont {Pacciani}, \citenamefont {Pellizzoni}, \citenamefont {Perotti}, \citenamefont {Piano}, \citenamefont {Picozza}, \citenamefont {Pilia}, \citenamefont {Pucella}, \citenamefont {Prest}, \citenamefont {Rapisarda}, \citenamefont {Rappoldi}, \citenamefont {Rossi}, \citenamefont {Rubini}, \citenamefont {Sabatini}, \citenamefont {Scalise}, \citenamefont {Soffitta}, \citenamefont {Striani}, \citenamefont {Vallazza}, \citenamefont {Vercellone}, \citenamefont {Zambra},\ and\ \citenamefont {Zanello}}]{Tavani2011}%
  \BibitemOpen
  \bibfield  {author} {\bibinfo {author} {\bibfnamefont {M.}~\bibnamefont {Tavani}}, \bibinfo {author} {\bibfnamefont {M.}~\bibnamefont {Marisaldi}}, \bibinfo {author} {\bibfnamefont {C.}~\bibnamefont {Labanti}}, \bibinfo {author} {\bibfnamefont {F.}~\bibnamefont {Fuschino}}, \bibinfo {author} {\bibfnamefont {A.}~\bibnamefont {Argan}}, \bibinfo {author} {\bibfnamefont {A.}~\bibnamefont {Trois}}, \bibinfo {author} {\bibfnamefont {P.}~\bibnamefont {Giommi}}, \bibinfo {author} {\bibfnamefont {S.}~\bibnamefont {Colafrancesco}}, \bibinfo {author} {\bibfnamefont {C.}~\bibnamefont {Pittori}}, \bibinfo {author} {\bibfnamefont {F.}~\bibnamefont {Palma}}, \bibinfo {author} {\bibfnamefont {M.}~\bibnamefont {Trifoglio}}, \bibinfo {author} {\bibfnamefont {F.}~\bibnamefont {Gianotti}}, \bibinfo {author} {\bibfnamefont {A.}~\bibnamefont {Bulgarelli}}, \bibinfo {author} {\bibfnamefont {V.}~\bibnamefont {Vittorini}}, \bibinfo {author} {\bibfnamefont {F.}~\bibnamefont {Verrecchia}}, \bibinfo {author} {\bibfnamefont
  {L.}~\bibnamefont {Salotti}}, \bibinfo {author} {\bibfnamefont {G.}~\bibnamefont {Barbiellini}}, \bibinfo {author} {\bibfnamefont {P.}~\bibnamefont {Caraveo}}, \bibinfo {author} {\bibfnamefont {P.~W.}\ \bibnamefont {Cattaneo}}, \bibinfo {author} {\bibfnamefont {A.}~\bibnamefont {Chen}}, \bibinfo {author} {\bibfnamefont {T.}~\bibnamefont {Contessi}}, \bibinfo {author} {\bibfnamefont {E.}~\bibnamefont {Costa}}, \bibinfo {author} {\bibfnamefont {F.}~\bibnamefont {D'Ammando}}, \bibinfo {author} {\bibfnamefont {E.}~\bibnamefont {Del~Monte}}, \bibinfo {author} {\bibfnamefont {G.}~\bibnamefont {De~Paris}}, \bibinfo {author} {\bibfnamefont {G.}~\bibnamefont {Di~Cocco}}, \bibinfo {author} {\bibfnamefont {G.}~\bibnamefont {Di~Persio}}, \bibinfo {author} {\bibfnamefont {I.}~\bibnamefont {Donnarumma}}, \bibinfo {author} {\bibfnamefont {Y.}~\bibnamefont {Evangelista}}, \bibinfo {author} {\bibfnamefont {M.}~\bibnamefont {Feroci}}, \bibinfo {author} {\bibfnamefont {A.}~\bibnamefont {Ferrari}}, \bibinfo {author}
  {\bibfnamefont {M.}~\bibnamefont {Galli}}, \bibinfo {author} {\bibfnamefont {A.}~\bibnamefont {Giuliani}}, \bibinfo {author} {\bibfnamefont {M.}~\bibnamefont {Giusti}}, \bibinfo {author} {\bibfnamefont {I.}~\bibnamefont {Lapshov}}, \bibinfo {author} {\bibfnamefont {F.}~\bibnamefont {Lazzarotto}}, \bibinfo {author} {\bibfnamefont {P.}~\bibnamefont {Lipari}}, \bibinfo {author} {\bibfnamefont {F.}~\bibnamefont {Longo}}, \bibinfo {author} {\bibfnamefont {S.}~\bibnamefont {Mereghetti}}, \bibinfo {author} {\bibfnamefont {E.}~\bibnamefont {Morelli}}, \bibinfo {author} {\bibfnamefont {E.}~\bibnamefont {Moretti}}, \bibinfo {author} {\bibfnamefont {A.}~\bibnamefont {Morselli}}, \bibinfo {author} {\bibfnamefont {L.}~\bibnamefont {Pacciani}}, \bibinfo {author} {\bibfnamefont {A.}~\bibnamefont {Pellizzoni}}, \bibinfo {author} {\bibfnamefont {F.}~\bibnamefont {Perotti}}, \bibinfo {author} {\bibfnamefont {G.}~\bibnamefont {Piano}}, \bibinfo {author} {\bibfnamefont {P.}~\bibnamefont {Picozza}}, \bibinfo {author}
  {\bibfnamefont {M.}~\bibnamefont {Pilia}}, \bibinfo {author} {\bibfnamefont {G.}~\bibnamefont {Pucella}}, \bibinfo {author} {\bibfnamefont {M.}~\bibnamefont {Prest}}, \bibinfo {author} {\bibfnamefont {M.}~\bibnamefont {Rapisarda}}, \bibinfo {author} {\bibfnamefont {A.}~\bibnamefont {Rappoldi}}, \bibinfo {author} {\bibfnamefont {E.}~\bibnamefont {Rossi}}, \bibinfo {author} {\bibfnamefont {A.}~\bibnamefont {Rubini}}, \bibinfo {author} {\bibfnamefont {S.}~\bibnamefont {Sabatini}}, \bibinfo {author} {\bibfnamefont {E.}~\bibnamefont {Scalise}}, \bibinfo {author} {\bibfnamefont {P.}~\bibnamefont {Soffitta}}, \bibinfo {author} {\bibfnamefont {E.}~\bibnamefont {Striani}}, \bibinfo {author} {\bibfnamefont {E.}~\bibnamefont {Vallazza}}, \bibinfo {author} {\bibfnamefont {S.}~\bibnamefont {Vercellone}}, \bibinfo {author} {\bibfnamefont {A.}~\bibnamefont {Zambra}},\ and\ \bibinfo {author} {\bibfnamefont {D.}~\bibnamefont {Zanello}},\ }\bibfield  {title} {\bibinfo {title} {Terrestrial {{Gamma-Ray Flashes}} as {{Powerful
  Particle Accelerators}}},\ }\href {https://doi.org/10.1103/PhysRevLett.106.018501} {\bibfield  {journal} {\bibinfo  {journal} {Phys. Rev. Lett.}\ }\textbf {\bibinfo {volume} {106}},\ \bibinfo {pages} {018501} (\bibinfo {year} {2011})}\BibitemShut {NoStop}%
\bibitem [{\citenamefont {Smith}\ \emph {et~al.}(2011)\citenamefont {Smith}, \citenamefont {Dwyer}, \citenamefont {Hazelton}, \citenamefont {Grefenstette}, \citenamefont {{Martinez-Mckinney}}, \citenamefont {Zhang}, \citenamefont {Lowell}, \citenamefont {Kelley}, \citenamefont {Splitt}, \citenamefont {Lazarus}, \citenamefont {Ulrich}, \citenamefont {Schaal}, \citenamefont {Saleh}, \citenamefont {Cramer}, \citenamefont {Rassoul}, \citenamefont {Cummer}, \citenamefont {Lu}, \citenamefont {Shao}, \citenamefont {Ho}, \citenamefont {Hamlin}, \citenamefont {Blakeslee},\ and\ \citenamefont {Heckman}}]{Smith2011a}%
  \BibitemOpen
  \bibfield  {author} {\bibinfo {author} {\bibfnamefont {D.~M.}\ \bibnamefont {Smith}}, \bibinfo {author} {\bibfnamefont {J.~R.}\ \bibnamefont {Dwyer}}, \bibinfo {author} {\bibfnamefont {B.~J.}\ \bibnamefont {Hazelton}}, \bibinfo {author} {\bibfnamefont {B.~W.}\ \bibnamefont {Grefenstette}}, \bibinfo {author} {\bibfnamefont {G.~F.}\ \bibnamefont {{Martinez-Mckinney}}}, \bibinfo {author} {\bibfnamefont {Z.~Y.}\ \bibnamefont {Zhang}}, \bibinfo {author} {\bibfnamefont {A.~W.}\ \bibnamefont {Lowell}}, \bibinfo {author} {\bibfnamefont {N.~A.}\ \bibnamefont {Kelley}}, \bibinfo {author} {\bibfnamefont {M.~E.}\ \bibnamefont {Splitt}}, \bibinfo {author} {\bibfnamefont {S.~M.}\ \bibnamefont {Lazarus}}, \bibinfo {author} {\bibfnamefont {W.}~\bibnamefont {Ulrich}}, \bibinfo {author} {\bibfnamefont {M.}~\bibnamefont {Schaal}}, \bibinfo {author} {\bibfnamefont {Z.~H.}\ \bibnamefont {Saleh}}, \bibinfo {author} {\bibfnamefont {E.}~\bibnamefont {Cramer}}, \bibinfo {author} {\bibfnamefont {H.}~\bibnamefont {Rassoul}}, \bibinfo
  {author} {\bibfnamefont {S.~A.}\ \bibnamefont {Cummer}}, \bibinfo {author} {\bibfnamefont {G.}~\bibnamefont {Lu}}, \bibinfo {author} {\bibfnamefont {X.~M.}\ \bibnamefont {Shao}}, \bibinfo {author} {\bibfnamefont {C.}~\bibnamefont {Ho}}, \bibinfo {author} {\bibfnamefont {T.}~\bibnamefont {Hamlin}}, \bibinfo {author} {\bibfnamefont {R.~J.}\ \bibnamefont {Blakeslee}},\ and\ \bibinfo {author} {\bibfnamefont {S.}~\bibnamefont {Heckman}},\ }\bibfield  {title} {\bibinfo {title} {A terrestrial gamma ray flash observed from an aircraft},\ }\href {https://doi.org/10.1029/2011JD016252} {\bibfield  {journal} {\bibinfo  {journal} {J. Geophys. Res. Atmospheres}\ }\textbf {\bibinfo {volume} {116}},\ \bibinfo {pages} {20124} (\bibinfo {year} {2011})}\BibitemShut {NoStop}%
\bibitem [{\citenamefont {Gjesteland}\ \emph {et~al.}(2015)\citenamefont {Gjesteland}, \citenamefont {{\O}stgaard}, \citenamefont {Laviola}, \citenamefont {Miglietta}, \citenamefont {Arnone}, \citenamefont {Marisaldi}, \citenamefont {Fuschino}, \citenamefont {Collier}, \citenamefont {Fabr{\'o}},\ and\ \citenamefont {Montanya}}]{Gjesteland2015}%
  \BibitemOpen
  \bibfield  {author} {\bibinfo {author} {\bibfnamefont {T.}~\bibnamefont {Gjesteland}}, \bibinfo {author} {\bibfnamefont {N.}~\bibnamefont {{\O}stgaard}}, \bibinfo {author} {\bibfnamefont {S.}~\bibnamefont {Laviola}}, \bibinfo {author} {\bibfnamefont {M.~M.}\ \bibnamefont {Miglietta}}, \bibinfo {author} {\bibfnamefont {E.}~\bibnamefont {Arnone}}, \bibinfo {author} {\bibfnamefont {M.}~\bibnamefont {Marisaldi}}, \bibinfo {author} {\bibfnamefont {F.}~\bibnamefont {Fuschino}}, \bibinfo {author} {\bibfnamefont {A.~B.}\ \bibnamefont {Collier}}, \bibinfo {author} {\bibfnamefont {F.}~\bibnamefont {Fabr{\'o}}},\ and\ \bibinfo {author} {\bibfnamefont {J.}~\bibnamefont {Montanya}},\ }\bibfield  {title} {\bibinfo {title} {Observation of intrinsically bright terrestrial gamma ray flashes from the {{Mediterranean}} basin},\ }\href {https://doi.org/10.1002/2015JD023704} {\bibfield  {journal} {\bibinfo  {journal} {J. Geophys. Res. Atmospheres}\ }\textbf {\bibinfo {volume} {120}},\ \bibinfo {pages} {12,143} (\bibinfo {year}
  {2015})}\BibitemShut {NoStop}%
\bibitem [{\citenamefont {Neubert}\ \emph {et~al.}(2020)\citenamefont {Neubert}, \citenamefont {{\O}stgaard}, \citenamefont {Reglero}, \citenamefont {Chanrion}, \citenamefont {Heumesser}, \citenamefont {Dimitriadou}, \citenamefont {Christiansen}, \citenamefont {{Budtz-J{\o}rgensen}}, \citenamefont {Kuvvetli}, \citenamefont {Rasmussen}, \citenamefont {Mezentsev}, \citenamefont {Marisaldi}, \citenamefont {Ullaland}, \citenamefont {Genov}, \citenamefont {Yang}, \citenamefont {Kochkin}, \citenamefont {{Navarro-Gonzalez}}, \citenamefont {Connell},\ and\ \citenamefont {Eyles}}]{Neubert2020a}%
  \BibitemOpen
  \bibfield  {author} {\bibinfo {author} {\bibfnamefont {T.}~\bibnamefont {Neubert}}, \bibinfo {author} {\bibfnamefont {N.}~\bibnamefont {{\O}stgaard}}, \bibinfo {author} {\bibfnamefont {V.}~\bibnamefont {Reglero}}, \bibinfo {author} {\bibfnamefont {O.}~\bibnamefont {Chanrion}}, \bibinfo {author} {\bibfnamefont {M.}~\bibnamefont {Heumesser}}, \bibinfo {author} {\bibfnamefont {K.}~\bibnamefont {Dimitriadou}}, \bibinfo {author} {\bibfnamefont {F.}~\bibnamefont {Christiansen}}, \bibinfo {author} {\bibfnamefont {C.}~\bibnamefont {{Budtz-J{\o}rgensen}}}, \bibinfo {author} {\bibfnamefont {I.}~\bibnamefont {Kuvvetli}}, \bibinfo {author} {\bibfnamefont {I.~L.}\ \bibnamefont {Rasmussen}}, \bibinfo {author} {\bibfnamefont {A.}~\bibnamefont {Mezentsev}}, \bibinfo {author} {\bibfnamefont {M.}~\bibnamefont {Marisaldi}}, \bibinfo {author} {\bibfnamefont {K.}~\bibnamefont {Ullaland}}, \bibinfo {author} {\bibfnamefont {G.}~\bibnamefont {Genov}}, \bibinfo {author} {\bibfnamefont {S.}~\bibnamefont {Yang}}, \bibinfo {author}
  {\bibfnamefont {P.}~\bibnamefont {Kochkin}}, \bibinfo {author} {\bibfnamefont {J.}~\bibnamefont {{Navarro-Gonzalez}}}, \bibinfo {author} {\bibfnamefont {P.~H.}\ \bibnamefont {Connell}},\ and\ \bibinfo {author} {\bibfnamefont {C.~J.}\ \bibnamefont {Eyles}},\ }\bibfield  {title} {\bibinfo {title} {A terrestrial gamma-ray flash and ionospheric ultraviolet emissions powered by lightning},\ }\href {https://doi.org/10.1126/science.aax3872} {\bibfield  {journal} {\bibinfo  {journal} {Science}\ }\textbf {\bibinfo {volume} {367}},\ \bibinfo {pages} {183} (\bibinfo {year} {2020})}\BibitemShut {NoStop}%
\bibitem [{\citenamefont {Appleton}\ \emph {et~al.}(2008)\citenamefont {Appleton}, \citenamefont {Miles}, \citenamefont {Green},\ and\ \citenamefont {Larmour}}]{Appleton2008a}%
  \BibitemOpen
  \bibfield  {author} {\bibinfo {author} {\bibfnamefont {J.~D.}\ \bibnamefont {Appleton}}, \bibinfo {author} {\bibfnamefont {J.~C.}\ \bibnamefont {Miles}}, \bibinfo {author} {\bibfnamefont {B.~M.}\ \bibnamefont {Green}},\ and\ \bibinfo {author} {\bibfnamefont {R.}~\bibnamefont {Larmour}},\ }\bibfield  {title} {\bibinfo {title} {Pilot study of the application of {{Tellus}} airborne radiometric and soil geochemical data for radon mapping},\ }\href {https://doi.org/10.1016/J.JENVRAD.2008.03.011} {\bibfield  {journal} {\bibinfo  {journal} {J. Environ. Radioact.}\ }\textbf {\bibinfo {volume} {99}},\ \bibinfo {pages} {1687} (\bibinfo {year} {2008})}\BibitemShut {NoStop}%
\bibitem [{\citenamefont {Sinclair}\ \emph {et~al.}(2011)\citenamefont {Sinclair}, \citenamefont {Seywerd}, \citenamefont {Fortin}, \citenamefont {Carson}, \citenamefont {Saull}, \citenamefont {Coyle}, \citenamefont {Van~Brabant}, \citenamefont {Buckle}, \citenamefont {Desjardins},\ and\ \citenamefont {Hall}}]{Sinclair2011a}%
  \BibitemOpen
  \bibfield  {author} {\bibinfo {author} {\bibfnamefont {L.~E.}\ \bibnamefont {Sinclair}}, \bibinfo {author} {\bibfnamefont {H.~C.}\ \bibnamefont {Seywerd}}, \bibinfo {author} {\bibfnamefont {R.}~\bibnamefont {Fortin}}, \bibinfo {author} {\bibfnamefont {J.~M.}\ \bibnamefont {Carson}}, \bibinfo {author} {\bibfnamefont {P.~R.}\ \bibnamefont {Saull}}, \bibinfo {author} {\bibfnamefont {M.~J.}\ \bibnamefont {Coyle}}, \bibinfo {author} {\bibfnamefont {R.~A.}\ \bibnamefont {Van~Brabant}}, \bibinfo {author} {\bibfnamefont {J.~L.}\ \bibnamefont {Buckle}}, \bibinfo {author} {\bibfnamefont {S.~M.}\ \bibnamefont {Desjardins}},\ and\ \bibinfo {author} {\bibfnamefont {R.~M.}\ \bibnamefont {Hall}},\ }\bibfield  {title} {\bibinfo {title} {Aerial measurement of radioxenon concentration off the west coast of vancouver island following the fukushima reactor accident},\ }\href {https://doi.org/10.1016/j.jenvrad.2011.06.008} {\bibfield  {journal} {\bibinfo  {journal} {J. Environ. Radioact.}\ }\textbf {\bibinfo {volume} {102}},\
  \bibinfo {pages} {1018} (\bibinfo {year} {2011})}\BibitemShut {NoStop}%
\bibitem [{\citenamefont {Baldoncini}\ \emph {et~al.}(2017)\citenamefont {Baldoncini}, \citenamefont {Alb{\'e}ri}, \citenamefont {Bottardi}, \citenamefont {Minty}, \citenamefont {Raptis}, \citenamefont {Strati},\ and\ \citenamefont {Mantovani}}]{Baldoncini2017a}%
  \BibitemOpen
  \bibfield  {author} {\bibinfo {author} {\bibfnamefont {M.}~\bibnamefont {Baldoncini}}, \bibinfo {author} {\bibfnamefont {M.}~\bibnamefont {Alb{\'e}ri}}, \bibinfo {author} {\bibfnamefont {C.}~\bibnamefont {Bottardi}}, \bibinfo {author} {\bibfnamefont {B.}~\bibnamefont {Minty}}, \bibinfo {author} {\bibfnamefont {K.~G.}\ \bibnamefont {Raptis}}, \bibinfo {author} {\bibfnamefont {V.}~\bibnamefont {Strati}},\ and\ \bibinfo {author} {\bibfnamefont {F.}~\bibnamefont {Mantovani}},\ }\bibfield  {title} {\bibinfo {title} {Exploring atmospheric radon with airborne gamma-ray spectroscopy},\ }\href {https://doi.org/10.1016/j.atmosenv.2017.09.048} {\bibfield  {journal} {\bibinfo  {journal} {Atmos. Environ.}\ }\textbf {\bibinfo {volume} {170}},\ \bibinfo {pages} {259} (\bibinfo {year} {2017})}\BibitemShut {NoStop}%
\bibitem [{\citenamefont {Prettyman}\ \emph {et~al.}(2006)\citenamefont {Prettyman}, \citenamefont {Hagerty}, \citenamefont {Elphic}, \citenamefont {Feldman}, \citenamefont {Lawrence}, \citenamefont {McKinney},\ and\ \citenamefont {Vaniman}}]{Prettyman2006a}%
  \BibitemOpen
  \bibfield  {author} {\bibinfo {author} {\bibfnamefont {T.~H.}\ \bibnamefont {Prettyman}}, \bibinfo {author} {\bibfnamefont {J.~J.}\ \bibnamefont {Hagerty}}, \bibinfo {author} {\bibfnamefont {R.~C.}\ \bibnamefont {Elphic}}, \bibinfo {author} {\bibfnamefont {W.~C.}\ \bibnamefont {Feldman}}, \bibinfo {author} {\bibfnamefont {D.~J.}\ \bibnamefont {Lawrence}}, \bibinfo {author} {\bibfnamefont {G.~W.}\ \bibnamefont {McKinney}},\ and\ \bibinfo {author} {\bibfnamefont {D.~T.}\ \bibnamefont {Vaniman}},\ }\bibfield  {title} {\bibinfo {title} {Elemental composition of the lunar surface: {{Analysis}} of gamma ray spectroscopy data from {{Lunar Prospector}}},\ }\bibfield  {journal} {\bibinfo  {journal} {J. Geophys. Res. Planets}\ }\textbf {\bibinfo {volume} {111}},\ \href {https://doi.org/10.1029/2005JE002656} {10.1029/2005JE002656} (\bibinfo {year} {2006})\BibitemShut {NoStop}%
\bibitem [{\citenamefont {Hahn}\ \emph {et~al.}(2007)\citenamefont {Hahn}, \citenamefont {McLennan}, \citenamefont {Taylor}, \citenamefont {Boynton}, \citenamefont {Dohm}, \citenamefont {Finch}, \citenamefont {Hamara}, \citenamefont {Janes}, \citenamefont {Karunatillake}, \citenamefont {Keller}, \citenamefont {Kerry}, \citenamefont {Metzger},\ and\ \citenamefont {Williams}}]{Hahn2007a}%
  \BibitemOpen
  \bibfield  {author} {\bibinfo {author} {\bibfnamefont {B.~C.}\ \bibnamefont {Hahn}}, \bibinfo {author} {\bibfnamefont {S.~M.}\ \bibnamefont {McLennan}}, \bibinfo {author} {\bibfnamefont {G.~J.}\ \bibnamefont {Taylor}}, \bibinfo {author} {\bibfnamefont {W.~V.}\ \bibnamefont {Boynton}}, \bibinfo {author} {\bibfnamefont {J.~M.}\ \bibnamefont {Dohm}}, \bibinfo {author} {\bibfnamefont {M.~J.}\ \bibnamefont {Finch}}, \bibinfo {author} {\bibfnamefont {D.~K.}\ \bibnamefont {Hamara}}, \bibinfo {author} {\bibfnamefont {D.~M.}\ \bibnamefont {Janes}}, \bibinfo {author} {\bibfnamefont {S.}~\bibnamefont {Karunatillake}}, \bibinfo {author} {\bibfnamefont {J.~M.}\ \bibnamefont {Keller}}, \bibinfo {author} {\bibfnamefont {K.~E.}\ \bibnamefont {Kerry}}, \bibinfo {author} {\bibfnamefont {A.~E.}\ \bibnamefont {Metzger}},\ and\ \bibinfo {author} {\bibfnamefont {R.~M.~S.}\ \bibnamefont {Williams}},\ }\bibfield  {title} {\bibinfo {title} {Mars {{Odyssey Gamma Ray Spectrometer}} elemental abundances and apparent relative surface
  age: {{Implications}} for {{Martian}} crustal evolution},\ }\bibfield  {journal} {\bibinfo  {journal} {J. Geophys. Res. Planets}\ }\textbf {\bibinfo {volume} {112}},\ \href {https://doi.org/10.1029/2006JE002821@10.1002/(ISSN)2169-9100.GAMMARAY1} {10.1029/2006JE002821@10.1002/(ISSN)2169-9100.GAMMARAY1} (\bibinfo {year} {2007})\BibitemShut {NoStop}%
\bibitem [{\citenamefont {Kobayashi}\ \emph {et~al.}(2010)\citenamefont {Kobayashi}, \citenamefont {Hasebe}, \citenamefont {Shibamura}, \citenamefont {Okudaira}, \citenamefont {Kobayashi}, \citenamefont {Yamashita}, \citenamefont {Karouji}, \citenamefont {Hareyama}, \citenamefont {Hayatsu}, \citenamefont {D'Uston}, \citenamefont {Maurice}, \citenamefont {Gasnault}, \citenamefont {Forni}, \citenamefont {Diez}, \citenamefont {Reedy},\ and\ \citenamefont {Kim}}]{Kobayashi2010a}%
  \BibitemOpen
  \bibfield  {author} {\bibinfo {author} {\bibfnamefont {S.}~\bibnamefont {Kobayashi}}, \bibinfo {author} {\bibfnamefont {N.}~\bibnamefont {Hasebe}}, \bibinfo {author} {\bibfnamefont {E.}~\bibnamefont {Shibamura}}, \bibinfo {author} {\bibfnamefont {O.}~\bibnamefont {Okudaira}}, \bibinfo {author} {\bibfnamefont {M.}~\bibnamefont {Kobayashi}}, \bibinfo {author} {\bibfnamefont {N.}~\bibnamefont {Yamashita}}, \bibinfo {author} {\bibfnamefont {Y.}~\bibnamefont {Karouji}}, \bibinfo {author} {\bibfnamefont {M.}~\bibnamefont {Hareyama}}, \bibinfo {author} {\bibfnamefont {K.}~\bibnamefont {Hayatsu}}, \bibinfo {author} {\bibfnamefont {C.}~\bibnamefont {D'Uston}}, \bibinfo {author} {\bibfnamefont {S.}~\bibnamefont {Maurice}}, \bibinfo {author} {\bibfnamefont {O.}~\bibnamefont {Gasnault}}, \bibinfo {author} {\bibfnamefont {O.}~\bibnamefont {Forni}}, \bibinfo {author} {\bibfnamefont {B.}~\bibnamefont {Diez}}, \bibinfo {author} {\bibfnamefont {R.~C.}\ \bibnamefont {Reedy}},\ and\ \bibinfo {author} {\bibfnamefont {K.~J.}\
  \bibnamefont {Kim}},\ }\bibfield  {title} {\bibinfo {title} {Determining the absolute abundances of natural radioactive elements on the lunar surface by the kaguya gamma-ray spectrometer},\ }\href {https://doi.org/10.1007/s11214-010-9650-2} {\bibfield  {journal} {\bibinfo  {journal} {Space Sci. Rev.}\ }\textbf {\bibinfo {volume} {154}},\ \bibinfo {pages} {193} (\bibinfo {year} {2010})}\BibitemShut {NoStop}%
\bibitem [{\citenamefont {Prettyman}\ \emph {et~al.}(2011)\citenamefont {Prettyman}, \citenamefont {Feldman}, \citenamefont {McSween}, \citenamefont {Dingler}, \citenamefont {Enemark}, \citenamefont {Patrick}, \citenamefont {Storms}, \citenamefont {Hendricks}, \citenamefont {Morgenthaler}, \citenamefont {Pitman},\ and\ \citenamefont {Reedy}}]{Prettyman2011a}%
  \BibitemOpen
  \bibfield  {author} {\bibinfo {author} {\bibfnamefont {T.~H.}\ \bibnamefont {Prettyman}}, \bibinfo {author} {\bibfnamefont {W.~C.}\ \bibnamefont {Feldman}}, \bibinfo {author} {\bibfnamefont {H.~Y.}\ \bibnamefont {McSween}}, \bibinfo {author} {\bibfnamefont {R.~D.}\ \bibnamefont {Dingler}}, \bibinfo {author} {\bibfnamefont {D.~C.}\ \bibnamefont {Enemark}}, \bibinfo {author} {\bibfnamefont {D.~E.}\ \bibnamefont {Patrick}}, \bibinfo {author} {\bibfnamefont {S.~A.}\ \bibnamefont {Storms}}, \bibinfo {author} {\bibfnamefont {J.~S.}\ \bibnamefont {Hendricks}}, \bibinfo {author} {\bibfnamefont {J.~P.}\ \bibnamefont {Morgenthaler}}, \bibinfo {author} {\bibfnamefont {K.~M.}\ \bibnamefont {Pitman}},\ and\ \bibinfo {author} {\bibfnamefont {R.~C.}\ \bibnamefont {Reedy}},\ }\bibfield  {title} {\bibinfo {title} {Dawn's gamma ray and neutron detector},\ }\href {https://doi.org/10.1007/s11214-011-9862-0} {\bibfield  {journal} {\bibinfo  {journal} {Space Sci. Rev.}\ }\textbf {\bibinfo {volume} {163}},\ \bibinfo {pages} {371}
  (\bibinfo {year} {2011})}\BibitemShut {NoStop}%
\bibitem [{\citenamefont {Peplowski}\ \emph {et~al.}(2011)\citenamefont {Peplowski}, \citenamefont {Evans}, \citenamefont {Hauck}, \citenamefont {McCoy}, \citenamefont {Boynton}, \citenamefont {{Gillis-Davis}}, \citenamefont {Ebel}, \citenamefont {Goldsten}, \citenamefont {Hamara}, \citenamefont {Lawrence}, \citenamefont {McNutt}, \citenamefont {Nittler}, \citenamefont {Solomon}, \citenamefont {Rhodes}, \citenamefont {Sprague}, \citenamefont {Starr},\ and\ \citenamefont {{Stockstill-Cahill}}}]{Peplowski2011b}%
  \BibitemOpen
  \bibfield  {author} {\bibinfo {author} {\bibfnamefont {P.~N.}\ \bibnamefont {Peplowski}}, \bibinfo {author} {\bibfnamefont {L.~G.}\ \bibnamefont {Evans}}, \bibinfo {author} {\bibfnamefont {S.~A.}\ \bibnamefont {Hauck}}, \bibinfo {author} {\bibfnamefont {T.~J.}\ \bibnamefont {McCoy}}, \bibinfo {author} {\bibfnamefont {W.~V.}\ \bibnamefont {Boynton}}, \bibinfo {author} {\bibfnamefont {J.~J.}\ \bibnamefont {{Gillis-Davis}}}, \bibinfo {author} {\bibfnamefont {D.~S.}\ \bibnamefont {Ebel}}, \bibinfo {author} {\bibfnamefont {J.~O.}\ \bibnamefont {Goldsten}}, \bibinfo {author} {\bibfnamefont {D.~K.}\ \bibnamefont {Hamara}}, \bibinfo {author} {\bibfnamefont {D.~J.}\ \bibnamefont {Lawrence}}, \bibinfo {author} {\bibfnamefont {R.~L.}\ \bibnamefont {McNutt}}, \bibinfo {author} {\bibfnamefont {L.~R.}\ \bibnamefont {Nittler}}, \bibinfo {author} {\bibfnamefont {S.~C.}\ \bibnamefont {Solomon}}, \bibinfo {author} {\bibfnamefont {E.~A.}\ \bibnamefont {Rhodes}}, \bibinfo {author} {\bibfnamefont {A.~L.}\ \bibnamefont
  {Sprague}}, \bibinfo {author} {\bibfnamefont {R.~D.}\ \bibnamefont {Starr}},\ and\ \bibinfo {author} {\bibfnamefont {K.~R.}\ \bibnamefont {{Stockstill-Cahill}}},\ }\bibfield  {title} {\bibinfo {title} {Radioactive elements on {{Mercury}}'s surface from {{MESSENGER}}: {{Implications}} for the planet's formation and evolution},\ }\href {https://doi.org/10.1126/science.1211576} {\bibfield  {journal} {\bibinfo  {journal} {Science}\ }\textbf {\bibinfo {volume} {333}},\ \bibinfo {pages} {1850} (\bibinfo {year} {2011})}\BibitemShut {NoStop}%
\bibitem [{\citenamefont {Peplowski}(2016)}]{Peplowski2016c}%
  \BibitemOpen
  \bibfield  {author} {\bibinfo {author} {\bibfnamefont {P.~N.}\ \bibnamefont {Peplowski}},\ }\bibfield  {title} {\bibinfo {title} {The global elemental composition of 433 {{Eros}}: {{First}} results from the {{NEAR}} gamma-ray spectrometer orbital dataset},\ }\href {https://doi.org/10.1016/j.pss.2016.10.006} {\bibfield  {journal} {\bibinfo  {journal} {Planet. Space Sci.}\ }\textbf {\bibinfo {volume} {134}},\ \bibinfo {pages} {36} (\bibinfo {year} {2016})}\BibitemShut {NoStop}%
\bibitem [{\citenamefont {Prettyman}\ \emph {et~al.}(2017)\citenamefont {Prettyman}, \citenamefont {Yamashita}, \citenamefont {Toplis}, \citenamefont {McSween}, \citenamefont {Sch{\"o}rghofer}, \citenamefont {Marchi}, \citenamefont {Feldman}, \citenamefont {{Castillo-Rogez}}, \citenamefont {Forni}, \citenamefont {Lawrence}, \citenamefont {Ammannito}, \citenamefont {Ehlmann}, \citenamefont {Sizemore}, \citenamefont {Joy}, \citenamefont {Polanskey}, \citenamefont {Rayman}, \citenamefont {Raymond},\ and\ \citenamefont {Russell}}]{Prettyman2017}%
  \BibitemOpen
  \bibfield  {author} {\bibinfo {author} {\bibfnamefont {T.~H.}\ \bibnamefont {Prettyman}}, \bibinfo {author} {\bibfnamefont {N.}~\bibnamefont {Yamashita}}, \bibinfo {author} {\bibfnamefont {M.~J.}\ \bibnamefont {Toplis}}, \bibinfo {author} {\bibfnamefont {H.~Y.}\ \bibnamefont {McSween}}, \bibinfo {author} {\bibfnamefont {N.}~\bibnamefont {Sch{\"o}rghofer}}, \bibinfo {author} {\bibfnamefont {S.}~\bibnamefont {Marchi}}, \bibinfo {author} {\bibfnamefont {W.~C.}\ \bibnamefont {Feldman}}, \bibinfo {author} {\bibfnamefont {J.}~\bibnamefont {{Castillo-Rogez}}}, \bibinfo {author} {\bibfnamefont {O.}~\bibnamefont {Forni}}, \bibinfo {author} {\bibfnamefont {D.~J.}\ \bibnamefont {Lawrence}}, \bibinfo {author} {\bibfnamefont {E.}~\bibnamefont {Ammannito}}, \bibinfo {author} {\bibfnamefont {B.~L.}\ \bibnamefont {Ehlmann}}, \bibinfo {author} {\bibfnamefont {H.~G.}\ \bibnamefont {Sizemore}}, \bibinfo {author} {\bibfnamefont {S.~P.}\ \bibnamefont {Joy}}, \bibinfo {author} {\bibfnamefont {C.~A.}\ \bibnamefont {Polanskey}},
  \bibinfo {author} {\bibfnamefont {M.~D.}\ \bibnamefont {Rayman}}, \bibinfo {author} {\bibfnamefont {C.~A.}\ \bibnamefont {Raymond}},\ and\ \bibinfo {author} {\bibfnamefont {C.~T.}\ \bibnamefont {Russell}},\ }\bibfield  {title} {\bibinfo {title} {Extensive water ice within {{Ceres}}' aqueously altered regolith: {{Evidence}} from nuclear spectroscopy},\ }\href {https://doi.org/10.1126/science.aah6765} {\bibfield  {journal} {\bibinfo  {journal} {Science}\ }\textbf {\bibinfo {volume} {355}},\ \bibinfo {pages} {55} (\bibinfo {year} {2017})}\BibitemShut {NoStop}%
\bibitem [{\citenamefont {Prettyman}\ \emph {et~al.}(2019)\citenamefont {Prettyman}, \citenamefont {Yamashita}, \citenamefont {Ammannito}, \citenamefont {Ehlmann}, \citenamefont {McSween}, \citenamefont {Mittlefehldt}, \citenamefont {Marchi}, \citenamefont {Sch{\"o}rghofer}, \citenamefont {Toplis}, \citenamefont {Li}, \citenamefont {Pieters}, \citenamefont {{Castillo-Rogez}}, \citenamefont {Raymond},\ and\ \citenamefont {Russell}}]{Prettyman2019a}%
  \BibitemOpen
  \bibfield  {author} {\bibinfo {author} {\bibfnamefont {T.~H.}\ \bibnamefont {Prettyman}}, \bibinfo {author} {\bibfnamefont {N.}~\bibnamefont {Yamashita}}, \bibinfo {author} {\bibfnamefont {E.}~\bibnamefont {Ammannito}}, \bibinfo {author} {\bibfnamefont {B.~L.}\ \bibnamefont {Ehlmann}}, \bibinfo {author} {\bibfnamefont {H.~Y.}\ \bibnamefont {McSween}}, \bibinfo {author} {\bibfnamefont {D.~W.}\ \bibnamefont {Mittlefehldt}}, \bibinfo {author} {\bibfnamefont {S.}~\bibnamefont {Marchi}}, \bibinfo {author} {\bibfnamefont {N.}~\bibnamefont {Sch{\"o}rghofer}}, \bibinfo {author} {\bibfnamefont {M.~J.}\ \bibnamefont {Toplis}}, \bibinfo {author} {\bibfnamefont {J.~Y.}\ \bibnamefont {Li}}, \bibinfo {author} {\bibfnamefont {C.~M.}\ \bibnamefont {Pieters}}, \bibinfo {author} {\bibfnamefont {J.~C.}\ \bibnamefont {{Castillo-Rogez}}}, \bibinfo {author} {\bibfnamefont {C.~A.}\ \bibnamefont {Raymond}},\ and\ \bibinfo {author} {\bibfnamefont {C.~T.}\ \bibnamefont {Russell}},\ }\bibfield  {title} {\bibinfo {title} {Elemental
  composition and mineralogy of {{Vesta}} and {{Ceres}}: {{Distribution}} and origins of hydrogen-bearing species},\ }\href {https://doi.org/10.1016/j.icarus.2018.04.032} {\bibfield  {journal} {\bibinfo  {journal} {Icarus}\ }\textbf {\bibinfo {volume} {318}},\ \bibinfo {pages} {42} (\bibinfo {year} {2019})}\BibitemShut {NoStop}%
\bibitem [{\citenamefont {Balaram}\ \emph {et~al.}(2021)\citenamefont {Balaram}, \citenamefont {Aung},\ and\ \citenamefont {Golombek}}]{Balaram2021}%
  \BibitemOpen
  \bibfield  {author} {\bibinfo {author} {\bibfnamefont {J.}~\bibnamefont {Balaram}}, \bibinfo {author} {\bibfnamefont {M.}~\bibnamefont {Aung}},\ and\ \bibinfo {author} {\bibfnamefont {M.~P.}\ \bibnamefont {Golombek}},\ }\bibfield  {title} {\bibinfo {title} {The {{Ingenuity Helicopter}} on the {{Perseverance Rover}}},\ }\href {https://doi.org/10.1007/s11214-021-00815-w} {\bibfield  {journal} {\bibinfo  {journal} {Space Sci Rev}\ }\textbf {\bibinfo {volume} {217}},\ \bibinfo {pages} {56} (\bibinfo {year} {2021})}\BibitemShut {NoStop}%
\bibitem [{\citenamefont {Tzanetos}\ \emph {et~al.}(2022)\citenamefont {Tzanetos}, \citenamefont {Aung}, \citenamefont {Balaram}, \citenamefont {Grip}, \citenamefont {Karras}, \citenamefont {Canham}, \citenamefont {Kubiak}, \citenamefont {Anderson}, \citenamefont {Merewether}, \citenamefont {Starch}, \citenamefont {Pauken}, \citenamefont {Cappucci}, \citenamefont {Chase}, \citenamefont {Golombek}, \citenamefont {Toupet}, \citenamefont {Smart}, \citenamefont {Dawson}, \citenamefont {Ramirez}, \citenamefont {Lam}, \citenamefont {Stern}, \citenamefont {Chahat}, \citenamefont {Ravich}, \citenamefont {Hogg}, \citenamefont {Pipenberg}, \citenamefont {Keennon},\ and\ \citenamefont {Williford}}]{Tzanetos2022}%
  \BibitemOpen
  \bibfield  {author} {\bibinfo {author} {\bibfnamefont {T.}~\bibnamefont {Tzanetos}}, \bibinfo {author} {\bibfnamefont {M.}~\bibnamefont {Aung}}, \bibinfo {author} {\bibfnamefont {J.}~\bibnamefont {Balaram}}, \bibinfo {author} {\bibfnamefont {H.~F.}\ \bibnamefont {Grip}}, \bibinfo {author} {\bibfnamefont {J.~T.}\ \bibnamefont {Karras}}, \bibinfo {author} {\bibfnamefont {T.~K.}\ \bibnamefont {Canham}}, \bibinfo {author} {\bibfnamefont {G.}~\bibnamefont {Kubiak}}, \bibinfo {author} {\bibfnamefont {J.}~\bibnamefont {Anderson}}, \bibinfo {author} {\bibfnamefont {G.}~\bibnamefont {Merewether}}, \bibinfo {author} {\bibfnamefont {M.}~\bibnamefont {Starch}}, \bibinfo {author} {\bibfnamefont {M.}~\bibnamefont {Pauken}}, \bibinfo {author} {\bibfnamefont {S.}~\bibnamefont {Cappucci}}, \bibinfo {author} {\bibfnamefont {M.}~\bibnamefont {Chase}}, \bibinfo {author} {\bibfnamefont {M.}~\bibnamefont {Golombek}}, \bibinfo {author} {\bibfnamefont {O.}~\bibnamefont {Toupet}}, \bibinfo {author} {\bibfnamefont {M.~C.}\
  \bibnamefont {Smart}}, \bibinfo {author} {\bibfnamefont {S.}~\bibnamefont {Dawson}}, \bibinfo {author} {\bibfnamefont {E.~B.}\ \bibnamefont {Ramirez}}, \bibinfo {author} {\bibfnamefont {J.}~\bibnamefont {Lam}}, \bibinfo {author} {\bibfnamefont {R.}~\bibnamefont {Stern}}, \bibinfo {author} {\bibfnamefont {N.}~\bibnamefont {Chahat}}, \bibinfo {author} {\bibfnamefont {J.}~\bibnamefont {Ravich}}, \bibinfo {author} {\bibfnamefont {R.}~\bibnamefont {Hogg}}, \bibinfo {author} {\bibfnamefont {B.}~\bibnamefont {Pipenberg}}, \bibinfo {author} {\bibfnamefont {M.}~\bibnamefont {Keennon}},\ and\ \bibinfo {author} {\bibfnamefont {K.~H.}\ \bibnamefont {Williford}},\ }\bibfield  {title} {\bibinfo {title} {Ingenuity {{Mars Helicopter}}: {{From Technology Demonstration}} to {{Extraterrestrial Scout}}},\ }in\ \href {https://doi.org/10.1109/AERO53065.2022.9843428} {\emph {\bibinfo {booktitle} {2022 {{IEEE Aerosp}}. {{Conf}}. {{AERO}}}}}\ (\bibinfo {year} {2022})\ pp.\ \bibinfo {pages} {01--19}\BibitemShut {NoStop}%
\bibitem [{\citenamefont {Grip}\ \emph {et~al.}(2022)\citenamefont {Grip}, \citenamefont {Conway}, \citenamefont {Lam}, \citenamefont {Williams}, \citenamefont {Golombek}, \citenamefont {Brockers}, \citenamefont {Mischna},\ and\ \citenamefont {Cacan}}]{Grip2022}%
  \BibitemOpen
  \bibfield  {author} {\bibinfo {author} {\bibfnamefont {H.~F.}\ \bibnamefont {Grip}}, \bibinfo {author} {\bibfnamefont {D.}~\bibnamefont {Conway}}, \bibinfo {author} {\bibfnamefont {J.}~\bibnamefont {Lam}}, \bibinfo {author} {\bibfnamefont {N.}~\bibnamefont {Williams}}, \bibinfo {author} {\bibfnamefont {M.~P.}\ \bibnamefont {Golombek}}, \bibinfo {author} {\bibfnamefont {R.}~\bibnamefont {Brockers}}, \bibinfo {author} {\bibfnamefont {M.}~\bibnamefont {Mischna}},\ and\ \bibinfo {author} {\bibfnamefont {M.~R.}\ \bibnamefont {Cacan}},\ }\bibfield  {title} {\bibinfo {title} {Flying a {{Helicopter}} on {{Mars}}: {{How Ingenuity}}'s {{Flights}} were {{Planned}}, {{Executed}}, and {{Analyzed}}},\ }in\ \href {https://doi.org/10.1109/AERO53065.2022.9843813} {\emph {\bibinfo {booktitle} {2022 {{IEEE Aerosp}}. {{Conf}}. {{AERO}}}}}\ (\bibinfo {year} {2022})\ pp.\ \bibinfo {pages} {1--17}\BibitemShut {NoStop}%
\bibitem [{\citenamefont {Grasty}\ \emph {et~al.}(1985)\citenamefont {Grasty}, \citenamefont {Glynn},\ and\ \citenamefont {Grant}}]{Grasty1985a}%
  \BibitemOpen
  \bibfield  {author} {\bibinfo {author} {\bibfnamefont {R.~L.}\ \bibnamefont {Grasty}}, \bibinfo {author} {\bibfnamefont {J.~E.}\ \bibnamefont {Glynn}},\ and\ \bibinfo {author} {\bibfnamefont {J.~A.}\ \bibnamefont {Grant}},\ }\bibfield  {title} {\bibinfo {title} {The analysis of multichannel airborne gamma-ray spectra},\ }\href {https://doi.org/10.1190/1.1441886} {\bibfield  {journal} {\bibinfo  {journal} {GEOPHYSICS}\ }\textbf {\bibinfo {volume} {50}},\ \bibinfo {pages} {2611} (\bibinfo {year} {1985})}\BibitemShut {NoStop}%
\bibitem [{\citenamefont {Minty}\ \emph {et~al.}(1998)\citenamefont {Minty}, \citenamefont {McFadden},\ and\ \citenamefont {Kennett}}]{Minty1998d}%
  \BibitemOpen
  \bibfield  {author} {\bibinfo {author} {\bibfnamefont {B.~R.}\ \bibnamefont {Minty}}, \bibinfo {author} {\bibfnamefont {P.}~\bibnamefont {McFadden}},\ and\ \bibinfo {author} {\bibfnamefont {B.~L.}\ \bibnamefont {Kennett}},\ }\bibfield  {title} {\bibinfo {title} {Multichannel processing for airborne gamma-ray spectrometry},\ }\href {https://doi.org/10.1190/1.1444491} {\bibfield  {journal} {\bibinfo  {journal} {Geophysics}\ }\textbf {\bibinfo {volume} {63}},\ \bibinfo {pages} {1971} (\bibinfo {year} {1998})}\BibitemShut {NoStop}%
\bibitem [{\citenamefont {Hendriks}\ \emph {et~al.}(2001)\citenamefont {Hendriks}, \citenamefont {Limburg},\ and\ \citenamefont {De~Meijer}}]{Hendriks2001a}%
  \BibitemOpen
  \bibfield  {author} {\bibinfo {author} {\bibfnamefont {P.~H.}\ \bibnamefont {Hendriks}}, \bibinfo {author} {\bibfnamefont {J.}~\bibnamefont {Limburg}},\ and\ \bibinfo {author} {\bibfnamefont {R.~J.}\ \bibnamefont {De~Meijer}},\ }\bibfield  {title} {\bibinfo {title} {Full-spectrum analysis of natural {$\gamma$}-ray spectra},\ }\href {https://doi.org/10.1016/S0265-931X(00)00142-9} {\bibfield  {journal} {\bibinfo  {journal} {J. Environ. Radioact.}\ }\textbf {\bibinfo {volume} {53}},\ \bibinfo {pages} {365} (\bibinfo {year} {2001})}\BibitemShut {NoStop}%
\bibitem [{\citenamefont {Bar{\'e}}\ and\ \citenamefont {Tondeur}(2011)}]{Bare2011a}%
  \BibitemOpen
  \bibfield  {author} {\bibinfo {author} {\bibfnamefont {J.}~\bibnamefont {Bar{\'e}}}\ and\ \bibinfo {author} {\bibfnamefont {F.}~\bibnamefont {Tondeur}},\ }\bibfield  {title} {\bibinfo {title} {Gamma spectrum unfolding for a {{NaI}} monitor of radioactivity in aquatic systems: {{Experimental}} evaluations of the minimal detectable activity},\ }\href {https://doi.org/10.1016/j.apradiso.2010.11.024} {\bibfield  {journal} {\bibinfo  {journal} {Appl. Radiat. Isot.}\ }\textbf {\bibinfo {volume} {69}},\ \bibinfo {pages} {1121} (\bibinfo {year} {2011})}\BibitemShut {NoStop}%
\bibitem [{\citenamefont {Caciolli}\ \emph {et~al.}(2012)\citenamefont {Caciolli}, \citenamefont {Baldoncini}, \citenamefont {Bezzon}, \citenamefont {Broggini}, \citenamefont {Buso}, \citenamefont {Callegari}, \citenamefont {Colonna}, \citenamefont {Fiorentini}, \citenamefont {Guastaldi}, \citenamefont {Mantovani}, \citenamefont {Massa}, \citenamefont {Menegazzo}, \citenamefont {Mou}, \citenamefont {Alvarez}, \citenamefont {Shyti}, \citenamefont {Zanon},\ and\ \citenamefont {Xhixha}}]{Caciolli2012a}%
  \BibitemOpen
  \bibfield  {author} {\bibinfo {author} {\bibfnamefont {A.}~\bibnamefont {Caciolli}}, \bibinfo {author} {\bibfnamefont {M.}~\bibnamefont {Baldoncini}}, \bibinfo {author} {\bibfnamefont {G.~P.}\ \bibnamefont {Bezzon}}, \bibinfo {author} {\bibfnamefont {C.}~\bibnamefont {Broggini}}, \bibinfo {author} {\bibfnamefont {G.~P.}\ \bibnamefont {Buso}}, \bibinfo {author} {\bibfnamefont {I.}~\bibnamefont {Callegari}}, \bibinfo {author} {\bibfnamefont {T.}~\bibnamefont {Colonna}}, \bibinfo {author} {\bibfnamefont {G.}~\bibnamefont {Fiorentini}}, \bibinfo {author} {\bibfnamefont {E.}~\bibnamefont {Guastaldi}}, \bibinfo {author} {\bibfnamefont {F.}~\bibnamefont {Mantovani}}, \bibinfo {author} {\bibfnamefont {G.}~\bibnamefont {Massa}}, \bibinfo {author} {\bibfnamefont {R.}~\bibnamefont {Menegazzo}}, \bibinfo {author} {\bibfnamefont {L.}~\bibnamefont {Mou}}, \bibinfo {author} {\bibfnamefont {C.~R.}\ \bibnamefont {Alvarez}}, \bibinfo {author} {\bibfnamefont {M.}~\bibnamefont {Shyti}}, \bibinfo {author} {\bibfnamefont
  {A.}~\bibnamefont {Zanon}},\ and\ \bibinfo {author} {\bibfnamefont {G.}~\bibnamefont {Xhixha}},\ }\bibfield  {title} {\bibinfo {title} {A new {{FSA}} approach for in situ {$\gamma$} ray spectroscopy},\ }\href {https://doi.org/10.1016/j.scitotenv.2011.10.071} {\bibfield  {journal} {\bibinfo  {journal} {Sci. Total Environ.}\ }\textbf {\bibinfo {volume} {414}},\ \bibinfo {pages} {639} (\bibinfo {year} {2012})}\BibitemShut {NoStop}%
\bibitem [{\citenamefont {Paradis}\ \emph {et~al.}(2020)\citenamefont {Paradis}, \citenamefont {Bobin}, \citenamefont {Bobin}, \citenamefont {Bouchard}, \citenamefont {Louren{\c c}o}, \citenamefont {Thiam}, \citenamefont {Andr{\'e}}, \citenamefont {Ferreux}, \citenamefont {{de Vismes Ott}},\ and\ \citenamefont {Th{\'e}venin}}]{Paradis2020a}%
  \BibitemOpen
  \bibfield  {author} {\bibinfo {author} {\bibfnamefont {H.}~\bibnamefont {Paradis}}, \bibinfo {author} {\bibfnamefont {C.}~\bibnamefont {Bobin}}, \bibinfo {author} {\bibfnamefont {J.}~\bibnamefont {Bobin}}, \bibinfo {author} {\bibfnamefont {J.}~\bibnamefont {Bouchard}}, \bibinfo {author} {\bibfnamefont {V.}~\bibnamefont {Louren{\c c}o}}, \bibinfo {author} {\bibfnamefont {C.}~\bibnamefont {Thiam}}, \bibinfo {author} {\bibfnamefont {R.}~\bibnamefont {Andr{\'e}}}, \bibinfo {author} {\bibfnamefont {L.}~\bibnamefont {Ferreux}}, \bibinfo {author} {\bibfnamefont {A.}~\bibnamefont {{de Vismes Ott}}},\ and\ \bibinfo {author} {\bibfnamefont {M.}~\bibnamefont {Th{\'e}venin}},\ }\bibfield  {title} {\bibinfo {title} {Spectral unmixing applied to fast identification of {$\gamma$}-emitting radionuclides using {{NaI}}({{Tl}}) detectors},\ }\href {https://doi.org/10.1016/j.apradiso.2020.109068} {\bibfield  {journal} {\bibinfo  {journal} {Appl. Radiat. Isot.}\ }\textbf {\bibinfo {volume} {158}},\ \bibinfo {pages} {109068}
  (\bibinfo {year} {2020})}\BibitemShut {NoStop}%
\bibitem [{\citenamefont {Andr{\'e}}\ \emph {et~al.}(2021)\citenamefont {Andr{\'e}}, \citenamefont {Bobin}, \citenamefont {Bobin}, \citenamefont {Xu},\ and\ \citenamefont {{de Vismes Ott}}}]{Andre2021a}%
  \BibitemOpen
  \bibfield  {author} {\bibinfo {author} {\bibfnamefont {R.}~\bibnamefont {Andr{\'e}}}, \bibinfo {author} {\bibfnamefont {C.}~\bibnamefont {Bobin}}, \bibinfo {author} {\bibfnamefont {J.}~\bibnamefont {Bobin}}, \bibinfo {author} {\bibfnamefont {J.}~\bibnamefont {Xu}},\ and\ \bibinfo {author} {\bibfnamefont {A.}~\bibnamefont {{de Vismes Ott}}},\ }\bibfield  {title} {\bibinfo {title} {Metrological approach of {$\gamma$}-emitting radionuclides identification at low statistics: {{Application}} of sparse spectral unmixing to scintillation detectors},\ }\href {https://doi.org/10.1088/1681-7575/abcc06} {\bibfield  {journal} {\bibinfo  {journal} {Metrologia}\ }\textbf {\bibinfo {volume} {58}},\ \bibinfo {pages} {15011} (\bibinfo {year} {2021})}\BibitemShut {NoStop}%
\bibitem [{\citenamefont {Dickson}\ \emph {et~al.}(1981)\citenamefont {Dickson}, \citenamefont {Bailey},\ and\ \citenamefont {Grasty}}]{Dickson1981a}%
  \BibitemOpen
  \bibfield  {author} {\bibinfo {author} {\bibfnamefont {B.~H.}\ \bibnamefont {Dickson}}, \bibinfo {author} {\bibfnamefont {R.~C.}\ \bibnamefont {Bailey}},\ and\ \bibinfo {author} {\bibfnamefont {R.~L.}\ \bibnamefont {Grasty}},\ }\bibfield  {title} {\bibinfo {title} {Utilizing multi-channel airborne gamma-ray spectra},\ }\href {https://doi.org/10.1139/E81-167} {\bibfield  {journal} {\bibinfo  {journal} {Can. J. Earth Sci.}\ }\textbf {\bibinfo {volume} {18}},\ \bibinfo {pages} {1793} (\bibinfo {year} {1981})}\BibitemShut {NoStop}%
\bibitem [{\citenamefont {Minty}\ \emph {et~al.}(1990)\citenamefont {Minty}, \citenamefont {Morse},\ and\ \citenamefont {Richardson}}]{Minty1990a}%
  \BibitemOpen
  \bibfield  {author} {\bibinfo {author} {\bibfnamefont {B.~R.~S.}\ \bibnamefont {Minty}}, \bibinfo {author} {\bibfnamefont {M.~P.}\ \bibnamefont {Morse}},\ and\ \bibinfo {author} {\bibfnamefont {L.~M.}\ \bibnamefont {Richardson}},\ }\bibfield  {title} {\bibinfo {title} {Portable {{Calibration Sources For Airborne Gamma-ray Spectrometers}}},\ }\href {https://doi.org/10.1071/EG990187} {\bibfield  {journal} {\bibinfo  {journal} {Explor. Geophys.}\ }\textbf {\bibinfo {volume} {21}},\ \bibinfo {pages} {187} (\bibinfo {year} {1990})}\BibitemShut {NoStop}%
\bibitem [{\citenamefont {Grasty}\ \emph {et~al.}(1991)\citenamefont {Grasty}, \citenamefont {Holman},\ and\ \citenamefont {Blanchard}}]{Grasty1991a}%
  \BibitemOpen
  \bibfield  {author} {\bibinfo {author} {\bibfnamefont {R.~L.}\ \bibnamefont {Grasty}}, \bibinfo {author} {\bibfnamefont {P.~B.}\ \bibnamefont {Holman}},\ and\ \bibinfo {author} {\bibfnamefont {Y.~B.}\ \bibnamefont {Blanchard}},\ }\bibfield  {title} {\bibinfo {title} {Transportable calibration pads for ground and airborne gamma-ray spectrometers},\ }\bibfield  {journal} {\bibinfo  {journal} {Geol. Surv. Can.}\ }\textbf {\bibinfo {volume} {90}},\ \href {https://doi.org/10.4095/132237} {10.4095/132237} (\bibinfo {year} {1991})\BibitemShut {NoStop}%
\bibitem [{\citenamefont {Allison}\ \emph {et~al.}(2016)\citenamefont {Allison}, \citenamefont {Amako}, \citenamefont {Apostolakis}, \citenamefont {Arce}, \citenamefont {Asai}, \citenamefont {Aso}, \citenamefont {Bagli}, \citenamefont {Bagulya}, \citenamefont {Banerjee}, \citenamefont {Barrand}, \citenamefont {Beck}, \citenamefont {Bogdanov}, \citenamefont {Brandt}, \citenamefont {Brown}, \citenamefont {Burkhardt}, \citenamefont {Canal}, \citenamefont {{Cano-Ott}}, \citenamefont {Chauvie}, \citenamefont {Cho}, \citenamefont {Cirrone}, \citenamefont {Cooperman}, \citenamefont {{Cort{\'e}s-Giraldo}}, \citenamefont {Cosmo}, \citenamefont {Cuttone}, \citenamefont {Depaola}, \citenamefont {Desorgher}, \citenamefont {Dong}, \citenamefont {Dotti}, \citenamefont {Elvira}, \citenamefont {Folger}, \citenamefont {Francis}, \citenamefont {Galoyan}, \citenamefont {Garnier}, \citenamefont {Gayer}, \citenamefont {Genser}, \citenamefont {Grichine}, \citenamefont {Guatelli}, \citenamefont {Gu{\`e}ye}, \citenamefont
  {Gumplinger}, \citenamefont {Howard}, \citenamefont {H{\v r}ivn{\'a}{\v c}ov{\'a}}, \citenamefont {Hwang}, \citenamefont {Incerti}, \citenamefont {Ivanchenko}, \citenamefont {Ivanchenko}, \citenamefont {Jones}, \citenamefont {Jun}, \citenamefont {Kaitaniemi}, \citenamefont {Karakatsanis}, \citenamefont {Karamitros}, \citenamefont {Kelsey}, \citenamefont {Kimura}, \citenamefont {Koi}, \citenamefont {Kurashige}, \citenamefont {Lechner}, \citenamefont {Lee}, \citenamefont {Longo}, \citenamefont {Maire}, \citenamefont {Mancusi}, \citenamefont {Mantero}, \citenamefont {Mendoza}, \citenamefont {Morgan}, \citenamefont {Murakami}, \citenamefont {Nikitina}, \citenamefont {Pandola}, \citenamefont {Paprocki}, \citenamefont {Perl}, \citenamefont {Petrovi{\'c}}, \citenamefont {Pia}, \citenamefont {Pokorski}, \citenamefont {Quesada}, \citenamefont {Raine}, \citenamefont {Reis}, \citenamefont {Ribon}, \citenamefont {Risti{\'c}~Fira}, \citenamefont {Romano}, \citenamefont {Russo}, \citenamefont {Santin}, \citenamefont
  {Sasaki}, \citenamefont {Sawkey}, \citenamefont {Shin}, \citenamefont {Strakovsky}, \citenamefont {Taborda}, \citenamefont {Tanaka}, \citenamefont {Tom{\'e}}, \citenamefont {Toshito}, \citenamefont {Tran}, \citenamefont {Truscott}, \citenamefont {Urban}, \citenamefont {Uzhinsky}, \citenamefont {Verbeke}, \citenamefont {Verderi}, \citenamefont {Wendt}, \citenamefont {Wenzel}, \citenamefont {Wright}, \citenamefont {Wright}, \citenamefont {Yamashita}, \citenamefont {Yarba},\ and\ \citenamefont {Yoshida}}]{Allison2016}%
  \BibitemOpen
  \bibfield  {author} {\bibinfo {author} {\bibfnamefont {J.}~\bibnamefont {Allison}}, \bibinfo {author} {\bibfnamefont {K.}~\bibnamefont {Amako}}, \bibinfo {author} {\bibfnamefont {J.}~\bibnamefont {Apostolakis}}, \bibinfo {author} {\bibfnamefont {P.}~\bibnamefont {Arce}}, \bibinfo {author} {\bibfnamefont {M.}~\bibnamefont {Asai}}, \bibinfo {author} {\bibfnamefont {T.}~\bibnamefont {Aso}}, \bibinfo {author} {\bibfnamefont {E.}~\bibnamefont {Bagli}}, \bibinfo {author} {\bibfnamefont {A.}~\bibnamefont {Bagulya}}, \bibinfo {author} {\bibfnamefont {S.}~\bibnamefont {Banerjee}}, \bibinfo {author} {\bibfnamefont {G.}~\bibnamefont {Barrand}}, \bibinfo {author} {\bibfnamefont {B.~R.}\ \bibnamefont {Beck}}, \bibinfo {author} {\bibfnamefont {A.~G.}\ \bibnamefont {Bogdanov}}, \bibinfo {author} {\bibfnamefont {D.}~\bibnamefont {Brandt}}, \bibinfo {author} {\bibfnamefont {J.~M.~C.}\ \bibnamefont {Brown}}, \bibinfo {author} {\bibfnamefont {H.}~\bibnamefont {Burkhardt}}, \bibinfo {author} {\bibfnamefont {{\relax
  Ph}.}~\bibnamefont {Canal}}, \bibinfo {author} {\bibfnamefont {D.}~\bibnamefont {{Cano-Ott}}}, \bibinfo {author} {\bibfnamefont {S.}~\bibnamefont {Chauvie}}, \bibinfo {author} {\bibfnamefont {K.}~\bibnamefont {Cho}}, \bibinfo {author} {\bibfnamefont {G.~A.~P.}\ \bibnamefont {Cirrone}}, \bibinfo {author} {\bibfnamefont {G.}~\bibnamefont {Cooperman}}, \bibinfo {author} {\bibfnamefont {M.~A.}\ \bibnamefont {{Cort{\'e}s-Giraldo}}}, \bibinfo {author} {\bibfnamefont {G.}~\bibnamefont {Cosmo}}, \bibinfo {author} {\bibfnamefont {G.}~\bibnamefont {Cuttone}}, \bibinfo {author} {\bibfnamefont {G.}~\bibnamefont {Depaola}}, \bibinfo {author} {\bibfnamefont {L.}~\bibnamefont {Desorgher}}, \bibinfo {author} {\bibfnamefont {X.}~\bibnamefont {Dong}}, \bibinfo {author} {\bibfnamefont {A.}~\bibnamefont {Dotti}}, \bibinfo {author} {\bibfnamefont {V.~D.}\ \bibnamefont {Elvira}}, \bibinfo {author} {\bibfnamefont {G.}~\bibnamefont {Folger}}, \bibinfo {author} {\bibfnamefont {Z.}~\bibnamefont {Francis}}, \bibinfo {author}
  {\bibfnamefont {A.}~\bibnamefont {Galoyan}}, \bibinfo {author} {\bibfnamefont {L.}~\bibnamefont {Garnier}}, \bibinfo {author} {\bibfnamefont {M.}~\bibnamefont {Gayer}}, \bibinfo {author} {\bibfnamefont {K.~L.}\ \bibnamefont {Genser}}, \bibinfo {author} {\bibfnamefont {V.~M.}\ \bibnamefont {Grichine}}, \bibinfo {author} {\bibfnamefont {S.}~\bibnamefont {Guatelli}}, \bibinfo {author} {\bibfnamefont {P.}~\bibnamefont {Gu{\`e}ye}}, \bibinfo {author} {\bibfnamefont {P.}~\bibnamefont {Gumplinger}}, \bibinfo {author} {\bibfnamefont {A.~S.}\ \bibnamefont {Howard}}, \bibinfo {author} {\bibfnamefont {I.}~\bibnamefont {H{\v r}ivn{\'a}{\v c}ov{\'a}}}, \bibinfo {author} {\bibfnamefont {S.}~\bibnamefont {Hwang}}, \bibinfo {author} {\bibfnamefont {S.}~\bibnamefont {Incerti}}, \bibinfo {author} {\bibfnamefont {A.}~\bibnamefont {Ivanchenko}}, \bibinfo {author} {\bibfnamefont {V.~N.}\ \bibnamefont {Ivanchenko}}, \bibinfo {author} {\bibfnamefont {F.~W.}\ \bibnamefont {Jones}}, \bibinfo {author} {\bibfnamefont {S.~Y.}\
  \bibnamefont {Jun}}, \bibinfo {author} {\bibfnamefont {P.}~\bibnamefont {Kaitaniemi}}, \bibinfo {author} {\bibfnamefont {N.}~\bibnamefont {Karakatsanis}}, \bibinfo {author} {\bibfnamefont {M.}~\bibnamefont {Karamitros}}, \bibinfo {author} {\bibfnamefont {M.}~\bibnamefont {Kelsey}}, \bibinfo {author} {\bibfnamefont {A.}~\bibnamefont {Kimura}}, \bibinfo {author} {\bibfnamefont {T.}~\bibnamefont {Koi}}, \bibinfo {author} {\bibfnamefont {H.}~\bibnamefont {Kurashige}}, \bibinfo {author} {\bibfnamefont {A.}~\bibnamefont {Lechner}}, \bibinfo {author} {\bibfnamefont {S.~B.}\ \bibnamefont {Lee}}, \bibinfo {author} {\bibfnamefont {F.}~\bibnamefont {Longo}}, \bibinfo {author} {\bibfnamefont {M.}~\bibnamefont {Maire}}, \bibinfo {author} {\bibfnamefont {D.}~\bibnamefont {Mancusi}}, \bibinfo {author} {\bibfnamefont {A.}~\bibnamefont {Mantero}}, \bibinfo {author} {\bibfnamefont {E.}~\bibnamefont {Mendoza}}, \bibinfo {author} {\bibfnamefont {B.}~\bibnamefont {Morgan}}, \bibinfo {author} {\bibfnamefont {K.}~\bibnamefont
  {Murakami}}, \bibinfo {author} {\bibfnamefont {T.}~\bibnamefont {Nikitina}}, \bibinfo {author} {\bibfnamefont {L.}~\bibnamefont {Pandola}}, \bibinfo {author} {\bibfnamefont {P.}~\bibnamefont {Paprocki}}, \bibinfo {author} {\bibfnamefont {J.}~\bibnamefont {Perl}}, \bibinfo {author} {\bibfnamefont {I.}~\bibnamefont {Petrovi{\'c}}}, \bibinfo {author} {\bibfnamefont {M.~G.}\ \bibnamefont {Pia}}, \bibinfo {author} {\bibfnamefont {W.}~\bibnamefont {Pokorski}}, \bibinfo {author} {\bibfnamefont {J.~M.}\ \bibnamefont {Quesada}}, \bibinfo {author} {\bibfnamefont {M.}~\bibnamefont {Raine}}, \bibinfo {author} {\bibfnamefont {M.~A.}\ \bibnamefont {Reis}}, \bibinfo {author} {\bibfnamefont {A.}~\bibnamefont {Ribon}}, \bibinfo {author} {\bibfnamefont {A.}~\bibnamefont {Risti{\'c}~Fira}}, \bibinfo {author} {\bibfnamefont {F.}~\bibnamefont {Romano}}, \bibinfo {author} {\bibfnamefont {G.}~\bibnamefont {Russo}}, \bibinfo {author} {\bibfnamefont {G.}~\bibnamefont {Santin}}, \bibinfo {author} {\bibfnamefont {T.}~\bibnamefont
  {Sasaki}}, \bibinfo {author} {\bibfnamefont {D.}~\bibnamefont {Sawkey}}, \bibinfo {author} {\bibfnamefont {J.~I.}\ \bibnamefont {Shin}}, \bibinfo {author} {\bibfnamefont {I.~I.}\ \bibnamefont {Strakovsky}}, \bibinfo {author} {\bibfnamefont {A.}~\bibnamefont {Taborda}}, \bibinfo {author} {\bibfnamefont {S.}~\bibnamefont {Tanaka}}, \bibinfo {author} {\bibfnamefont {B.}~\bibnamefont {Tom{\'e}}}, \bibinfo {author} {\bibfnamefont {T.}~\bibnamefont {Toshito}}, \bibinfo {author} {\bibfnamefont {H.~N.}\ \bibnamefont {Tran}}, \bibinfo {author} {\bibfnamefont {P.~R.}\ \bibnamefont {Truscott}}, \bibinfo {author} {\bibfnamefont {L.}~\bibnamefont {Urban}}, \bibinfo {author} {\bibfnamefont {V.}~\bibnamefont {Uzhinsky}}, \bibinfo {author} {\bibfnamefont {J.~M.}\ \bibnamefont {Verbeke}}, \bibinfo {author} {\bibfnamefont {M.}~\bibnamefont {Verderi}}, \bibinfo {author} {\bibfnamefont {B.~L.}\ \bibnamefont {Wendt}}, \bibinfo {author} {\bibfnamefont {H.}~\bibnamefont {Wenzel}}, \bibinfo {author} {\bibfnamefont {D.~H.}\
  \bibnamefont {Wright}}, \bibinfo {author} {\bibfnamefont {D.~M.}\ \bibnamefont {Wright}}, \bibinfo {author} {\bibfnamefont {T.}~\bibnamefont {Yamashita}}, \bibinfo {author} {\bibfnamefont {J.}~\bibnamefont {Yarba}},\ and\ \bibinfo {author} {\bibfnamefont {H.}~\bibnamefont {Yoshida}},\ }\bibfield  {title} {\bibinfo {title} {Recent developments in {{Geant4}}},\ }\href {https://doi.org/10.1016/j.nima.2016.06.125} {\bibfield  {journal} {\bibinfo  {journal} {Nuclear Instruments and Methods in Physics Research Section A: Accelerators, Spectrometers, Detectors and Associated Equipment}\ }\textbf {\bibinfo {volume} {835}},\ \bibinfo {pages} {186} (\bibinfo {year} {2016})}\BibitemShut {NoStop}%
\bibitem [{\citenamefont {Goorley}\ \emph {et~al.}(2016)\citenamefont {Goorley}, \citenamefont {James}, \citenamefont {Booth}, \citenamefont {Brown}, \citenamefont {Bull}, \citenamefont {Cox}, \citenamefont {Durkee}, \citenamefont {Elson}, \citenamefont {Fensin}, \citenamefont {Forster}, \citenamefont {Hendricks}, \citenamefont {Hughes}, \citenamefont {Johns}, \citenamefont {Kiedrowski}, \citenamefont {Martz}, \citenamefont {Mashnik}, \citenamefont {McKinney}, \citenamefont {Pelowitz}, \citenamefont {Prael}, \citenamefont {Sweezy}, \citenamefont {Waters}, \citenamefont {Wilcox},\ and\ \citenamefont {Zukaitis}}]{Goorley2016}%
  \BibitemOpen
  \bibfield  {author} {\bibinfo {author} {\bibfnamefont {T.}~\bibnamefont {Goorley}}, \bibinfo {author} {\bibfnamefont {M.}~\bibnamefont {James}}, \bibinfo {author} {\bibfnamefont {T.}~\bibnamefont {Booth}}, \bibinfo {author} {\bibfnamefont {F.}~\bibnamefont {Brown}}, \bibinfo {author} {\bibfnamefont {J.}~\bibnamefont {Bull}}, \bibinfo {author} {\bibfnamefont {L.~J.}\ \bibnamefont {Cox}}, \bibinfo {author} {\bibfnamefont {J.}~\bibnamefont {Durkee}}, \bibinfo {author} {\bibfnamefont {J.}~\bibnamefont {Elson}}, \bibinfo {author} {\bibfnamefont {M.}~\bibnamefont {Fensin}}, \bibinfo {author} {\bibfnamefont {R.~A.}\ \bibnamefont {Forster}}, \bibinfo {author} {\bibfnamefont {J.}~\bibnamefont {Hendricks}}, \bibinfo {author} {\bibfnamefont {H.~G.}\ \bibnamefont {Hughes}}, \bibinfo {author} {\bibfnamefont {R.}~\bibnamefont {Johns}}, \bibinfo {author} {\bibfnamefont {B.}~\bibnamefont {Kiedrowski}}, \bibinfo {author} {\bibfnamefont {R.}~\bibnamefont {Martz}}, \bibinfo {author} {\bibfnamefont {S.}~\bibnamefont
  {Mashnik}}, \bibinfo {author} {\bibfnamefont {G.}~\bibnamefont {McKinney}}, \bibinfo {author} {\bibfnamefont {D.}~\bibnamefont {Pelowitz}}, \bibinfo {author} {\bibfnamefont {R.}~\bibnamefont {Prael}}, \bibinfo {author} {\bibfnamefont {J.}~\bibnamefont {Sweezy}}, \bibinfo {author} {\bibfnamefont {L.}~\bibnamefont {Waters}}, \bibinfo {author} {\bibfnamefont {T.}~\bibnamefont {Wilcox}},\ and\ \bibinfo {author} {\bibfnamefont {T.}~\bibnamefont {Zukaitis}},\ }\bibfield  {title} {\bibinfo {title} {Features of {{MCNP6}}},\ }\href {https://doi.org/10.1016/j.anucene.2015.02.020} {\bibfield  {journal} {\bibinfo  {journal} {Annals of Nuclear Energy}\ }\textbf {\bibinfo {volume} {87}},\ \bibinfo {pages} {772} (\bibinfo {year} {2016})}\BibitemShut {NoStop}%
\bibitem [{\citenamefont {Ahdida}\ \emph {et~al.}(2022)\citenamefont {Ahdida}, \citenamefont {Bozzato}, \citenamefont {Calzolari}, \citenamefont {Cerutti}, \citenamefont {Charitonidis}, \citenamefont {Cimmino}, \citenamefont {Coronetti}, \citenamefont {D'Alessandro}, \citenamefont {Donadon~Servelle}, \citenamefont {Esposito}, \citenamefont {Froeschl}, \citenamefont {Garc{\'i}a~Al{\'i}a}, \citenamefont {Gerbershagen}, \citenamefont {Gilardoni}, \citenamefont {Horv{\'a}th}, \citenamefont {Hugo}, \citenamefont {Infantino}, \citenamefont {Kouskoura}, \citenamefont {Lechner}, \citenamefont {Lefebvre}, \citenamefont {Lerner}, \citenamefont {Magistris}, \citenamefont {Manousos}, \citenamefont {Moryc}, \citenamefont {Ogallar~Ruiz}, \citenamefont {Pozzi}, \citenamefont {Prelipcean}, \citenamefont {Roesler}, \citenamefont {Rossi}, \citenamefont {Sabat{\'e}~Gilarte}, \citenamefont {Salvat~Pujol}, \citenamefont {Schoofs}, \citenamefont {Str{\'a}nsk{\'y}}, \citenamefont {Theis}, \citenamefont {Tsinganis}, \citenamefont
  {Versaci}, \citenamefont {Vlachoudis}, \citenamefont {Waets},\ and\ \citenamefont {Widorski}}]{Ahdida2022a}%
  \BibitemOpen
  \bibfield  {author} {\bibinfo {author} {\bibfnamefont {C.}~\bibnamefont {Ahdida}}, \bibinfo {author} {\bibfnamefont {D.}~\bibnamefont {Bozzato}}, \bibinfo {author} {\bibfnamefont {D.}~\bibnamefont {Calzolari}}, \bibinfo {author} {\bibfnamefont {F.}~\bibnamefont {Cerutti}}, \bibinfo {author} {\bibfnamefont {N.}~\bibnamefont {Charitonidis}}, \bibinfo {author} {\bibfnamefont {A.}~\bibnamefont {Cimmino}}, \bibinfo {author} {\bibfnamefont {A.}~\bibnamefont {Coronetti}}, \bibinfo {author} {\bibfnamefont {G.~L.}\ \bibnamefont {D'Alessandro}}, \bibinfo {author} {\bibfnamefont {A.}~\bibnamefont {Donadon~Servelle}}, \bibinfo {author} {\bibfnamefont {L.~S.}\ \bibnamefont {Esposito}}, \bibinfo {author} {\bibfnamefont {R.}~\bibnamefont {Froeschl}}, \bibinfo {author} {\bibfnamefont {R.}~\bibnamefont {Garc{\'i}a~Al{\'i}a}}, \bibinfo {author} {\bibfnamefont {A.}~\bibnamefont {Gerbershagen}}, \bibinfo {author} {\bibfnamefont {S.}~\bibnamefont {Gilardoni}}, \bibinfo {author} {\bibfnamefont {D.}~\bibnamefont {Horv{\'a}th}},
  \bibinfo {author} {\bibfnamefont {G.}~\bibnamefont {Hugo}}, \bibinfo {author} {\bibfnamefont {A.}~\bibnamefont {Infantino}}, \bibinfo {author} {\bibfnamefont {V.}~\bibnamefont {Kouskoura}}, \bibinfo {author} {\bibfnamefont {A.}~\bibnamefont {Lechner}}, \bibinfo {author} {\bibfnamefont {B.}~\bibnamefont {Lefebvre}}, \bibinfo {author} {\bibfnamefont {G.}~\bibnamefont {Lerner}}, \bibinfo {author} {\bibfnamefont {M.}~\bibnamefont {Magistris}}, \bibinfo {author} {\bibfnamefont {A.}~\bibnamefont {Manousos}}, \bibinfo {author} {\bibfnamefont {G.}~\bibnamefont {Moryc}}, \bibinfo {author} {\bibfnamefont {F.}~\bibnamefont {Ogallar~Ruiz}}, \bibinfo {author} {\bibfnamefont {F.}~\bibnamefont {Pozzi}}, \bibinfo {author} {\bibfnamefont {D.}~\bibnamefont {Prelipcean}}, \bibinfo {author} {\bibfnamefont {S.}~\bibnamefont {Roesler}}, \bibinfo {author} {\bibfnamefont {R.}~\bibnamefont {Rossi}}, \bibinfo {author} {\bibfnamefont {M.}~\bibnamefont {Sabat{\'e}~Gilarte}}, \bibinfo {author} {\bibfnamefont {F.}~\bibnamefont
  {Salvat~Pujol}}, \bibinfo {author} {\bibfnamefont {P.}~\bibnamefont {Schoofs}}, \bibinfo {author} {\bibfnamefont {V.}~\bibnamefont {Str{\'a}nsk{\'y}}}, \bibinfo {author} {\bibfnamefont {C.}~\bibnamefont {Theis}}, \bibinfo {author} {\bibfnamefont {A.}~\bibnamefont {Tsinganis}}, \bibinfo {author} {\bibfnamefont {R.}~\bibnamefont {Versaci}}, \bibinfo {author} {\bibfnamefont {V.}~\bibnamefont {Vlachoudis}}, \bibinfo {author} {\bibfnamefont {A.}~\bibnamefont {Waets}},\ and\ \bibinfo {author} {\bibfnamefont {M.}~\bibnamefont {Widorski}},\ }\bibfield  {title} {\bibinfo {title} {New {{Capabilities}} of the {{FLUKA Multi-Purpose Code}}},\ }\href {https://doi.org/10.3389/fphy.2021.788253} {\bibfield  {journal} {\bibinfo  {journal} {Front. Phys.}\ }\textbf {\bibinfo {volume} {9}},\ \bibinfo {pages} {788253} (\bibinfo {year} {2022})}\BibitemShut {NoStop}%
\bibitem [{\citenamefont {Sato}\ \emph {et~al.}(2024)\citenamefont {Sato}, \citenamefont {Iwamoto}, \citenamefont {Hashimoto}, \citenamefont {Ogawa}, \citenamefont {Furuta}, \citenamefont {Abe}, \citenamefont {Kai}, \citenamefont {Matsuya}, \citenamefont {Matsuda}, \citenamefont {Hirata}, \citenamefont {Sekikawa}, \citenamefont {Yao}, \citenamefont {Tsai}, \citenamefont {Ratliff}, \citenamefont {Iwase}, \citenamefont {Sakaki}, \citenamefont {Sugihara}, \citenamefont {Shigyo}, \citenamefont {Sihver},\ and\ \citenamefont {Niita}}]{Sato2024}%
  \BibitemOpen
  \bibfield  {author} {\bibinfo {author} {\bibfnamefont {T.}~\bibnamefont {Sato}}, \bibinfo {author} {\bibfnamefont {Y.}~\bibnamefont {Iwamoto}}, \bibinfo {author} {\bibfnamefont {S.}~\bibnamefont {Hashimoto}}, \bibinfo {author} {\bibfnamefont {T.}~\bibnamefont {Ogawa}}, \bibinfo {author} {\bibfnamefont {T.}~\bibnamefont {Furuta}}, \bibinfo {author} {\bibfnamefont {S.-I.}\ \bibnamefont {Abe}}, \bibinfo {author} {\bibfnamefont {T.}~\bibnamefont {Kai}}, \bibinfo {author} {\bibfnamefont {Y.}~\bibnamefont {Matsuya}}, \bibinfo {author} {\bibfnamefont {N.}~\bibnamefont {Matsuda}}, \bibinfo {author} {\bibfnamefont {Y.}~\bibnamefont {Hirata}}, \bibinfo {author} {\bibfnamefont {T.}~\bibnamefont {Sekikawa}}, \bibinfo {author} {\bibfnamefont {L.}~\bibnamefont {Yao}}, \bibinfo {author} {\bibfnamefont {P.-E.}\ \bibnamefont {Tsai}}, \bibinfo {author} {\bibfnamefont {H.~N.}\ \bibnamefont {Ratliff}}, \bibinfo {author} {\bibfnamefont {H.}~\bibnamefont {Iwase}}, \bibinfo {author} {\bibfnamefont {Y.}~\bibnamefont {Sakaki}},
  \bibinfo {author} {\bibfnamefont {K.}~\bibnamefont {Sugihara}}, \bibinfo {author} {\bibfnamefont {N.}~\bibnamefont {Shigyo}}, \bibinfo {author} {\bibfnamefont {L.}~\bibnamefont {Sihver}},\ and\ \bibinfo {author} {\bibfnamefont {K.}~\bibnamefont {Niita}},\ }\bibfield  {title} {\bibinfo {title} {Recent improvements of the particle and heavy ion transport code system -- {{PHITS}} version 3.33},\ }\href {https://doi.org/10.1080/00223131.2023.2275736} {\bibfield  {journal} {\bibinfo  {journal} {J. Nucl. Sci. Technol.}\ }\textbf {\bibinfo {volume} {61}},\ \bibinfo {pages} {127} (\bibinfo {year} {2024})}\BibitemShut {NoStop}%
\bibitem [{\citenamefont {{Van der Graaf}}\ \emph {et~al.}(2011)\citenamefont {{Van der Graaf}}, \citenamefont {Limburg}, \citenamefont {Koomans},\ and\ \citenamefont {Tijs}}]{VanderGraaf2011a}%
  \BibitemOpen
  \bibfield  {author} {\bibinfo {author} {\bibfnamefont {E.~R.}\ \bibnamefont {{Van der Graaf}}}, \bibinfo {author} {\bibfnamefont {J.}~\bibnamefont {Limburg}}, \bibinfo {author} {\bibfnamefont {R.~L.}\ \bibnamefont {Koomans}},\ and\ \bibinfo {author} {\bibfnamefont {M.}~\bibnamefont {Tijs}},\ }\bibfield  {title} {\bibinfo {title} {Monte {{Carlo}} based calibration of scintillation detectors for laboratory and in situ gamma ray measurements},\ }\href {https://doi.org/10.1016/j.jenvrad.2010.12.001} {\bibfield  {journal} {\bibinfo  {journal} {J. Environ. Radioact.}\ }\textbf {\bibinfo {volume} {102}},\ \bibinfo {pages} {270} (\bibinfo {year} {2011})}\BibitemShut {NoStop}%
\bibitem [{\citenamefont {Kulisek}\ \emph {et~al.}(2018)\citenamefont {Kulisek}, \citenamefont {Wittman}, \citenamefont {Miller}, \citenamefont {Kernan}, \citenamefont {McCall}, \citenamefont {McConn}, \citenamefont {Schweppe}, \citenamefont {Seifert}, \citenamefont {Stave},\ and\ \citenamefont {Stewart}}]{Kulisek2018a}%
  \BibitemOpen
  \bibfield  {author} {\bibinfo {author} {\bibfnamefont {J.~A.}\ \bibnamefont {Kulisek}}, \bibinfo {author} {\bibfnamefont {R.~S.}\ \bibnamefont {Wittman}}, \bibinfo {author} {\bibfnamefont {E.~A.}\ \bibnamefont {Miller}}, \bibinfo {author} {\bibfnamefont {W.~J.}\ \bibnamefont {Kernan}}, \bibinfo {author} {\bibfnamefont {J.~D.}\ \bibnamefont {McCall}}, \bibinfo {author} {\bibfnamefont {R.~J.}\ \bibnamefont {McConn}}, \bibinfo {author} {\bibfnamefont {J.~E.}\ \bibnamefont {Schweppe}}, \bibinfo {author} {\bibfnamefont {C.~E.}\ \bibnamefont {Seifert}}, \bibinfo {author} {\bibfnamefont {S.~C.}\ \bibnamefont {Stave}},\ and\ \bibinfo {author} {\bibfnamefont {T.~N.}\ \bibnamefont {Stewart}},\ }\bibfield  {title} {\bibinfo {title} {A {{3D}} simulation look-up library for real-time airborne gamma-ray spectroscopy},\ }\href {https://doi.org/10.1016/j.nima.2017.10.030} {\bibfield  {journal} {\bibinfo  {journal} {Nucl. Instrum. Methods Phys. Res. Sect. Accel. Spectrometers Detect. Assoc. Equip.}\ }\textbf {\bibinfo
  {volume} {879}},\ \bibinfo {pages} {84} (\bibinfo {year} {2018})}\BibitemShut {NoStop}%
\bibitem [{\citenamefont {Androulakaki}\ \emph {et~al.}(2016)\citenamefont {Androulakaki}, \citenamefont {Kokkoris}, \citenamefont {Skordis}, \citenamefont {Fatsea}, \citenamefont {Patiris}, \citenamefont {Tsabaris},\ and\ \citenamefont {Vlastou}}]{Androulakaki2016b}%
  \BibitemOpen
  \bibfield  {author} {\bibinfo {author} {\bibfnamefont {E.~G.}\ \bibnamefont {Androulakaki}}, \bibinfo {author} {\bibfnamefont {M.}~\bibnamefont {Kokkoris}}, \bibinfo {author} {\bibfnamefont {E.}~\bibnamefont {Skordis}}, \bibinfo {author} {\bibfnamefont {E.}~\bibnamefont {Fatsea}}, \bibinfo {author} {\bibfnamefont {D.~L.}\ \bibnamefont {Patiris}}, \bibinfo {author} {\bibfnamefont {C.}~\bibnamefont {Tsabaris}},\ and\ \bibinfo {author} {\bibfnamefont {R.}~\bibnamefont {Vlastou}},\ }\bibfield  {title} {\bibinfo {title} {Implementation of {{FLUKA}} for {$\gamma$}-ray applications in the marine environment},\ }\href {https://doi.org/10.1016/j.jenvrad.2016.08.008} {\bibfield  {journal} {\bibinfo  {journal} {J. Environ. Radioact.}\ }\textbf {\bibinfo {volume} {164}},\ \bibinfo {pages} {253} (\bibinfo {year} {2016})}\BibitemShut {NoStop}%
\bibitem [{\citenamefont {Breitenmoser}\ \emph {et~al.}(2025)\citenamefont {Breitenmoser}, \citenamefont {Stabilini}, \citenamefont {Kasprzak},\ and\ \citenamefont {Mayer}}]{Breitenmoser2025a}%
  \BibitemOpen
  \bibfield  {author} {\bibinfo {author} {\bibfnamefont {D.}~\bibnamefont {Breitenmoser}}, \bibinfo {author} {\bibfnamefont {A.}~\bibnamefont {Stabilini}}, \bibinfo {author} {\bibfnamefont {M.~M.}\ \bibnamefont {Kasprzak}},\ and\ \bibinfo {author} {\bibfnamefont {S.}~\bibnamefont {Mayer}},\ }\bibfield  {title} {\bibinfo {title} {Development and validation of a high-fidelity full-spectrum {{Monte Carlo}} model for the {{Swiss}} airborne gamma-ray spectrometry system},\ }\href {https://doi.org/10.1016/j.nima.2025.170512} {\bibfield  {journal} {\bibinfo  {journal} {Nuclear Instruments and Methods in Physics Research Section A: Accelerators, Spectrometers, Detectors and Associated Equipment}\ }\textbf {\bibinfo {volume} {1077}},\ \bibinfo {pages} {170512} (\bibinfo {year} {2025})}\BibitemShut {NoStop}%
\bibitem [{\citenamefont {Reedy}\ \emph {et~al.}(1973)\citenamefont {Reedy}, \citenamefont {Arnold},\ and\ \citenamefont {Trombka}}]{Reedy1973a}%
  \BibitemOpen
  \bibfield  {author} {\bibinfo {author} {\bibfnamefont {R.~C.}\ \bibnamefont {Reedy}}, \bibinfo {author} {\bibfnamefont {J.~R.}\ \bibnamefont {Arnold}},\ and\ \bibinfo {author} {\bibfnamefont {J.~I.}\ \bibnamefont {Trombka}},\ }\bibfield  {title} {\bibinfo {title} {Expected {$\gamma$} ray emission spectra from the lunar surface as a function of chemical composition},\ }\href {https://doi.org/10.1029/jb078i026p05847} {\bibfield  {journal} {\bibinfo  {journal} {J. Geophys. Res.}\ }\textbf {\bibinfo {volume} {78}},\ \bibinfo {pages} {5847} (\bibinfo {year} {1973})}\BibitemShut {NoStop}%
\bibitem [{\citenamefont {Knoll}(2010)}]{Knoll2010a}%
  \BibitemOpen
  \bibfield  {author} {\bibinfo {author} {\bibfnamefont {G.~F.}\ \bibnamefont {Knoll}},\ }\href@noop {} {\emph {\bibinfo {title} {Radiation {{Detection}} and {{Measurement}}}}}\ (\bibinfo  {publisher} {John Wiley \& Sons},\ \bibinfo {address} {New York, USA},\ \bibinfo {year} {2010})\BibitemShut {NoStop}%
\bibitem [{\citenamefont {Kluso{\v n}}(2010)}]{Kluson2010c}%
  \BibitemOpen
  \bibfield  {author} {\bibinfo {author} {\bibfnamefont {J.}~\bibnamefont {Kluso{\v n}}},\ }\bibfield  {title} {\bibinfo {title} {In-situ gamma spectrometry in environmental monitoring},\ }\href {https://doi.org/10.1016/j.apradiso.2009.11.041} {\bibfield  {journal} {\bibinfo  {journal} {Appl. Radiat. Isot.}\ }\textbf {\bibinfo {volume} {68}},\ \bibinfo {pages} {529} (\bibinfo {year} {2010})}\BibitemShut {NoStop}%
\bibitem [{\citenamefont {Kaastra}\ and\ \citenamefont {Bleeker}(2016)}]{Kaastra2016a}%
  \BibitemOpen
  \bibfield  {author} {\bibinfo {author} {\bibfnamefont {J.~S.}\ \bibnamefont {Kaastra}}\ and\ \bibinfo {author} {\bibfnamefont {J.~A.~M.}\ \bibnamefont {Bleeker}},\ }\bibfield  {title} {\bibinfo {title} {Optimal binning of {{X-ray}} spectra and response matrix design},\ }\href {https://doi.org/10.1051/0004-6361/201527395} {\bibfield  {journal} {\bibinfo  {journal} {Astron. Astrophys.}\ }\textbf {\bibinfo {volume} {587}},\ \bibinfo {pages} {A151} (\bibinfo {year} {2016})}\BibitemShut {NoStop}%
\bibitem [{\citenamefont {Sjoden}\ and\ \citenamefont {Haghighat}(1996)}]{Sjoden1996}%
  \BibitemOpen
  \bibfield  {author} {\bibinfo {author} {\bibfnamefont {G.~E.}\ \bibnamefont {Sjoden}}\ and\ \bibinfo {author} {\bibfnamefont {A.}~\bibnamefont {Haghighat}},\ }\bibfield  {title} {\bibinfo {title} {Simplified multigrid acceleration in the {{PENTRAN}} 3-{{D}} parallel code},\ }in\ \href@noop {} {\emph {\bibinfo {booktitle} {Trans. {{Am}}. {{Nucl}}. {{Soc}}.}}},\ Vol.~\bibinfo {volume} {75}\ (\bibinfo {address} {Washington, US},\ \bibinfo {year} {1996})\BibitemShut {NoStop}%
\bibitem [{\citenamefont {Wareing}\ \emph {et~al.}(1996)\citenamefont {Wareing}, \citenamefont {McGhee},\ and\ \citenamefont {Morel}}]{Wareing1996}%
  \BibitemOpen
  \bibfield  {author} {\bibinfo {author} {\bibfnamefont {T.~A.}\ \bibnamefont {Wareing}}, \bibinfo {author} {\bibfnamefont {J.~M.}\ \bibnamefont {McGhee}},\ and\ \bibinfo {author} {\bibfnamefont {J.~E.}\ \bibnamefont {Morel}},\ }\bibfield  {title} {\bibinfo {title} {{{ATTILA}}: {{A}} three-dimensional, unstructured tetrahedral mesh discrete ordinates transport code},\ }in\ \href@noop {} {\emph {\bibinfo {booktitle} {Trans. {{Am}}. {{Nucl}}. {{Soc}}.}}},\ Vol.~\bibinfo {volume} {75}\ (\bibinfo {address} {Washington, US},\ \bibinfo {year} {1996})\BibitemShut {NoStop}%
\bibitem [{\citenamefont {Alcouffe}(2001)}]{Alcouffe2001}%
  \BibitemOpen
  \bibfield  {author} {\bibinfo {author} {\bibfnamefont {R.~E.}\ \bibnamefont {Alcouffe}},\ }\bibfield  {title} {\bibinfo {title} {Partisn calculations of {{3D}} radiation transport benchmarks for simple geometries with void regions},\ }\href {https://doi.org/10.1016/S0149-1970(01)00011-7} {\bibfield  {journal} {\bibinfo  {journal} {Progress in Nuclear Energy}\ }\textbf {\bibinfo {volume} {39}},\ \bibinfo {pages} {181} (\bibinfo {year} {2001})}\BibitemShut {NoStop}%
\bibitem [{\citenamefont {Breitenmoser}(2024)}]{Breitenmoser2024}%
  \BibitemOpen
  \bibfield  {author} {\bibinfo {author} {\bibfnamefont {D.}~\bibnamefont {Breitenmoser}},\ }\emph {\bibinfo {title} {Towards {{Monte Carlo}} Based {{Full Spectrum Modeling}} of {{Airborne Gamma-Ray Spectrometry Systems}}}},\ \href {https://doi.org/10.3929/ethz-b-000694094} {\bibinfo {type} {Doctoral {{Thesis}}}},\ \bibinfo  {school} {ETH Zurich} (\bibinfo {year} {2024}),\ \Eprint {https://arxiv.org/abs/2411.02606} {arXiv:2411.02606} \BibitemShut {NoStop}%
\bibitem [{\citenamefont {Breitenmoser}\ \emph {et~al.}(2022)\citenamefont {Breitenmoser}, \citenamefont {Butterweck}, \citenamefont {Kasprzak}, \citenamefont {Yukihara},\ and\ \citenamefont {Mayer}}]{Breitenmoser2022e}%
  \BibitemOpen
  \bibfield  {author} {\bibinfo {author} {\bibfnamefont {D.}~\bibnamefont {Breitenmoser}}, \bibinfo {author} {\bibfnamefont {G.}~\bibnamefont {Butterweck}}, \bibinfo {author} {\bibfnamefont {M.~M.}\ \bibnamefont {Kasprzak}}, \bibinfo {author} {\bibfnamefont {E.~G.}\ \bibnamefont {Yukihara}},\ and\ \bibinfo {author} {\bibfnamefont {S.}~\bibnamefont {Mayer}},\ }\bibfield  {title} {\bibinfo {title} {Experimental and {{Simulated Spectral Gamma-Ray Response}} of a {{NaI}}({{Tl}}) {{Scintillation Detector}} used in {{Airborne Gamma-Ray Spectrometry}}},\ }\href {https://doi.org/10.5194/ADGEO-57-89-2022} {\bibfield  {journal} {\bibinfo  {journal} {Adv. Geosci.}\ }\textbf {\bibinfo {volume} {57}},\ \bibinfo {pages} {89} (\bibinfo {year} {2022})}\BibitemShut {NoStop}%
\bibitem [{\citenamefont {Calura}\ \emph {et~al.}(2000)\citenamefont {Calura}, \citenamefont {Rapisarda}, \citenamefont {Frontera}, \citenamefont {Montanari}, \citenamefont {Guidorzi}, \citenamefont {Amati}, \citenamefont {Feroci}, \citenamefont {Costa},\ and\ \citenamefont {Collina}}]{Calura2000}%
  \BibitemOpen
  \bibfield  {author} {\bibinfo {author} {\bibfnamefont {F.}~\bibnamefont {Calura}}, \bibinfo {author} {\bibfnamefont {M.}~\bibnamefont {Rapisarda}}, \bibinfo {author} {\bibfnamefont {F.}~\bibnamefont {Frontera}}, \bibinfo {author} {\bibfnamefont {E.}~\bibnamefont {Montanari}}, \bibinfo {author} {\bibfnamefont {C.}~\bibnamefont {Guidorzi}}, \bibinfo {author} {\bibfnamefont {L.}~\bibnamefont {Amati}}, \bibinfo {author} {\bibfnamefont {M.}~\bibnamefont {Feroci}}, \bibinfo {author} {\bibfnamefont {E.}~\bibnamefont {Costa}},\ and\ \bibinfo {author} {\bibfnamefont {P.}~\bibnamefont {Collina}},\ }\bibfield  {title} {\bibinfo {title} {Response function of the {{Gamma-Ray Burst Monitor}} ({{GRBM}}) onboard the {{BeppoSAX}} satellite},\ }in\ \href {https://doi.org/10.1063/1.1361629} {\emph {\bibinfo {booktitle} {{{AIP Conf}}. {{Proc}}.}}},\ Vol.\ \bibinfo {volume} {526}\ (\bibinfo  {publisher} {American Institute of PhysicsAIP},\ \bibinfo {address} {Huntsville, US},\ \bibinfo {year} {2000})\ pp.\ \bibinfo {pages}
  {721--725}\BibitemShut {NoStop}%
\bibitem [{\citenamefont {Chen}\ \emph {et~al.}(2013)\citenamefont {Chen}, \citenamefont {Argan}, \citenamefont {Bulgarelli}, \citenamefont {Cattaneo}, \citenamefont {Contessi}, \citenamefont {Giuliani}, \citenamefont {Pittori}, \citenamefont {Pucella}, \citenamefont {Tavani}, \citenamefont {Trois}, \citenamefont {Verrecchia}, \citenamefont {Barbiellini}, \citenamefont {Caraveo}, \citenamefont {Colafrancesco}, \citenamefont {Costa}, \citenamefont {De~Paris}, \citenamefont {Del~Monte}, \citenamefont {Di~Cocco}, \citenamefont {Donnarumma}, \citenamefont {Evangelista}, \citenamefont {Ferrari}, \citenamefont {Feroci}, \citenamefont {Fioretti}, \citenamefont {Fiorini}, \citenamefont {Fuschino}, \citenamefont {Galli}, \citenamefont {Gianotti}, \citenamefont {Giommi}, \citenamefont {Giusti}, \citenamefont {Labanti}, \citenamefont {Lapshov}, \citenamefont {Lazzarotto}, \citenamefont {Lipari}, \citenamefont {Longo}, \citenamefont {Lucarelli}, \citenamefont {Marisaldi}, \citenamefont {Mereghetti}, \citenamefont
  {Morelli}, \citenamefont {Moretti}, \citenamefont {Morselli}, \citenamefont {Pacciani}, \citenamefont {Pellizzoni}, \citenamefont {Perotti}, \citenamefont {Piano}, \citenamefont {Picozza}, \citenamefont {Pilia}, \citenamefont {Prest}, \citenamefont {Rapisarda}, \citenamefont {Rappoldi}, \citenamefont {Rubini}, \citenamefont {Sabatini}, \citenamefont {Santolamazza}, \citenamefont {Soffitta}, \citenamefont {Striani}, \citenamefont {Trifoglio}, \citenamefont {Valentini}, \citenamefont {Vallazza}, \citenamefont {Vercellone}, \citenamefont {Vittorini},\ and\ \citenamefont {Zanello}}]{Chen2013a}%
  \BibitemOpen
  \bibfield  {author} {\bibinfo {author} {\bibfnamefont {A.~W.}\ \bibnamefont {Chen}}, \bibinfo {author} {\bibfnamefont {A.}~\bibnamefont {Argan}}, \bibinfo {author} {\bibfnamefont {A.}~\bibnamefont {Bulgarelli}}, \bibinfo {author} {\bibfnamefont {P.~W.}\ \bibnamefont {Cattaneo}}, \bibinfo {author} {\bibfnamefont {T.}~\bibnamefont {Contessi}}, \bibinfo {author} {\bibfnamefont {A.}~\bibnamefont {Giuliani}}, \bibinfo {author} {\bibfnamefont {C.}~\bibnamefont {Pittori}}, \bibinfo {author} {\bibfnamefont {G.}~\bibnamefont {Pucella}}, \bibinfo {author} {\bibfnamefont {M.}~\bibnamefont {Tavani}}, \bibinfo {author} {\bibfnamefont {A.}~\bibnamefont {Trois}}, \bibinfo {author} {\bibfnamefont {F.}~\bibnamefont {Verrecchia}}, \bibinfo {author} {\bibfnamefont {G.}~\bibnamefont {Barbiellini}}, \bibinfo {author} {\bibfnamefont {P.}~\bibnamefont {Caraveo}}, \bibinfo {author} {\bibfnamefont {S.}~\bibnamefont {Colafrancesco}}, \bibinfo {author} {\bibfnamefont {E.}~\bibnamefont {Costa}}, \bibinfo {author} {\bibfnamefont
  {G.}~\bibnamefont {De~Paris}}, \bibinfo {author} {\bibfnamefont {E.}~\bibnamefont {Del~Monte}}, \bibinfo {author} {\bibfnamefont {G.}~\bibnamefont {Di~Cocco}}, \bibinfo {author} {\bibfnamefont {I.}~\bibnamefont {Donnarumma}}, \bibinfo {author} {\bibfnamefont {Y.}~\bibnamefont {Evangelista}}, \bibinfo {author} {\bibfnamefont {A.}~\bibnamefont {Ferrari}}, \bibinfo {author} {\bibfnamefont {M.}~\bibnamefont {Feroci}}, \bibinfo {author} {\bibfnamefont {V.}~\bibnamefont {Fioretti}}, \bibinfo {author} {\bibfnamefont {M.}~\bibnamefont {Fiorini}}, \bibinfo {author} {\bibfnamefont {F.}~\bibnamefont {Fuschino}}, \bibinfo {author} {\bibfnamefont {M.}~\bibnamefont {Galli}}, \bibinfo {author} {\bibfnamefont {F.}~\bibnamefont {Gianotti}}, \bibinfo {author} {\bibfnamefont {P.}~\bibnamefont {Giommi}}, \bibinfo {author} {\bibfnamefont {M.}~\bibnamefont {Giusti}}, \bibinfo {author} {\bibfnamefont {C.}~\bibnamefont {Labanti}}, \bibinfo {author} {\bibfnamefont {I.}~\bibnamefont {Lapshov}}, \bibinfo {author} {\bibfnamefont
  {F.}~\bibnamefont {Lazzarotto}}, \bibinfo {author} {\bibfnamefont {P.}~\bibnamefont {Lipari}}, \bibinfo {author} {\bibfnamefont {F.}~\bibnamefont {Longo}}, \bibinfo {author} {\bibfnamefont {F.}~\bibnamefont {Lucarelli}}, \bibinfo {author} {\bibfnamefont {M.}~\bibnamefont {Marisaldi}}, \bibinfo {author} {\bibfnamefont {S.}~\bibnamefont {Mereghetti}}, \bibinfo {author} {\bibfnamefont {E.}~\bibnamefont {Morelli}}, \bibinfo {author} {\bibfnamefont {E.}~\bibnamefont {Moretti}}, \bibinfo {author} {\bibfnamefont {A.}~\bibnamefont {Morselli}}, \bibinfo {author} {\bibfnamefont {L.}~\bibnamefont {Pacciani}}, \bibinfo {author} {\bibfnamefont {A.}~\bibnamefont {Pellizzoni}}, \bibinfo {author} {\bibfnamefont {F.}~\bibnamefont {Perotti}}, \bibinfo {author} {\bibfnamefont {G.}~\bibnamefont {Piano}}, \bibinfo {author} {\bibfnamefont {P.}~\bibnamefont {Picozza}}, \bibinfo {author} {\bibfnamefont {M.}~\bibnamefont {Pilia}}, \bibinfo {author} {\bibfnamefont {M.}~\bibnamefont {Prest}}, \bibinfo {author} {\bibfnamefont
  {M.}~\bibnamefont {Rapisarda}}, \bibinfo {author} {\bibfnamefont {A.}~\bibnamefont {Rappoldi}}, \bibinfo {author} {\bibfnamefont {A.}~\bibnamefont {Rubini}}, \bibinfo {author} {\bibfnamefont {S.}~\bibnamefont {Sabatini}}, \bibinfo {author} {\bibfnamefont {P.}~\bibnamefont {Santolamazza}}, \bibinfo {author} {\bibfnamefont {P.}~\bibnamefont {Soffitta}}, \bibinfo {author} {\bibfnamefont {E.}~\bibnamefont {Striani}}, \bibinfo {author} {\bibfnamefont {M.}~\bibnamefont {Trifoglio}}, \bibinfo {author} {\bibfnamefont {G.}~\bibnamefont {Valentini}}, \bibinfo {author} {\bibfnamefont {E.}~\bibnamefont {Vallazza}}, \bibinfo {author} {\bibfnamefont {S.}~\bibnamefont {Vercellone}}, \bibinfo {author} {\bibfnamefont {V.}~\bibnamefont {Vittorini}},\ and\ \bibinfo {author} {\bibfnamefont {D.}~\bibnamefont {Zanello}},\ }\bibfield  {title} {\bibinfo {title} {Calibration of {{AGILE-GRID}} with in-flight data and {{Monte Carlo}} simulations},\ }\href {https://doi.org/10.1051/0004-6361/201321767} {\bibfield  {journal} {\bibinfo
  {journal} {Astron. Astrophys.}\ }\textbf {\bibinfo {volume} {558}},\ \bibinfo {pages} {A37} (\bibinfo {year} {2013})}\BibitemShut {NoStop}%
\bibitem [{\citenamefont {Ackermann}\ \emph {et~al.}(2012)\citenamefont {Ackermann}, \citenamefont {Ajello}, \citenamefont {Albert}, \citenamefont {Allafort}, \citenamefont {Atwood}, \citenamefont {Axelsson}, \citenamefont {Baldini}, \citenamefont {Ballet}, \citenamefont {Barbiellini}, \citenamefont {Bastieri}, \citenamefont {Bechtol}, \citenamefont {Bellazzini}, \citenamefont {Bissaldi}, \citenamefont {Blandford}, \citenamefont {Bloom}, \citenamefont {Bogart}, \citenamefont {Bonamente}, \citenamefont {Borgland}, \citenamefont {Bottacini}, \citenamefont {Bouvier}, \citenamefont {Brandt}, \citenamefont {Bregeon}, \citenamefont {Brigida}, \citenamefont {Bruel}, \citenamefont {Buehler}, \citenamefont {Burnett}, \citenamefont {Buson}, \citenamefont {Caliandro}, \citenamefont {Cameron}, \citenamefont {Caraveo}, \citenamefont {Casandjian}, \citenamefont {Cavazzuti}, \citenamefont {Cecchi}, \citenamefont {{\c C}elik}, \citenamefont {Charles}, \citenamefont {Chaves}, \citenamefont {Chekhtman}, \citenamefont {Cheung},
  \citenamefont {Chiang}, \citenamefont {Ciprini}, \citenamefont {Claus}, \citenamefont {{Cohen-Tanugi}}, \citenamefont {Conrad}, \citenamefont {Corbet}, \citenamefont {Cutini}, \citenamefont {D'Ammando}, \citenamefont {Davis}, \citenamefont {De~Angelis}, \citenamefont {Deklotz}, \citenamefont {De~Palma}, \citenamefont {Dermer}, \citenamefont {Digel}, \citenamefont {Do~Couto E~Silva}, \citenamefont {Drell}, \citenamefont {{Drlica-Wagner}}, \citenamefont {Dubois}, \citenamefont {Favuzzi}, \citenamefont {Fegan}, \citenamefont {Ferrara}, \citenamefont {Focke}, \citenamefont {Fortin}, \citenamefont {Fukazawa}, \citenamefont {Funk}, \citenamefont {Fusco}, \citenamefont {Gargano}, \citenamefont {Gasparrini}, \citenamefont {Gehrels}, \citenamefont {Giebels}, \citenamefont {Giglietto}, \citenamefont {Giordano}, \citenamefont {Giroletti}, \citenamefont {Glanzman}, \citenamefont {Godfrey}, \citenamefont {Grenier}, \citenamefont {Grove}, \citenamefont {Guiriec}, \citenamefont {Hadasch}, \citenamefont {Hayashida},
  \citenamefont {Hays}, \citenamefont {Horan}, \citenamefont {Hou}, \citenamefont {Hughes}, \citenamefont {Jackson}, \citenamefont {Jogler}, \citenamefont {J{\'o}hannesson}, \citenamefont {Johnson}, \citenamefont {Johnson}, \citenamefont {Johnson}, \citenamefont {Kamae}, \citenamefont {Katagiri}, \citenamefont {Kataoka}, \citenamefont {Kerr}, \citenamefont {Kn{\"o}dlseder}, \citenamefont {Kuss}, \citenamefont {Lande}, \citenamefont {Larsson}, \citenamefont {Latronico}, \citenamefont {Lavalley}, \citenamefont {{Lemoine-Goumard}}, \citenamefont {Longo}, \citenamefont {Loparco}, \citenamefont {Lott}, \citenamefont {Lovellette}, \citenamefont {Lubrano}, \citenamefont {Mazziotta}, \citenamefont {McConville}, \citenamefont {McEnery}, \citenamefont {Mehault}, \citenamefont {Michelson}, \citenamefont {Mitthumsiri}, \citenamefont {Mizuno}, \citenamefont {Moiseev}, \citenamefont {Monte}, \citenamefont {Monzani}, \citenamefont {Morselli}, \citenamefont {Moskalenko}, \citenamefont {Murgia}, \citenamefont
  {{Naumann-Godo}}, \citenamefont {Nemmen}, \citenamefont {Nishino}, \citenamefont {Norris}, \citenamefont {Nuss}, \citenamefont {Ohno}, \citenamefont {Ohsugi}, \citenamefont {Okumura}, \citenamefont {Omodei}, \citenamefont {Orienti}, \citenamefont {Orlando}, \citenamefont {Ormes}, \citenamefont {Paneque}, \citenamefont {Panetta}, \citenamefont {Perkins}, \citenamefont {{Pesce-Rollins}}, \citenamefont {Pierbattista}, \citenamefont {Piron}, \citenamefont {Pivato}, \citenamefont {Porter}, \citenamefont {Racusin}, \citenamefont {Rain{\`o}}, \citenamefont {Rando}, \citenamefont {Razzano}, \citenamefont {Razzaque}, \citenamefont {Reimer}, \citenamefont {Reimer}, \citenamefont {Reposeur}, \citenamefont {Reyes}, \citenamefont {Ritz}, \citenamefont {Rochester}, \citenamefont {Romoli}, \citenamefont {Roth}, \citenamefont {Sadrozinski}, \citenamefont {Sanchez}, \citenamefont {Saz~Parkinson}, \citenamefont {Sbarra}, \citenamefont {Scargle}, \citenamefont {Sgr{\`o}}, \citenamefont {{Siegal-Gaskins}}, \citenamefont
  {Siskind}, \citenamefont {Spandre}, \citenamefont {Spinelli}, \citenamefont {Stephens}, \citenamefont {Suson}, \citenamefont {Tajima}, \citenamefont {Takahashi}, \citenamefont {Tanaka}, \citenamefont {Thayer}, \citenamefont {Thayer}, \citenamefont {Thompson}, \citenamefont {Tibaldo}, \citenamefont {Tinivella}, \citenamefont {Tosti}, \citenamefont {Troja}, \citenamefont {Usher}, \citenamefont {Vandenbroucke}, \citenamefont {Van~Klaveren}, \citenamefont {Vasileiou}, \citenamefont {Vianello}, \citenamefont {Vitale}, \citenamefont {Waite}, \citenamefont {Wallace}, \citenamefont {Winer}, \citenamefont {Wood}, \citenamefont {Wood}, \citenamefont {Wood}, \citenamefont {Yang},\ and\ \citenamefont {Zimmer}}]{Ackermann2012a}%
  \BibitemOpen
  \bibfield  {author} {\bibinfo {author} {\bibfnamefont {M.}~\bibnamefont {Ackermann}}, \bibinfo {author} {\bibfnamefont {M.}~\bibnamefont {Ajello}}, \bibinfo {author} {\bibfnamefont {A.}~\bibnamefont {Albert}}, \bibinfo {author} {\bibfnamefont {A.}~\bibnamefont {Allafort}}, \bibinfo {author} {\bibfnamefont {W.~B.}\ \bibnamefont {Atwood}}, \bibinfo {author} {\bibfnamefont {M.}~\bibnamefont {Axelsson}}, \bibinfo {author} {\bibfnamefont {L.}~\bibnamefont {Baldini}}, \bibinfo {author} {\bibfnamefont {J.}~\bibnamefont {Ballet}}, \bibinfo {author} {\bibfnamefont {G.}~\bibnamefont {Barbiellini}}, \bibinfo {author} {\bibfnamefont {D.}~\bibnamefont {Bastieri}}, \bibinfo {author} {\bibfnamefont {K.}~\bibnamefont {Bechtol}}, \bibinfo {author} {\bibfnamefont {R.}~\bibnamefont {Bellazzini}}, \bibinfo {author} {\bibfnamefont {E.}~\bibnamefont {Bissaldi}}, \bibinfo {author} {\bibfnamefont {R.~D.}\ \bibnamefont {Blandford}}, \bibinfo {author} {\bibfnamefont {E.~D.}\ \bibnamefont {Bloom}}, \bibinfo {author} {\bibfnamefont
  {J.~R.}\ \bibnamefont {Bogart}}, \bibinfo {author} {\bibfnamefont {E.}~\bibnamefont {Bonamente}}, \bibinfo {author} {\bibfnamefont {A.~W.}\ \bibnamefont {Borgland}}, \bibinfo {author} {\bibfnamefont {E.}~\bibnamefont {Bottacini}}, \bibinfo {author} {\bibfnamefont {A.}~\bibnamefont {Bouvier}}, \bibinfo {author} {\bibfnamefont {T.~J.}\ \bibnamefont {Brandt}}, \bibinfo {author} {\bibfnamefont {J.}~\bibnamefont {Bregeon}}, \bibinfo {author} {\bibfnamefont {M.}~\bibnamefont {Brigida}}, \bibinfo {author} {\bibfnamefont {P.}~\bibnamefont {Bruel}}, \bibinfo {author} {\bibfnamefont {R.}~\bibnamefont {Buehler}}, \bibinfo {author} {\bibfnamefont {T.~H.}\ \bibnamefont {Burnett}}, \bibinfo {author} {\bibfnamefont {S.}~\bibnamefont {Buson}}, \bibinfo {author} {\bibfnamefont {G.~A.}\ \bibnamefont {Caliandro}}, \bibinfo {author} {\bibfnamefont {R.~A.}\ \bibnamefont {Cameron}}, \bibinfo {author} {\bibfnamefont {P.~A.}\ \bibnamefont {Caraveo}}, \bibinfo {author} {\bibfnamefont {J.~M.}\ \bibnamefont {Casandjian}}, \bibinfo
  {author} {\bibfnamefont {E.}~\bibnamefont {Cavazzuti}}, \bibinfo {author} {\bibfnamefont {C.}~\bibnamefont {Cecchi}}, \bibinfo {author} {\bibfnamefont {{\"O}.}~\bibnamefont {{\c C}elik}}, \bibinfo {author} {\bibfnamefont {E.}~\bibnamefont {Charles}}, \bibinfo {author} {\bibfnamefont {R.~C.}\ \bibnamefont {Chaves}}, \bibinfo {author} {\bibfnamefont {A.}~\bibnamefont {Chekhtman}}, \bibinfo {author} {\bibfnamefont {C.~C.}\ \bibnamefont {Cheung}}, \bibinfo {author} {\bibfnamefont {J.}~\bibnamefont {Chiang}}, \bibinfo {author} {\bibfnamefont {S.}~\bibnamefont {Ciprini}}, \bibinfo {author} {\bibfnamefont {R.}~\bibnamefont {Claus}}, \bibinfo {author} {\bibfnamefont {J.}~\bibnamefont {{Cohen-Tanugi}}}, \bibinfo {author} {\bibfnamefont {J.}~\bibnamefont {Conrad}}, \bibinfo {author} {\bibfnamefont {R.}~\bibnamefont {Corbet}}, \bibinfo {author} {\bibfnamefont {S.}~\bibnamefont {Cutini}}, \bibinfo {author} {\bibfnamefont {F.}~\bibnamefont {D'Ammando}}, \bibinfo {author} {\bibfnamefont {D.~S.}\ \bibnamefont {Davis}},
  \bibinfo {author} {\bibfnamefont {A.}~\bibnamefont {De~Angelis}}, \bibinfo {author} {\bibfnamefont {M.}~\bibnamefont {Deklotz}}, \bibinfo {author} {\bibfnamefont {F.}~\bibnamefont {De~Palma}}, \bibinfo {author} {\bibfnamefont {C.~D.}\ \bibnamefont {Dermer}}, \bibinfo {author} {\bibfnamefont {S.~W.}\ \bibnamefont {Digel}}, \bibinfo {author} {\bibfnamefont {E.}~\bibnamefont {Do~Couto E~Silva}}, \bibinfo {author} {\bibfnamefont {P.~S.}\ \bibnamefont {Drell}}, \bibinfo {author} {\bibfnamefont {A.}~\bibnamefont {{Drlica-Wagner}}}, \bibinfo {author} {\bibfnamefont {R.}~\bibnamefont {Dubois}}, \bibinfo {author} {\bibfnamefont {C.}~\bibnamefont {Favuzzi}}, \bibinfo {author} {\bibfnamefont {S.~J.}\ \bibnamefont {Fegan}}, \bibinfo {author} {\bibfnamefont {E.~C.}\ \bibnamefont {Ferrara}}, \bibinfo {author} {\bibfnamefont {W.~B.}\ \bibnamefont {Focke}}, \bibinfo {author} {\bibfnamefont {P.}~\bibnamefont {Fortin}}, \bibinfo {author} {\bibfnamefont {Y.}~\bibnamefont {Fukazawa}}, \bibinfo {author} {\bibfnamefont
  {S.}~\bibnamefont {Funk}}, \bibinfo {author} {\bibfnamefont {P.}~\bibnamefont {Fusco}}, \bibinfo {author} {\bibfnamefont {F.}~\bibnamefont {Gargano}}, \bibinfo {author} {\bibfnamefont {D.}~\bibnamefont {Gasparrini}}, \bibinfo {author} {\bibfnamefont {N.}~\bibnamefont {Gehrels}}, \bibinfo {author} {\bibfnamefont {B.}~\bibnamefont {Giebels}}, \bibinfo {author} {\bibfnamefont {N.}~\bibnamefont {Giglietto}}, \bibinfo {author} {\bibfnamefont {F.}~\bibnamefont {Giordano}}, \bibinfo {author} {\bibfnamefont {M.}~\bibnamefont {Giroletti}}, \bibinfo {author} {\bibfnamefont {T.}~\bibnamefont {Glanzman}}, \bibinfo {author} {\bibfnamefont {G.}~\bibnamefont {Godfrey}}, \bibinfo {author} {\bibfnamefont {I.~A.}\ \bibnamefont {Grenier}}, \bibinfo {author} {\bibfnamefont {J.~E.}\ \bibnamefont {Grove}}, \bibinfo {author} {\bibfnamefont {S.}~\bibnamefont {Guiriec}}, \bibinfo {author} {\bibfnamefont {D.}~\bibnamefont {Hadasch}}, \bibinfo {author} {\bibfnamefont {M.}~\bibnamefont {Hayashida}}, \bibinfo {author} {\bibfnamefont
  {E.}~\bibnamefont {Hays}}, \bibinfo {author} {\bibfnamefont {D.}~\bibnamefont {Horan}}, \bibinfo {author} {\bibfnamefont {X.}~\bibnamefont {Hou}}, \bibinfo {author} {\bibfnamefont {R.~E.}\ \bibnamefont {Hughes}}, \bibinfo {author} {\bibfnamefont {M.~S.}\ \bibnamefont {Jackson}}, \bibinfo {author} {\bibfnamefont {T.}~\bibnamefont {Jogler}}, \bibinfo {author} {\bibfnamefont {G.}~\bibnamefont {J{\'o}hannesson}}, \bibinfo {author} {\bibfnamefont {R.~P.}\ \bibnamefont {Johnson}}, \bibinfo {author} {\bibfnamefont {T.~J.}\ \bibnamefont {Johnson}}, \bibinfo {author} {\bibfnamefont {W.~N.}\ \bibnamefont {Johnson}}, \bibinfo {author} {\bibfnamefont {T.}~\bibnamefont {Kamae}}, \bibinfo {author} {\bibfnamefont {H.}~\bibnamefont {Katagiri}}, \bibinfo {author} {\bibfnamefont {J.}~\bibnamefont {Kataoka}}, \bibinfo {author} {\bibfnamefont {M.}~\bibnamefont {Kerr}}, \bibinfo {author} {\bibfnamefont {J.}~\bibnamefont {Kn{\"o}dlseder}}, \bibinfo {author} {\bibfnamefont {M.}~\bibnamefont {Kuss}}, \bibinfo {author}
  {\bibfnamefont {J.}~\bibnamefont {Lande}}, \bibinfo {author} {\bibfnamefont {S.}~\bibnamefont {Larsson}}, \bibinfo {author} {\bibfnamefont {L.}~\bibnamefont {Latronico}}, \bibinfo {author} {\bibfnamefont {C.}~\bibnamefont {Lavalley}}, \bibinfo {author} {\bibfnamefont {M.}~\bibnamefont {{Lemoine-Goumard}}}, \bibinfo {author} {\bibfnamefont {F.}~\bibnamefont {Longo}}, \bibinfo {author} {\bibfnamefont {F.}~\bibnamefont {Loparco}}, \bibinfo {author} {\bibfnamefont {B.}~\bibnamefont {Lott}}, \bibinfo {author} {\bibfnamefont {M.~N.}\ \bibnamefont {Lovellette}}, \bibinfo {author} {\bibfnamefont {P.}~\bibnamefont {Lubrano}}, \bibinfo {author} {\bibfnamefont {M.~N.}\ \bibnamefont {Mazziotta}}, \bibinfo {author} {\bibfnamefont {W.}~\bibnamefont {McConville}}, \bibinfo {author} {\bibfnamefont {J.~E.}\ \bibnamefont {McEnery}}, \bibinfo {author} {\bibfnamefont {J.}~\bibnamefont {Mehault}}, \bibinfo {author} {\bibfnamefont {P.~F.}\ \bibnamefont {Michelson}}, \bibinfo {author} {\bibfnamefont {W.}~\bibnamefont
  {Mitthumsiri}}, \bibinfo {author} {\bibfnamefont {T.}~\bibnamefont {Mizuno}}, \bibinfo {author} {\bibfnamefont {A.~A.}\ \bibnamefont {Moiseev}}, \bibinfo {author} {\bibfnamefont {C.}~\bibnamefont {Monte}}, \bibinfo {author} {\bibfnamefont {M.~E.}\ \bibnamefont {Monzani}}, \bibinfo {author} {\bibfnamefont {A.}~\bibnamefont {Morselli}}, \bibinfo {author} {\bibfnamefont {I.~V.}\ \bibnamefont {Moskalenko}}, \bibinfo {author} {\bibfnamefont {S.}~\bibnamefont {Murgia}}, \bibinfo {author} {\bibfnamefont {M.}~\bibnamefont {{Naumann-Godo}}}, \bibinfo {author} {\bibfnamefont {R.}~\bibnamefont {Nemmen}}, \bibinfo {author} {\bibfnamefont {S.}~\bibnamefont {Nishino}}, \bibinfo {author} {\bibfnamefont {J.~P.}\ \bibnamefont {Norris}}, \bibinfo {author} {\bibfnamefont {E.}~\bibnamefont {Nuss}}, \bibinfo {author} {\bibfnamefont {M.}~\bibnamefont {Ohno}}, \bibinfo {author} {\bibfnamefont {T.}~\bibnamefont {Ohsugi}}, \bibinfo {author} {\bibfnamefont {A.}~\bibnamefont {Okumura}}, \bibinfo {author} {\bibfnamefont
  {N.}~\bibnamefont {Omodei}}, \bibinfo {author} {\bibfnamefont {M.}~\bibnamefont {Orienti}}, \bibinfo {author} {\bibfnamefont {E.}~\bibnamefont {Orlando}}, \bibinfo {author} {\bibfnamefont {J.~F.}\ \bibnamefont {Ormes}}, \bibinfo {author} {\bibfnamefont {D.}~\bibnamefont {Paneque}}, \bibinfo {author} {\bibfnamefont {J.~H.}\ \bibnamefont {Panetta}}, \bibinfo {author} {\bibfnamefont {J.~S.}\ \bibnamefont {Perkins}}, \bibinfo {author} {\bibfnamefont {M.}~\bibnamefont {{Pesce-Rollins}}}, \bibinfo {author} {\bibfnamefont {M.}~\bibnamefont {Pierbattista}}, \bibinfo {author} {\bibfnamefont {F.}~\bibnamefont {Piron}}, \bibinfo {author} {\bibfnamefont {G.}~\bibnamefont {Pivato}}, \bibinfo {author} {\bibfnamefont {T.~A.}\ \bibnamefont {Porter}}, \bibinfo {author} {\bibfnamefont {J.~L.}\ \bibnamefont {Racusin}}, \bibinfo {author} {\bibfnamefont {S.}~\bibnamefont {Rain{\`o}}}, \bibinfo {author} {\bibfnamefont {R.}~\bibnamefont {Rando}}, \bibinfo {author} {\bibfnamefont {M.}~\bibnamefont {Razzano}}, \bibinfo {author}
  {\bibfnamefont {S.}~\bibnamefont {Razzaque}}, \bibinfo {author} {\bibfnamefont {A.}~\bibnamefont {Reimer}}, \bibinfo {author} {\bibfnamefont {O.}~\bibnamefont {Reimer}}, \bibinfo {author} {\bibfnamefont {T.}~\bibnamefont {Reposeur}}, \bibinfo {author} {\bibfnamefont {L.~C.}\ \bibnamefont {Reyes}}, \bibinfo {author} {\bibfnamefont {S.}~\bibnamefont {Ritz}}, \bibinfo {author} {\bibfnamefont {L.~S.}\ \bibnamefont {Rochester}}, \bibinfo {author} {\bibfnamefont {C.}~\bibnamefont {Romoli}}, \bibinfo {author} {\bibfnamefont {M.}~\bibnamefont {Roth}}, \bibinfo {author} {\bibfnamefont {H.~F.}\ \bibnamefont {Sadrozinski}}, \bibinfo {author} {\bibfnamefont {D.~A.}\ \bibnamefont {Sanchez}}, \bibinfo {author} {\bibfnamefont {P.~M.}\ \bibnamefont {Saz~Parkinson}}, \bibinfo {author} {\bibfnamefont {C.}~\bibnamefont {Sbarra}}, \bibinfo {author} {\bibfnamefont {J.~D.}\ \bibnamefont {Scargle}}, \bibinfo {author} {\bibfnamefont {C.}~\bibnamefont {Sgr{\`o}}}, \bibinfo {author} {\bibfnamefont {J.}~\bibnamefont
  {{Siegal-Gaskins}}}, \bibinfo {author} {\bibfnamefont {E.~J.}\ \bibnamefont {Siskind}}, \bibinfo {author} {\bibfnamefont {G.}~\bibnamefont {Spandre}}, \bibinfo {author} {\bibfnamefont {P.}~\bibnamefont {Spinelli}}, \bibinfo {author} {\bibfnamefont {T.~E.}\ \bibnamefont {Stephens}}, \bibinfo {author} {\bibfnamefont {D.~J.}\ \bibnamefont {Suson}}, \bibinfo {author} {\bibfnamefont {H.}~\bibnamefont {Tajima}}, \bibinfo {author} {\bibfnamefont {H.}~\bibnamefont {Takahashi}}, \bibinfo {author} {\bibfnamefont {T.}~\bibnamefont {Tanaka}}, \bibinfo {author} {\bibfnamefont {J.~G.}\ \bibnamefont {Thayer}}, \bibinfo {author} {\bibfnamefont {J.~B.}\ \bibnamefont {Thayer}}, \bibinfo {author} {\bibfnamefont {D.~J.}\ \bibnamefont {Thompson}}, \bibinfo {author} {\bibfnamefont {L.}~\bibnamefont {Tibaldo}}, \bibinfo {author} {\bibfnamefont {M.}~\bibnamefont {Tinivella}}, \bibinfo {author} {\bibfnamefont {G.}~\bibnamefont {Tosti}}, \bibinfo {author} {\bibfnamefont {E.}~\bibnamefont {Troja}}, \bibinfo {author} {\bibfnamefont
  {T.~L.}\ \bibnamefont {Usher}}, \bibinfo {author} {\bibfnamefont {J.}~\bibnamefont {Vandenbroucke}}, \bibinfo {author} {\bibfnamefont {B.}~\bibnamefont {Van~Klaveren}}, \bibinfo {author} {\bibfnamefont {V.}~\bibnamefont {Vasileiou}}, \bibinfo {author} {\bibfnamefont {G.}~\bibnamefont {Vianello}}, \bibinfo {author} {\bibfnamefont {V.}~\bibnamefont {Vitale}}, \bibinfo {author} {\bibfnamefont {A.~P.}\ \bibnamefont {Waite}}, \bibinfo {author} {\bibfnamefont {E.}~\bibnamefont {Wallace}}, \bibinfo {author} {\bibfnamefont {B.~L.}\ \bibnamefont {Winer}}, \bibinfo {author} {\bibfnamefont {D.~L.}\ \bibnamefont {Wood}}, \bibinfo {author} {\bibfnamefont {K.~S.}\ \bibnamefont {Wood}}, \bibinfo {author} {\bibfnamefont {M.}~\bibnamefont {Wood}}, \bibinfo {author} {\bibfnamefont {Z.}~\bibnamefont {Yang}},\ and\ \bibinfo {author} {\bibfnamefont {S.}~\bibnamefont {Zimmer}},\ }\bibfield  {title} {\bibinfo {title} {{{THE FERMI LARGE AREA TELESCOPE ON ORBIT}}: {{EVENT CLASSIFICATION}}, {{INSTRUMENT RESPONSE FUNCTIONS}}, {{AND
  CALIBRATION}}},\ }\href {https://doi.org/10.1088/0067-0049/203/1/4} {\bibfield  {journal} {\bibinfo  {journal} {Astrophys. J. Suppl. Ser.}\ }\textbf {\bibinfo {volume} {203}},\ \bibinfo {pages} {4} (\bibinfo {year} {2012})},\ \Eprint {https://arxiv.org/abs/1206.1896} {arXiv:1206.1896} \BibitemShut {NoStop}%
\bibitem [{\citenamefont {Vlachoudis}(2009)}]{Vlachoudis2009a}%
  \BibitemOpen
  \bibfield  {author} {\bibinfo {author} {\bibfnamefont {V.}~\bibnamefont {Vlachoudis}},\ }\bibfield  {title} {\bibinfo {title} {Flair: {{A}} powerful but user friendly graphical interface for {{FLUKA}}},\ }\href@noop {} {\bibfield  {journal} {\bibinfo  {journal} {Int. Conf. Math. Comput. Methods React. Phys. MC 2009}\ } (\bibinfo {year} {2009})}\BibitemShut {NoStop}%
\bibitem [{zot()}]{zotero-item-4600}%
  \BibitemOpen
  \href@noop {} {\bibinfo {title} {See the {{Supplemental Material}} at {$<$}{{TBD}}{$>$} for more detailed information on the plane wave radius sensitivity analysis, transport threshold benchmark, additional photon energy evaluations of the instrument response function (energy and angular dispersion) including uncertainty quantification results, as well as double differential gamma-ray flux estimates adopted for the {{Monte Carlo}} benchmark.}}\BibitemShut {Stop}%
\bibitem [{\citenamefont {B{\"o}hlen}\ \emph {et~al.}(2014)\citenamefont {B{\"o}hlen}, \citenamefont {Cerutti}, \citenamefont {Chin}, \citenamefont {Fass{\`o}}, \citenamefont {Ferrari}, \citenamefont {Ortega}, \citenamefont {Mairani}, \citenamefont {Sala}, \citenamefont {Smirnov},\ and\ \citenamefont {Vlachoudis}}]{Bohlen2014a}%
  \BibitemOpen
  \bibfield  {author} {\bibinfo {author} {\bibfnamefont {T.~T.}\ \bibnamefont {B{\"o}hlen}}, \bibinfo {author} {\bibfnamefont {F.}~\bibnamefont {Cerutti}}, \bibinfo {author} {\bibfnamefont {M.~P.}\ \bibnamefont {Chin}}, \bibinfo {author} {\bibfnamefont {A.}~\bibnamefont {Fass{\`o}}}, \bibinfo {author} {\bibfnamefont {A.}~\bibnamefont {Ferrari}}, \bibinfo {author} {\bibfnamefont {P.~G.}\ \bibnamefont {Ortega}}, \bibinfo {author} {\bibfnamefont {A.}~\bibnamefont {Mairani}}, \bibinfo {author} {\bibfnamefont {P.~R.}\ \bibnamefont {Sala}}, \bibinfo {author} {\bibfnamefont {G.}~\bibnamefont {Smirnov}},\ and\ \bibinfo {author} {\bibfnamefont {V.}~\bibnamefont {Vlachoudis}},\ }\bibfield  {title} {\bibinfo {title} {The {{FLUKA Code}}: {{Developments}} and challenges for high energy and medical applications},\ }\href {https://doi.org/10.1016/j.nds.2014.07.049} {\bibfield  {journal} {\bibinfo  {journal} {Nucl. Data Sheets}\ }\textbf {\bibinfo {volume} {120}},\ \bibinfo {pages} {211} (\bibinfo {year} {2014})}\BibitemShut
  {NoStop}%
\bibitem [{\citenamefont {Battistoni}\ \emph {et~al.}(2015)\citenamefont {Battistoni}, \citenamefont {Boehlen}, \citenamefont {Cerutti}, \citenamefont {Chin}, \citenamefont {Esposito}, \citenamefont {Fass{\`o}}, \citenamefont {Ferrari}, \citenamefont {Lechner}, \citenamefont {Empl}, \citenamefont {Mairani}, \citenamefont {Mereghetti}, \citenamefont {Ortega}, \citenamefont {Ranft}, \citenamefont {Roesler}, \citenamefont {Sala}, \citenamefont {Vlachoudis},\ and\ \citenamefont {Smirnov}}]{Battistoni2015a}%
  \BibitemOpen
  \bibfield  {author} {\bibinfo {author} {\bibfnamefont {G.}~\bibnamefont {Battistoni}}, \bibinfo {author} {\bibfnamefont {T.}~\bibnamefont {Boehlen}}, \bibinfo {author} {\bibfnamefont {F.}~\bibnamefont {Cerutti}}, \bibinfo {author} {\bibfnamefont {P.~W.}\ \bibnamefont {Chin}}, \bibinfo {author} {\bibfnamefont {L.~S.}\ \bibnamefont {Esposito}}, \bibinfo {author} {\bibfnamefont {A.}~\bibnamefont {Fass{\`o}}}, \bibinfo {author} {\bibfnamefont {A.}~\bibnamefont {Ferrari}}, \bibinfo {author} {\bibfnamefont {A.}~\bibnamefont {Lechner}}, \bibinfo {author} {\bibfnamefont {A.}~\bibnamefont {Empl}}, \bibinfo {author} {\bibfnamefont {A.}~\bibnamefont {Mairani}}, \bibinfo {author} {\bibfnamefont {A.}~\bibnamefont {Mereghetti}}, \bibinfo {author} {\bibfnamefont {P.~G.}\ \bibnamefont {Ortega}}, \bibinfo {author} {\bibfnamefont {J.}~\bibnamefont {Ranft}}, \bibinfo {author} {\bibfnamefont {S.}~\bibnamefont {Roesler}}, \bibinfo {author} {\bibfnamefont {P.~R.}\ \bibnamefont {Sala}}, \bibinfo {author} {\bibfnamefont
  {V.}~\bibnamefont {Vlachoudis}},\ and\ \bibinfo {author} {\bibfnamefont {G.}~\bibnamefont {Smirnov}},\ }\bibfield  {title} {\bibinfo {title} {Overview of the {{FLUKA}} code},\ }\href {https://doi.org/10.1016/j.anucene.2014.11.007} {\bibfield  {journal} {\bibinfo  {journal} {Ann. Nucl. Energy}\ }\textbf {\bibinfo {volume} {82}},\ \bibinfo {pages} {10} (\bibinfo {year} {2015})}\BibitemShut {NoStop}%
\bibitem [{\citenamefont {Payne}\ \emph {et~al.}(2009)\citenamefont {Payne}, \citenamefont {Cherepy}, \citenamefont {Hull}, \citenamefont {Valentine}, \citenamefont {Moses},\ and\ \citenamefont {Choong}}]{Payne2009a}%
  \BibitemOpen
  \bibfield  {author} {\bibinfo {author} {\bibfnamefont {S.~A.}\ \bibnamefont {Payne}}, \bibinfo {author} {\bibfnamefont {N.~J.}\ \bibnamefont {Cherepy}}, \bibinfo {author} {\bibfnamefont {G.}~\bibnamefont {Hull}}, \bibinfo {author} {\bibfnamefont {J.~D.}\ \bibnamefont {Valentine}}, \bibinfo {author} {\bibfnamefont {W.~W.}\ \bibnamefont {Moses}},\ and\ \bibinfo {author} {\bibfnamefont {W.~S.}\ \bibnamefont {Choong}},\ }\bibfield  {title} {\bibinfo {title} {Nonproportionality of scintillator detectors: {{Theory}} and experiment},\ }\href {https://doi.org/10.1109/TNS.2009.2023657} {\bibfield  {journal} {\bibinfo  {journal} {IEEE Trans. Nucl. Sci.}\ }\textbf {\bibinfo {volume} {56}},\ \bibinfo {pages} {2506} (\bibinfo {year} {2009})}\BibitemShut {NoStop}%
\bibitem [{\citenamefont {Payne}\ \emph {et~al.}(2011)\citenamefont {Payne}, \citenamefont {Moses}, \citenamefont {Sheets}, \citenamefont {Ahle}, \citenamefont {Cherepy}, \citenamefont {Sturm}, \citenamefont {Dazeley}, \citenamefont {Bizarri},\ and\ \citenamefont {Choong}}]{Payne2011a}%
  \BibitemOpen
  \bibfield  {author} {\bibinfo {author} {\bibfnamefont {S.~A.}\ \bibnamefont {Payne}}, \bibinfo {author} {\bibfnamefont {W.~W.}\ \bibnamefont {Moses}}, \bibinfo {author} {\bibfnamefont {S.}~\bibnamefont {Sheets}}, \bibinfo {author} {\bibfnamefont {L.}~\bibnamefont {Ahle}}, \bibinfo {author} {\bibfnamefont {N.~J.}\ \bibnamefont {Cherepy}}, \bibinfo {author} {\bibfnamefont {B.}~\bibnamefont {Sturm}}, \bibinfo {author} {\bibfnamefont {S.}~\bibnamefont {Dazeley}}, \bibinfo {author} {\bibfnamefont {G.}~\bibnamefont {Bizarri}},\ and\ \bibinfo {author} {\bibfnamefont {W.~S.}\ \bibnamefont {Choong}},\ }\bibfield  {title} {\bibinfo {title} {Nonproportionality of scintillator detectors: {{Theory}} and experiment. {{II}}},\ }\href {https://doi.org/10.1109/TNS.2011.2167687} {\bibfield  {journal} {\bibinfo  {journal} {IEEE Trans. Nucl. Sci.}\ }\textbf {\bibinfo {volume} {58}},\ \bibinfo {pages} {3392} (\bibinfo {year} {2011})}\BibitemShut {NoStop}%
\bibitem [{\citenamefont {Breitenmoser}\ \emph {et~al.}(2023{\natexlab{a}})\citenamefont {Breitenmoser}, \citenamefont {Cerutti}, \citenamefont {Butterweck}, \citenamefont {Kasprzak},\ and\ \citenamefont {Mayer}}]{Breitenmoser2023i}%
  \BibitemOpen
  \bibfield  {author} {\bibinfo {author} {\bibfnamefont {D.}~\bibnamefont {Breitenmoser}}, \bibinfo {author} {\bibfnamefont {F.}~\bibnamefont {Cerutti}}, \bibinfo {author} {\bibfnamefont {G.}~\bibnamefont {Butterweck}}, \bibinfo {author} {\bibfnamefont {M.~M.}\ \bibnamefont {Kasprzak}},\ and\ \bibinfo {author} {\bibfnamefont {S.}~\bibnamefont {Mayer}},\ }\href {https://doi.org/10.3929/ETHZ-B-000595727} {\bibinfo {title} {{{FLUKA}} user routines for non-proportional scintillation simulations}} (\bibinfo {year} {2023}{\natexlab{a}})\BibitemShut {NoStop}%
\bibitem [{\citenamefont {Breitenmoser}\ \emph {et~al.}(2023{\natexlab{b}})\citenamefont {Breitenmoser}, \citenamefont {Cerutti}, \citenamefont {Butterweck}, \citenamefont {Kasprzak},\ and\ \citenamefont {Mayer}}]{Breitenmoser2023f}%
  \BibitemOpen
  \bibfield  {author} {\bibinfo {author} {\bibfnamefont {D.}~\bibnamefont {Breitenmoser}}, \bibinfo {author} {\bibfnamefont {F.}~\bibnamefont {Cerutti}}, \bibinfo {author} {\bibfnamefont {G.}~\bibnamefont {Butterweck}}, \bibinfo {author} {\bibfnamefont {M.~M.}\ \bibnamefont {Kasprzak}},\ and\ \bibinfo {author} {\bibfnamefont {S.}~\bibnamefont {Mayer}},\ }\bibfield  {title} {\bibinfo {title} {Emulator-based {{Bayesian}} inference on non-proportional scintillation models by compton-edge probing},\ }\href {https://doi.org/10.1038/s41467-023-42574-y} {\bibfield  {journal} {\bibinfo  {journal} {Nat. Commun.}\ }\textbf {\bibinfo {volume} {14}},\ \bibinfo {pages} {7790} (\bibinfo {year} {2023}{\natexlab{b}})},\ \Eprint {https://arxiv.org/abs/2302.05641} {arXiv:2302.05641} \BibitemShut {NoStop}%
\bibitem [{\citenamefont {McConn}\ \emph {et~al.}(2011)\citenamefont {McConn}, \citenamefont {Gesh}, \citenamefont {Pagh},\ and\ \citenamefont {Rucker}}]{Mcconn2011a}%
  \BibitemOpen
  \bibfield  {author} {\bibinfo {author} {\bibfnamefont {R.~J.}\ \bibnamefont {McConn}}, \bibinfo {author} {\bibfnamefont {C.~J.}\ \bibnamefont {Gesh}}, \bibinfo {author} {\bibfnamefont {R.~T.}\ \bibnamefont {Pagh}},\ and\ \bibinfo {author} {\bibfnamefont {R.~A.}\ \bibnamefont {Rucker}},\ }\href@noop {} {\emph {\bibinfo {title} {Compendium of {{Material Composition Data}} for {{Radiation Transport Modeling}}}}},\ \bibinfo {type} {Tech. Rep.}\ (\bibinfo  {institution} {Pacific Northwest National Laboratory},\ \bibinfo {address} {Richland},\ \bibinfo {year} {2011})\BibitemShut {NoStop}%
\bibitem [{\citenamefont {Jandel}\ \emph {et~al.}(2004)\citenamefont {Jandel}, \citenamefont {Morh{\'a}{\v c}}, \citenamefont {Kliman}, \citenamefont {Krupa}, \citenamefont {Matou{\v s}ek}, \citenamefont {Hamilton},\ and\ \citenamefont {Ramayya}}]{Jandel2004a}%
  \BibitemOpen
  \bibfield  {author} {\bibinfo {author} {\bibfnamefont {M.}~\bibnamefont {Jandel}}, \bibinfo {author} {\bibfnamefont {M.}~\bibnamefont {Morh{\'a}{\v c}}}, \bibinfo {author} {\bibfnamefont {J.}~\bibnamefont {Kliman}}, \bibinfo {author} {\bibfnamefont {L.}~\bibnamefont {Krupa}}, \bibinfo {author} {\bibfnamefont {V.}~\bibnamefont {Matou{\v s}ek}}, \bibinfo {author} {\bibfnamefont {J.~H.}\ \bibnamefont {Hamilton}},\ and\ \bibinfo {author} {\bibfnamefont {A.~V.}\ \bibnamefont {Ramayya}},\ }\bibfield  {title} {\bibinfo {title} {Decomposition of continuum {$\gamma$}-ray spectra using synthesized response matrix},\ }\href {https://doi.org/10.1016/J.NIMA.2003.07.047} {\bibfield  {journal} {\bibinfo  {journal} {Nucl. Instrum. Methods Phys. Res. Sect. Accel. Spectrometers Detect. Assoc. Equip.}\ }\textbf {\bibinfo {volume} {516}},\ \bibinfo {pages} {172} (\bibinfo {year} {2004})}\BibitemShut {NoStop}%
\bibitem [{\citenamefont {Bevington}\ and\ \citenamefont {Robinson}(2003)}]{Bevington2003a}%
  \BibitemOpen
  \bibfield  {author} {\bibinfo {author} {\bibfnamefont {P.~R.}\ \bibnamefont {Bevington}}\ and\ \bibinfo {author} {\bibfnamefont {D.~K.}\ \bibnamefont {Robinson}},\ }\href@noop {} {\emph {\bibinfo {title} {Data Reduction and Error Analysis for the Physical Sciences}}}\ (\bibinfo  {publisher} {McGraw-Hill},\ \bibinfo {year} {2003})\BibitemShut {NoStop}%
\bibitem [{\citenamefont {{Joint Committee for Guides in Metrology}}(2008)}]{JointCommitteeForGuidesInMetrology2008}%
  \BibitemOpen
  \bibfield  {author} {\bibinfo {author} {\bibnamefont {{Joint Committee for Guides in Metrology}}},\ }\href@noop {} {\emph {\bibinfo {title} {Evaluation of Measurement Data --- {{Guide}} to the Expression of Uncertainty in Measurement}}},\ \bibinfo {type} {Tech. Rep.}\ (\bibinfo  {institution} {Bureau International des Poids et Mesures},\ \bibinfo {year} {2008})\BibitemShut {NoStop}%
\bibitem [{\citenamefont {{Particle Data Group}}\ \emph {et~al.}(2022)\citenamefont {{Particle Data Group}}, \citenamefont {Workman}, \citenamefont {Burkert}, \citenamefont {Crede}, \citenamefont {Klempt}, \citenamefont {Thoma}, \citenamefont {Tiator}, \citenamefont {Agashe}, \citenamefont {Aielli}, \citenamefont {Allanach}, \citenamefont {Amsler}, \citenamefont {Antonelli}, \citenamefont {Aschenauer}, \citenamefont {Asner}, \citenamefont {Baer}, \citenamefont {Banerjee}, \citenamefont {Barnett}, \citenamefont {Baudis}, \citenamefont {Bauer}, \citenamefont {Beatty}, \citenamefont {Belousov}, \citenamefont {Beringer}, \citenamefont {Bettini}, \citenamefont {Biebel}, \citenamefont {Black}, \citenamefont {Blucher}, \citenamefont {Bonventre}, \citenamefont {Bryzgalov}, \citenamefont {Buchmuller}, \citenamefont {Bychkov}, \citenamefont {Cahn}, \citenamefont {Carena}, \citenamefont {Ceccucci}, \citenamefont {Cerri}, \citenamefont {Chivukula}, \citenamefont {Cowan}, \citenamefont {Cranmer}, \citenamefont
  {Cremonesi}, \citenamefont {D'Ambrosio}, \citenamefont {Damour}, \citenamefont {{de Florian}}, \citenamefont {{de Gouv{\^e}a}}, \citenamefont {DeGrand}, \citenamefont {{de Jong}}, \citenamefont {Demers}, \citenamefont {Dobrescu}, \citenamefont {D'Onofrio}, \citenamefont {Doser}, \citenamefont {Dreiner}, \citenamefont {Eerola}, \citenamefont {Egede}, \citenamefont {Eidelman}, \citenamefont {{El-Khadra}}, \citenamefont {Ellis}, \citenamefont {Eno}, \citenamefont {Erler}, \citenamefont {Ezhela}, \citenamefont {Fetscher}, \citenamefont {Fields}, \citenamefont {Freitas}, \citenamefont {Gallagher}, \citenamefont {Gershtein}, \citenamefont {Gherghetta}, \citenamefont {{Gonzalez-Garcia}}, \citenamefont {Goodman}, \citenamefont {Grab}, \citenamefont {Gritsan}, \citenamefont {Grojean}, \citenamefont {Groom}, \citenamefont {Gr{\"u}newald}, \citenamefont {Gurtu}, \citenamefont {Gutsche}, \citenamefont {Haber}, \citenamefont {Hamel}, \citenamefont {Hanhart}, \citenamefont {Hashimoto}, \citenamefont {Hayato},
  \citenamefont {Hebecker}, \citenamefont {Heinemeyer}, \citenamefont {{Hern{\'a}ndez-Rey}}, \citenamefont {Hikasa}, \citenamefont {Hisano}, \citenamefont {H{\"o}cker}, \citenamefont {Holder}, \citenamefont {Hsu}, \citenamefont {Huston}, \citenamefont {Hyodo}, \citenamefont {Ianni}, \citenamefont {Kado}, \citenamefont {Karliner}, \citenamefont {Katz}, \citenamefont {Kenzie}, \citenamefont {Khoze}, \citenamefont {Klein}, \citenamefont {Krauss}, \citenamefont {Kreps}, \citenamefont {Kri{\v z}an}, \citenamefont {Krusche}, \citenamefont {Kwon}, \citenamefont {Lahav}, \citenamefont {Laiho}, \citenamefont {Lellouch}, \citenamefont {Lesgourgues}, \citenamefont {Liddle}, \citenamefont {Ligeti}, \citenamefont {Lin}, \citenamefont {Lippmann}, \citenamefont {Liss}, \citenamefont {Littenberg}, \citenamefont {Louren{\c c}o}, \citenamefont {Lugovsky}, \citenamefont {Lugovsky}, \citenamefont {Lusiani}, \citenamefont {Makida}, \citenamefont {Maltoni}, \citenamefont {Mannel}, \citenamefont {Manohar}, \citenamefont {Marciano},
  \citenamefont {Masoni}, \citenamefont {Matthews}, \citenamefont {Mei{\ss}ner}, \citenamefont {{Melzer-Pellmann}}, \citenamefont {Mikhasenko}, \citenamefont {Miller}, \citenamefont {Milstead}, \citenamefont {Mitchell}, \citenamefont {M{\"o}nig}, \citenamefont {Molaro}, \citenamefont {Moortgat}, \citenamefont {Moskovic}, \citenamefont {Nakamura}, \citenamefont {Narain}, \citenamefont {Nason}, \citenamefont {Navas}, \citenamefont {Nelles}, \citenamefont {Neubert}, \citenamefont {Nevski}, \citenamefont {Nir}, \citenamefont {Olive}, \citenamefont {Patrignani}, \citenamefont {Peacock}, \citenamefont {Petrov}, \citenamefont {Pianori}, \citenamefont {Pich}, \citenamefont {Piepke}, \citenamefont {Pietropaolo}, \citenamefont {Pomarol}, \citenamefont {Pordes}, \citenamefont {Profumo}, \citenamefont {Quadt}, \citenamefont {Rabbertz}, \citenamefont {Rademacker}, \citenamefont {Raffelt}, \citenamefont {{Ramsey-Musolf}}, \citenamefont {Ratcliff}, \citenamefont {Richardson}, \citenamefont {Ringwald}, \citenamefont
  {Robinson}, \citenamefont {Roesler}, \citenamefont {Rolli}, \citenamefont {Romaniouk}, \citenamefont {Rosenberg}, \citenamefont {Rosner}, \citenamefont {Rybka}, \citenamefont {Ryskin}, \citenamefont {Ryutin}, \citenamefont {Sakai}, \citenamefont {Sarkar}, \citenamefont {Sauli}, \citenamefont {Schneider}, \citenamefont {Sch{\"o}nert}, \citenamefont {Scholberg}, \citenamefont {Schwartz}, \citenamefont {Schwiening}, \citenamefont {Scott}, \citenamefont {Sefkow}, \citenamefont {Seljak}, \citenamefont {Sharma}, \citenamefont {Sharpe}, \citenamefont {Shiltsev}, \citenamefont {Signorelli}, \citenamefont {Silari}, \citenamefont {Simon}, \citenamefont {Sj{\"o}strand}, \citenamefont {Skands}, \citenamefont {Skwarnicki}, \citenamefont {Smoot}, \citenamefont {Soffer}, \citenamefont {Sozzi}, \citenamefont {Spanier}, \citenamefont {Spiering}, \citenamefont {Stahl}, \citenamefont {Stone}, \citenamefont {Sumino}, \citenamefont {Syphers}, \citenamefont {Takahashi}, \citenamefont {Tanabashi}, \citenamefont {Tanaka},
  \citenamefont {Ta{\v s}evsk{\'y}}, \citenamefont {Terao}, \citenamefont {Terashi}, \citenamefont {Terning}, \citenamefont {Thorne}, \citenamefont {Titov}, \citenamefont {Tkachenko}, \citenamefont {Tovey}, \citenamefont {Trabelsi}, \citenamefont {Urquijo}, \citenamefont {Valencia}, \citenamefont {{Van de Water}}, \citenamefont {Varelas}, \citenamefont {Venanzoni}, \citenamefont {Verde}, \citenamefont {Vivarelli}, \citenamefont {Vogel}, \citenamefont {Vogelsang}, \citenamefont {Vorobyev}, \citenamefont {Wakely}, \citenamefont {Walkowiak}, \citenamefont {Walter}, \citenamefont {Wands}, \citenamefont {Weinberg}, \citenamefont {Weinberg}, \citenamefont {Wermes}, \citenamefont {White}, \citenamefont {Wiencke}, \citenamefont {Willocq}, \citenamefont {Wohl}, \citenamefont {Woody}, \citenamefont {Yao}, \citenamefont {Yokoyama}, \citenamefont {Yoshida}, \citenamefont {Zanderighi}, \citenamefont {Zeller}, \citenamefont {Zenin}, \citenamefont {Zhu}, \citenamefont {Zhu}, \citenamefont {Zimmermann},\ and\ \citenamefont
  {Zyla}}]{ParticleDataGroup2022}%
  \BibitemOpen
  \bibfield  {author} {\bibinfo {author} {\bibnamefont {{Particle Data Group}}}, \bibinfo {author} {\bibfnamefont {R.~L.}\ \bibnamefont {Workman}}, \bibinfo {author} {\bibfnamefont {V.~D.}\ \bibnamefont {Burkert}}, \bibinfo {author} {\bibfnamefont {V.}~\bibnamefont {Crede}}, \bibinfo {author} {\bibfnamefont {E.}~\bibnamefont {Klempt}}, \bibinfo {author} {\bibfnamefont {U.}~\bibnamefont {Thoma}}, \bibinfo {author} {\bibfnamefont {L.}~\bibnamefont {Tiator}}, \bibinfo {author} {\bibfnamefont {K.}~\bibnamefont {Agashe}}, \bibinfo {author} {\bibfnamefont {G.}~\bibnamefont {Aielli}}, \bibinfo {author} {\bibfnamefont {B.~C.}\ \bibnamefont {Allanach}}, \bibinfo {author} {\bibfnamefont {C.}~\bibnamefont {Amsler}}, \bibinfo {author} {\bibfnamefont {M.}~\bibnamefont {Antonelli}}, \bibinfo {author} {\bibfnamefont {E.~C.}\ \bibnamefont {Aschenauer}}, \bibinfo {author} {\bibfnamefont {D.~M.}\ \bibnamefont {Asner}}, \bibinfo {author} {\bibfnamefont {H.}~\bibnamefont {Baer}}, \bibinfo {author} {\bibfnamefont
  {S.}~\bibnamefont {Banerjee}}, \bibinfo {author} {\bibfnamefont {R.~M.}\ \bibnamefont {Barnett}}, \bibinfo {author} {\bibfnamefont {L.}~\bibnamefont {Baudis}}, \bibinfo {author} {\bibfnamefont {C.~W.}\ \bibnamefont {Bauer}}, \bibinfo {author} {\bibfnamefont {J.~J.}\ \bibnamefont {Beatty}}, \bibinfo {author} {\bibfnamefont {V.~I.}\ \bibnamefont {Belousov}}, \bibinfo {author} {\bibfnamefont {J.}~\bibnamefont {Beringer}}, \bibinfo {author} {\bibfnamefont {A.}~\bibnamefont {Bettini}}, \bibinfo {author} {\bibfnamefont {O.}~\bibnamefont {Biebel}}, \bibinfo {author} {\bibfnamefont {K.~M.}\ \bibnamefont {Black}}, \bibinfo {author} {\bibfnamefont {E.}~\bibnamefont {Blucher}}, \bibinfo {author} {\bibfnamefont {R.}~\bibnamefont {Bonventre}}, \bibinfo {author} {\bibfnamefont {V.~V.}\ \bibnamefont {Bryzgalov}}, \bibinfo {author} {\bibfnamefont {O.}~\bibnamefont {Buchmuller}}, \bibinfo {author} {\bibfnamefont {M.~A.}\ \bibnamefont {Bychkov}}, \bibinfo {author} {\bibfnamefont {R.~N.}\ \bibnamefont {Cahn}}, \bibinfo
  {author} {\bibfnamefont {M.}~\bibnamefont {Carena}}, \bibinfo {author} {\bibfnamefont {A.}~\bibnamefont {Ceccucci}}, \bibinfo {author} {\bibfnamefont {A.}~\bibnamefont {Cerri}}, \bibinfo {author} {\bibfnamefont {R.~S.}\ \bibnamefont {Chivukula}}, \bibinfo {author} {\bibfnamefont {G.}~\bibnamefont {Cowan}}, \bibinfo {author} {\bibfnamefont {K.}~\bibnamefont {Cranmer}}, \bibinfo {author} {\bibfnamefont {O.}~\bibnamefont {Cremonesi}}, \bibinfo {author} {\bibfnamefont {G.}~\bibnamefont {D'Ambrosio}}, \bibinfo {author} {\bibfnamefont {T.}~\bibnamefont {Damour}}, \bibinfo {author} {\bibfnamefont {D.}~\bibnamefont {{de Florian}}}, \bibinfo {author} {\bibfnamefont {A.}~\bibnamefont {{de Gouv{\^e}a}}}, \bibinfo {author} {\bibfnamefont {T.}~\bibnamefont {DeGrand}}, \bibinfo {author} {\bibfnamefont {P.}~\bibnamefont {{de Jong}}}, \bibinfo {author} {\bibfnamefont {S.}~\bibnamefont {Demers}}, \bibinfo {author} {\bibfnamefont {B.~A.}\ \bibnamefont {Dobrescu}}, \bibinfo {author} {\bibfnamefont {M.}~\bibnamefont
  {D'Onofrio}}, \bibinfo {author} {\bibfnamefont {M.}~\bibnamefont {Doser}}, \bibinfo {author} {\bibfnamefont {H.~K.}\ \bibnamefont {Dreiner}}, \bibinfo {author} {\bibfnamefont {P.}~\bibnamefont {Eerola}}, \bibinfo {author} {\bibfnamefont {U.}~\bibnamefont {Egede}}, \bibinfo {author} {\bibfnamefont {S.}~\bibnamefont {Eidelman}}, \bibinfo {author} {\bibfnamefont {A.~X.}\ \bibnamefont {{El-Khadra}}}, \bibinfo {author} {\bibfnamefont {J.}~\bibnamefont {Ellis}}, \bibinfo {author} {\bibfnamefont {S.~C.}\ \bibnamefont {Eno}}, \bibinfo {author} {\bibfnamefont {J.}~\bibnamefont {Erler}}, \bibinfo {author} {\bibfnamefont {V.~V.}\ \bibnamefont {Ezhela}}, \bibinfo {author} {\bibfnamefont {W.}~\bibnamefont {Fetscher}}, \bibinfo {author} {\bibfnamefont {B.~D.}\ \bibnamefont {Fields}}, \bibinfo {author} {\bibfnamefont {A.}~\bibnamefont {Freitas}}, \bibinfo {author} {\bibfnamefont {H.}~\bibnamefont {Gallagher}}, \bibinfo {author} {\bibfnamefont {Y.}~\bibnamefont {Gershtein}}, \bibinfo {author} {\bibfnamefont
  {T.}~\bibnamefont {Gherghetta}}, \bibinfo {author} {\bibfnamefont {M.~C.}\ \bibnamefont {{Gonzalez-Garcia}}}, \bibinfo {author} {\bibfnamefont {M.}~\bibnamefont {Goodman}}, \bibinfo {author} {\bibfnamefont {C.}~\bibnamefont {Grab}}, \bibinfo {author} {\bibfnamefont {A.~V.}\ \bibnamefont {Gritsan}}, \bibinfo {author} {\bibfnamefont {C.}~\bibnamefont {Grojean}}, \bibinfo {author} {\bibfnamefont {D.~E.}\ \bibnamefont {Groom}}, \bibinfo {author} {\bibfnamefont {M.}~\bibnamefont {Gr{\"u}newald}}, \bibinfo {author} {\bibfnamefont {A.}~\bibnamefont {Gurtu}}, \bibinfo {author} {\bibfnamefont {T.}~\bibnamefont {Gutsche}}, \bibinfo {author} {\bibfnamefont {H.~E.}\ \bibnamefont {Haber}}, \bibinfo {author} {\bibfnamefont {M.}~\bibnamefont {Hamel}}, \bibinfo {author} {\bibfnamefont {C.}~\bibnamefont {Hanhart}}, \bibinfo {author} {\bibfnamefont {S.}~\bibnamefont {Hashimoto}}, \bibinfo {author} {\bibfnamefont {Y.}~\bibnamefont {Hayato}}, \bibinfo {author} {\bibfnamefont {A.}~\bibnamefont {Hebecker}}, \bibinfo {author}
  {\bibfnamefont {S.}~\bibnamefont {Heinemeyer}}, \bibinfo {author} {\bibfnamefont {J.~J.}\ \bibnamefont {{Hern{\'a}ndez-Rey}}}, \bibinfo {author} {\bibfnamefont {K.}~\bibnamefont {Hikasa}}, \bibinfo {author} {\bibfnamefont {J.}~\bibnamefont {Hisano}}, \bibinfo {author} {\bibfnamefont {A.}~\bibnamefont {H{\"o}cker}}, \bibinfo {author} {\bibfnamefont {J.}~\bibnamefont {Holder}}, \bibinfo {author} {\bibfnamefont {L.}~\bibnamefont {Hsu}}, \bibinfo {author} {\bibfnamefont {J.}~\bibnamefont {Huston}}, \bibinfo {author} {\bibfnamefont {T.}~\bibnamefont {Hyodo}}, \bibinfo {author} {\bibfnamefont {A.}~\bibnamefont {Ianni}}, \bibinfo {author} {\bibfnamefont {M.}~\bibnamefont {Kado}}, \bibinfo {author} {\bibfnamefont {M.}~\bibnamefont {Karliner}}, \bibinfo {author} {\bibfnamefont {U.~F.}\ \bibnamefont {Katz}}, \bibinfo {author} {\bibfnamefont {M.}~\bibnamefont {Kenzie}}, \bibinfo {author} {\bibfnamefont {V.~A.}\ \bibnamefont {Khoze}}, \bibinfo {author} {\bibfnamefont {S.~R.}\ \bibnamefont {Klein}}, \bibinfo {author}
  {\bibfnamefont {F.}~\bibnamefont {Krauss}}, \bibinfo {author} {\bibfnamefont {M.}~\bibnamefont {Kreps}}, \bibinfo {author} {\bibfnamefont {P.}~\bibnamefont {Kri{\v z}an}}, \bibinfo {author} {\bibfnamefont {B.}~\bibnamefont {Krusche}}, \bibinfo {author} {\bibfnamefont {Y.}~\bibnamefont {Kwon}}, \bibinfo {author} {\bibfnamefont {O.}~\bibnamefont {Lahav}}, \bibinfo {author} {\bibfnamefont {J.}~\bibnamefont {Laiho}}, \bibinfo {author} {\bibfnamefont {L.~P.}\ \bibnamefont {Lellouch}}, \bibinfo {author} {\bibfnamefont {J.}~\bibnamefont {Lesgourgues}}, \bibinfo {author} {\bibfnamefont {A.~R.}\ \bibnamefont {Liddle}}, \bibinfo {author} {\bibfnamefont {Z.}~\bibnamefont {Ligeti}}, \bibinfo {author} {\bibfnamefont {C.-J.}\ \bibnamefont {Lin}}, \bibinfo {author} {\bibfnamefont {C.}~\bibnamefont {Lippmann}}, \bibinfo {author} {\bibfnamefont {T.~M.}\ \bibnamefont {Liss}}, \bibinfo {author} {\bibfnamefont {L.}~\bibnamefont {Littenberg}}, \bibinfo {author} {\bibfnamefont {C.}~\bibnamefont {Louren{\c c}o}}, \bibinfo
  {author} {\bibfnamefont {K.~S.}\ \bibnamefont {Lugovsky}}, \bibinfo {author} {\bibfnamefont {S.~B.}\ \bibnamefont {Lugovsky}}, \bibinfo {author} {\bibfnamefont {A.}~\bibnamefont {Lusiani}}, \bibinfo {author} {\bibfnamefont {Y.}~\bibnamefont {Makida}}, \bibinfo {author} {\bibfnamefont {F.}~\bibnamefont {Maltoni}}, \bibinfo {author} {\bibfnamefont {T.}~\bibnamefont {Mannel}}, \bibinfo {author} {\bibfnamefont {A.~V.}\ \bibnamefont {Manohar}}, \bibinfo {author} {\bibfnamefont {W.~J.}\ \bibnamefont {Marciano}}, \bibinfo {author} {\bibfnamefont {A.}~\bibnamefont {Masoni}}, \bibinfo {author} {\bibfnamefont {J.}~\bibnamefont {Matthews}}, \bibinfo {author} {\bibfnamefont {U.-G.}\ \bibnamefont {Mei{\ss}ner}}, \bibinfo {author} {\bibfnamefont {I.-A.}\ \bibnamefont {{Melzer-Pellmann}}}, \bibinfo {author} {\bibfnamefont {M.}~\bibnamefont {Mikhasenko}}, \bibinfo {author} {\bibfnamefont {D.~J.}\ \bibnamefont {Miller}}, \bibinfo {author} {\bibfnamefont {D.}~\bibnamefont {Milstead}}, \bibinfo {author} {\bibfnamefont
  {R.~E.}\ \bibnamefont {Mitchell}}, \bibinfo {author} {\bibfnamefont {K.}~\bibnamefont {M{\"o}nig}}, \bibinfo {author} {\bibfnamefont {P.}~\bibnamefont {Molaro}}, \bibinfo {author} {\bibfnamefont {F.}~\bibnamefont {Moortgat}}, \bibinfo {author} {\bibfnamefont {M.}~\bibnamefont {Moskovic}}, \bibinfo {author} {\bibfnamefont {K.}~\bibnamefont {Nakamura}}, \bibinfo {author} {\bibfnamefont {M.}~\bibnamefont {Narain}}, \bibinfo {author} {\bibfnamefont {P.}~\bibnamefont {Nason}}, \bibinfo {author} {\bibfnamefont {S.}~\bibnamefont {Navas}}, \bibinfo {author} {\bibfnamefont {A.}~\bibnamefont {Nelles}}, \bibinfo {author} {\bibfnamefont {M.}~\bibnamefont {Neubert}}, \bibinfo {author} {\bibfnamefont {P.}~\bibnamefont {Nevski}}, \bibinfo {author} {\bibfnamefont {Y.}~\bibnamefont {Nir}}, \bibinfo {author} {\bibfnamefont {K.~A.}\ \bibnamefont {Olive}}, \bibinfo {author} {\bibfnamefont {C.}~\bibnamefont {Patrignani}}, \bibinfo {author} {\bibfnamefont {J.~A.}\ \bibnamefont {Peacock}}, \bibinfo {author} {\bibfnamefont
  {V.~A.}\ \bibnamefont {Petrov}}, \bibinfo {author} {\bibfnamefont {E.}~\bibnamefont {Pianori}}, \bibinfo {author} {\bibfnamefont {A.}~\bibnamefont {Pich}}, \bibinfo {author} {\bibfnamefont {A.}~\bibnamefont {Piepke}}, \bibinfo {author} {\bibfnamefont {F.}~\bibnamefont {Pietropaolo}}, \bibinfo {author} {\bibfnamefont {A.}~\bibnamefont {Pomarol}}, \bibinfo {author} {\bibfnamefont {S.}~\bibnamefont {Pordes}}, \bibinfo {author} {\bibfnamefont {S.}~\bibnamefont {Profumo}}, \bibinfo {author} {\bibfnamefont {A.}~\bibnamefont {Quadt}}, \bibinfo {author} {\bibfnamefont {K.}~\bibnamefont {Rabbertz}}, \bibinfo {author} {\bibfnamefont {J.}~\bibnamefont {Rademacker}}, \bibinfo {author} {\bibfnamefont {G.}~\bibnamefont {Raffelt}}, \bibinfo {author} {\bibfnamefont {M.}~\bibnamefont {{Ramsey-Musolf}}}, \bibinfo {author} {\bibfnamefont {B.~N.}\ \bibnamefont {Ratcliff}}, \bibinfo {author} {\bibfnamefont {P.}~\bibnamefont {Richardson}}, \bibinfo {author} {\bibfnamefont {A.}~\bibnamefont {Ringwald}}, \bibinfo {author}
  {\bibfnamefont {D.~J.}\ \bibnamefont {Robinson}}, \bibinfo {author} {\bibfnamefont {S.}~\bibnamefont {Roesler}}, \bibinfo {author} {\bibfnamefont {S.}~\bibnamefont {Rolli}}, \bibinfo {author} {\bibfnamefont {A.}~\bibnamefont {Romaniouk}}, \bibinfo {author} {\bibfnamefont {L.~J.}\ \bibnamefont {Rosenberg}}, \bibinfo {author} {\bibfnamefont {J.~L.}\ \bibnamefont {Rosner}}, \bibinfo {author} {\bibfnamefont {G.}~\bibnamefont {Rybka}}, \bibinfo {author} {\bibfnamefont {M.~G.}\ \bibnamefont {Ryskin}}, \bibinfo {author} {\bibfnamefont {R.~A.}\ \bibnamefont {Ryutin}}, \bibinfo {author} {\bibfnamefont {Y.}~\bibnamefont {Sakai}}, \bibinfo {author} {\bibfnamefont {S.}~\bibnamefont {Sarkar}}, \bibinfo {author} {\bibfnamefont {F.}~\bibnamefont {Sauli}}, \bibinfo {author} {\bibfnamefont {O.}~\bibnamefont {Schneider}}, \bibinfo {author} {\bibfnamefont {S.}~\bibnamefont {Sch{\"o}nert}}, \bibinfo {author} {\bibfnamefont {K.}~\bibnamefont {Scholberg}}, \bibinfo {author} {\bibfnamefont {A.~J.}\ \bibnamefont {Schwartz}},
  \bibinfo {author} {\bibfnamefont {J.}~\bibnamefont {Schwiening}}, \bibinfo {author} {\bibfnamefont {D.}~\bibnamefont {Scott}}, \bibinfo {author} {\bibfnamefont {F.}~\bibnamefont {Sefkow}}, \bibinfo {author} {\bibfnamefont {U.}~\bibnamefont {Seljak}}, \bibinfo {author} {\bibfnamefont {V.}~\bibnamefont {Sharma}}, \bibinfo {author} {\bibfnamefont {S.~R.}\ \bibnamefont {Sharpe}}, \bibinfo {author} {\bibfnamefont {V.}~\bibnamefont {Shiltsev}}, \bibinfo {author} {\bibfnamefont {G.}~\bibnamefont {Signorelli}}, \bibinfo {author} {\bibfnamefont {M.}~\bibnamefont {Silari}}, \bibinfo {author} {\bibfnamefont {F.}~\bibnamefont {Simon}}, \bibinfo {author} {\bibfnamefont {T.}~\bibnamefont {Sj{\"o}strand}}, \bibinfo {author} {\bibfnamefont {P.}~\bibnamefont {Skands}}, \bibinfo {author} {\bibfnamefont {T.}~\bibnamefont {Skwarnicki}}, \bibinfo {author} {\bibfnamefont {G.~F.}\ \bibnamefont {Smoot}}, \bibinfo {author} {\bibfnamefont {A.}~\bibnamefont {Soffer}}, \bibinfo {author} {\bibfnamefont {M.~S.}\ \bibnamefont {Sozzi}},
  \bibinfo {author} {\bibfnamefont {S.}~\bibnamefont {Spanier}}, \bibinfo {author} {\bibfnamefont {C.}~\bibnamefont {Spiering}}, \bibinfo {author} {\bibfnamefont {A.}~\bibnamefont {Stahl}}, \bibinfo {author} {\bibfnamefont {S.~L.}\ \bibnamefont {Stone}}, \bibinfo {author} {\bibfnamefont {Y.}~\bibnamefont {Sumino}}, \bibinfo {author} {\bibfnamefont {M.~J.}\ \bibnamefont {Syphers}}, \bibinfo {author} {\bibfnamefont {F.}~\bibnamefont {Takahashi}}, \bibinfo {author} {\bibfnamefont {M.}~\bibnamefont {Tanabashi}}, \bibinfo {author} {\bibfnamefont {J.}~\bibnamefont {Tanaka}}, \bibinfo {author} {\bibfnamefont {M.}~\bibnamefont {Ta{\v s}evsk{\'y}}}, \bibinfo {author} {\bibfnamefont {K.}~\bibnamefont {Terao}}, \bibinfo {author} {\bibfnamefont {K.}~\bibnamefont {Terashi}}, \bibinfo {author} {\bibfnamefont {J.}~\bibnamefont {Terning}}, \bibinfo {author} {\bibfnamefont {R.~S.}\ \bibnamefont {Thorne}}, \bibinfo {author} {\bibfnamefont {M.}~\bibnamefont {Titov}}, \bibinfo {author} {\bibfnamefont {N.~P.}\ \bibnamefont
  {Tkachenko}}, \bibinfo {author} {\bibfnamefont {D.~R.}\ \bibnamefont {Tovey}}, \bibinfo {author} {\bibfnamefont {K.}~\bibnamefont {Trabelsi}}, \bibinfo {author} {\bibfnamefont {P.}~\bibnamefont {Urquijo}}, \bibinfo {author} {\bibfnamefont {G.}~\bibnamefont {Valencia}}, \bibinfo {author} {\bibfnamefont {R.}~\bibnamefont {{Van de Water}}}, \bibinfo {author} {\bibfnamefont {N.}~\bibnamefont {Varelas}}, \bibinfo {author} {\bibfnamefont {G.}~\bibnamefont {Venanzoni}}, \bibinfo {author} {\bibfnamefont {L.}~\bibnamefont {Verde}}, \bibinfo {author} {\bibfnamefont {I.}~\bibnamefont {Vivarelli}}, \bibinfo {author} {\bibfnamefont {P.}~\bibnamefont {Vogel}}, \bibinfo {author} {\bibfnamefont {W.}~\bibnamefont {Vogelsang}}, \bibinfo {author} {\bibfnamefont {V.}~\bibnamefont {Vorobyev}}, \bibinfo {author} {\bibfnamefont {S.~P.}\ \bibnamefont {Wakely}}, \bibinfo {author} {\bibfnamefont {W.}~\bibnamefont {Walkowiak}}, \bibinfo {author} {\bibfnamefont {C.~W.}\ \bibnamefont {Walter}}, \bibinfo {author} {\bibfnamefont
  {D.}~\bibnamefont {Wands}}, \bibinfo {author} {\bibfnamefont {D.~H.}\ \bibnamefont {Weinberg}}, \bibinfo {author} {\bibfnamefont {E.~J.}\ \bibnamefont {Weinberg}}, \bibinfo {author} {\bibfnamefont {N.}~\bibnamefont {Wermes}}, \bibinfo {author} {\bibfnamefont {M.}~\bibnamefont {White}}, \bibinfo {author} {\bibfnamefont {L.~R.}\ \bibnamefont {Wiencke}}, \bibinfo {author} {\bibfnamefont {S.}~\bibnamefont {Willocq}}, \bibinfo {author} {\bibfnamefont {C.~G.}\ \bibnamefont {Wohl}}, \bibinfo {author} {\bibfnamefont {C.~L.}\ \bibnamefont {Woody}}, \bibinfo {author} {\bibfnamefont {W.-M.}\ \bibnamefont {Yao}}, \bibinfo {author} {\bibfnamefont {M.}~\bibnamefont {Yokoyama}}, \bibinfo {author} {\bibfnamefont {R.}~\bibnamefont {Yoshida}}, \bibinfo {author} {\bibfnamefont {G.}~\bibnamefont {Zanderighi}}, \bibinfo {author} {\bibfnamefont {G.~P.}\ \bibnamefont {Zeller}}, \bibinfo {author} {\bibfnamefont {O.~V.}\ \bibnamefont {Zenin}}, \bibinfo {author} {\bibfnamefont {R.-Y.}\ \bibnamefont {Zhu}}, \bibinfo {author}
  {\bibfnamefont {S.-L.}\ \bibnamefont {Zhu}}, \bibinfo {author} {\bibfnamefont {F.}~\bibnamefont {Zimmermann}},\ and\ \bibinfo {author} {\bibfnamefont {P.~A.}\ \bibnamefont {Zyla}},\ }\bibfield  {title} {\bibinfo {title} {Review of {{Particle Physics}}},\ }\href {https://doi.org/10.1093/ptep/ptac097} {\bibfield  {journal} {\bibinfo  {journal} {Progress of Theoretical and Experimental Physics}\ }\textbf {\bibinfo {volume} {2022}},\ \bibinfo {pages} {083C01} (\bibinfo {year} {2022})}\BibitemShut {NoStop}%
\bibitem [{\citenamefont {Roithmayr}\ and\ \citenamefont {Hodges}(2016)}]{Roithmayr2016}%
  \BibitemOpen
  \bibfield  {author} {\bibinfo {author} {\bibfnamefont {C.~M.}\ \bibnamefont {Roithmayr}}\ and\ \bibinfo {author} {\bibfnamefont {D.~H.}\ \bibnamefont {Hodges}},\ }\href {https://doi.org/10.1017/CBO9781139047524} {\emph {\bibinfo {title} {Dynamics: {{Theory}} and {{Application}} of {{Kane}}'s {{Method}}}}}\ (\bibinfo  {publisher} {Cambridge University Press},\ \bibinfo {address} {Cambridge},\ \bibinfo {year} {2016})\BibitemShut {NoStop}%
\end{thebibliography}
\end{document}